\documentstyle[emulateapj,apjfonts,psfig,natbib,amsmath]{article}

\newcommand{\lsim}{\lesssim}

\begin{document}

\def\cfa{1}
\def\churchill{2}
\def\weizmann{3}
\def\ucb{4}
\def\psu{5}
\def\sdsu{6}
\def\lco{7}
\def\ucsb{8}
\def\ut{9}

\title{\large The First Systematic Study of Type Ibc Supernova 
multi-band light curves}

\author{Maria~R.~Drout\altaffilmark{\cfa}$^,$\altaffilmark{2}, Alicia~M.~Soderberg\altaffilmark{\cfa}, Avishay Gal-Yam\altaffilmark{\weizmann}, S.~Bradley Cenko\altaffilmark{\ucb}, Derek~B.~Fox\altaffilmark{\psu}, Douglas~C.~Leonard\altaffilmark{\sdsu}, David~J.~Sand\altaffilmark{\lco}$^,$\altaffilmark{8}, Dae-Sik~Moon\altaffilmark{\ut}, Iair~Arcavi\altaffilmark{\weizmann}, and Yoav~Green\altaffilmark{\weizmann}}

\altaffiltext{\cfa}{Harvard-Smithsonian Center for Astrophysics, 60
  Garden Street, MS-51, Cambridge, MA 02138, USA}
\altaffiltext{\churchill}{Churchill College, Cambridge, CB3 0DS, UK}
\altaffiltext{\weizmann}{Benoziyo Center for Astrophysics, Faculty of Physics, The Weizmann Institute of Science, Rehovot 76100, Israel}
\altaffiltext{\ucb}{Department of Astronomy, University of California, Berkeley, CA 94720-3411, USA}
\altaffiltext{\psu}{Department of Astronomy and Astrophysics, Pennsylvania State University, 525 Davey Lab, University Park, PA 16802, USA}
\altaffiltext{\sdsu}{Department of Astronomy, San Diego State University, San Diego, CA 92182, USA}
\altaffiltext{\lco}{Las Cumbres Observatory Global Telescope Network, 6740
Cortona Drive, Suite 102, Santa Barbara, CA 93117, USA}
\altaffiltext{\ucsb}{Department of Physics, Broida Hall, University of
California, Santa Barbara, CA 93106, USA}
\altaffiltext{\ut}{Department of Astronomy and Astrophysics, University of Toronto, Toronto, ON M5S 3H4}

\begin{abstract} 
We present detailed optical photometry for 25 Type Ibc supernovae (SNe
Ibc) within $d\approx 150$ Mpc obtained with the robotic Palomar
60-inch telescope in 2004-2007.  This study represents the first
uniform, systematic, and statistical sample of multi-band SNe Ibc
light curves available to date.  We correct the light curves for host
galaxy extinction using a new technique based on the photometric color
evolution, namely, we show that the $(V-R)$ color of
extinction-corrected SNe Ibc at $\Delta t\approx 10$ d after $V-$band
maximum is tightly distributed, $\langle (V-R)_{V10}\rangle =0.26\pm
0.06$ mag.  Using this technique, we find that SNe Ibc typically
suffer from significant host galaxy extinction, $\langle
E(B-V)\rangle \approx 0.4$ mag.  A comparison of the
extinction-corrected light curves for helium-rich (Type Ib) and
helium-poor (Type Ic) SNe reveals that they are statistically
indistinguishable, both in luminosity and decline rate.  We report
peak absolute magnitudes of $\langle M_R\rangle =-17.9\pm 0.9$ mag and
$\langle M_R\rangle =-18.3\pm 0.6$ mag for SNe Ib and Ic,
respectively.  Focusing on the broad-lined (BL) SNe Ic, we find that
they are more luminous than the normal SNe Ibc sample, $\langle
M_R\rangle =-19.0\pm 1.1$ mag, with a probability of only 1.6\% that
they are drawn from the same population of explosions.  By comparing
the peak absolute magnitudes of SNe Ic-BL with those inferred for
local engine-driven explosions (GRB-SN\,1998bw, XRF-SN\,2006aj, and
SN\,2009bb) we find a $25\%$ probability that relativistic SNe are
drawn from the overall SNe Ic-BL population.  Finally, we fit analytic
models to the light curves to derive typical $^{56}$Ni masses of
$M_{\rm Ni}\approx 0.2$ and $0.5~M_{\odot}$ for SNe Ibc and SNe Ic-BL,
respectively.  With reasonable assumptions for the photospheric
velocities, we further extract kinetic energy and ejecta mass values
of $M_{\rm ej}\approx 2~M_{\odot}$ and $E_K\approx 10^{51}$ erg for
SNe Ibc, while for SNe Ic-BL we find higher values, $M_{\rm ej}\approx
5~M_{\odot}$ and $E_K\approx 10^{52}$ erg.  We discuss the
implications for the progenitors of SNe Ibc and their relation to
those of engine-driven explosions.
\end{abstract}


\section{\label{Intro} Introduction}

Type Ibc supernovae (SNe Ibc) are a distinct class of catastrophic
stellar explosions, identified by the lack of hydrogen features and
the presence of weak silicon features in their optical spectra (see
\citealt{fil97} for a review).  In the traditional classification
scheme, the presence and strength of helium absorption features {\it
may} further differentiate the class of SNe Ibc into two sub-classes:
Ib (He-rich) and Ic (He-poor; see \citealt{mfl+01}).  Based on their
proximity to star-forming regions \citep{pf87,vhf96}, and their often
strong radio/X-ray emission attributed to circumstellar interaction
(\citealt{wsp+86,che98,bkf+03,cf06,ams07}), SNe Ibc are now recognized
as core-collapse explosions of massive stars\footnote{There are
notable exceptions, however, which comprise only a small fraction of the
bulk SN Ibc population and include SN\,2005E-like white dwarf
events \citep{pgm+10}, pair-instability SNe \citep{gmo+09} and
pulsational pair-instability SNe \citep{qkm+09}.} stripped of their
hydrogen (and often helium) layers
\citep{emn+85,fs85,wl85,uk85,cwb+96,whw02}.

Unlike other core-collapse SNe, however, the progenitor systems of SNe
Ibc have yet to be directly identified in pre-explosion images (see
\citealt{sma09} for a review).  Motivated by their observed properties, 
two primary progenitor channels have been proposed: (i) isolated and
massive Wolf-Rayet (WR) stars that ejected their outer envelope
through radiation-driven stellar winds
\citep{bs86,wlw95}, and (ii) lower mass helium stars in close binary
systems characterized by mass transfer \citep{wl85,pjh92,ywl10}.
It remains open to debate whether one or both of these progenitor channels
gives rise to the observed SN Ibc population.  

Recent host galaxy studies suggest dissimilar progenitor channels for
SNe Ib and Ic based on the properties (e.g.,~luminosity, chemistry,
star-formation rate) of the explosion sites
(\citealt{kkp08,psb08,agk+10,mbf+10} but see \citealt{acj+10}).  These
results indicate that progenitor age, mass and/or metallicity drive
the observed optical spectroscopic diversity seen for SNe Ibc, namely
the presence/strength of the He I features.  Furthermore, the
discovery that some SNe Ic produce relativistic outflows that give
rise to long-duration gamma-ray bursts (GRBs) points to a unique
progenitor channel for a small fraction of these events (see
\citealt{Summary} for a review).  These relativistic outflows are
powered by an additional energy source, a central engine, commonly
thought to be a rapidly rotating and accreting compact object
\citep{mwh01} or a magnetar \citep{tcq04}. We refer to these
explosions as ``engine-driven''.

In this paper we search for additional progenitor clues based on the
optical light curves of SNe Ibc.  Free from the massive envelopes
and/or circumstellar shells associated with other core-collapse
explosions, the photospheric emission of ``naked'' SNe Ibc is powered
by the radioactive decay of freshly synthesized $^{56}$Ni. This
optical signal reaches maximum intensity within a month of the
explosion, producing parabola-shaped light curves that resemble those
of SNe Ia.  In the simple picture, three physical parameters determine
the shape of the light curves and peak luminosity: the mass of newly
synthesized $^{56}$Ni ($M_{\rm Ni}$), the total ejecta mass ($M_{\rm
ej}$), and the kinetic energy of the explosion ($E_K$; see
\citealt{arn82}).  In addition, spectroscopic measurements of the 
photospheric velocity, $v_{\rm ph}$,directly constrain the quantity,
$\sqrt{E_K/M_{\rm ej}}$.  Since the explosion parameters are imprinted
on the optical emission, their detailed study can reveal clues as to
the properties and diversity within the progenitor channel(s).

Literature studies of the photospheric emission from SNe Ibc are
relatively few \citep{rbb06} and tend to focus on detailed photometric
and/or spectroscopic observations of single explosions.  These studies
often target those SNe Ibc with extreme properties, such as the 5-10\%
of the population showing evidence for unusually broad absorption
features (``broad-lined'', BL; \citealt{pmn+04}) and photospheric
velocities of $v_{\rm ph}\gtrsim 20,000~\rm km~s^{-1}$
(e.g.,~SN\,2003jd, \citealt{vbc+08}). Such events are
spectroscopically similar to GRB-SNe.  Preliminary comparisons of the
optical luminosities of ordinary SNe Ibc and GRB-SNe revealed overlap
between the two samples, with GRB-SNe populating the upper end of the
SN luminosity distribution \citep{graph,Summary,ric09}.  However, a
more comprehensive and uniform sample of ordinary SNe Ibc is required
for a robust comparison.

More recently, a set of two dozen unfiltered light curves for SNe Ibc
was reported by the Lick Observatory Supernova Search (LOSS) within an
estimated completeness limit of $d\approx 60$ Mpc \citep{llc+10}. The
LOSS light curves reveal an overall dispersion in decay rate and peak
magnitude with some evidence that SNe Ib are more luminous than SNe
Ic.  However, due to the lack of photometric color information and
often poor light curve coverage of the LOSS SNe Ibc, a robust and
systematic study of their multi-band light curves and host
galaxy extinction-corrected peak luminosities, and an extraction of their
explosion parameters remain outstanding projects.

%GRB-associated SNe (hereafter, GRB-SNe) are distinguished from
%ordinary core-collapse explosions by the presense of an energetic,
%relativistic outflow that is powered by a central engine --  an accreting
%compact object \citep{coll3} or a magnetar (e.g.~\citealt{tcq04}).
%Such events are instrically rare, as evidenced by radio studies of
%optically-selected SNe Ibc which find relativistic outflows in 
%$0.7\%$ of the bulk population of SNe Ibc and at most 30\%$ of
%all SNe Ic-BL \citep{bkf+03,skn+06,snb+06,scp+10}.  

Motivated thus, between 2004 and 2007 we conducted a dedicated optical
monitoring campaign of nearby SNe Ibc with the robotic Palomar 60-inch
telescope (P60; \citealt{cfm+06}). The primary goal of this effort was
to observationally constrain the physical parameters of the explosion
($M_{\rm Ni}$, $M_{\rm ej}$, and $E_K$).  In this paper we present the
P60 light curves for a sample of $25$ SNe Ibc, representing the first
uniform and systematic multi-band study to date.  We present a novel
approach to correct the light curves for (often significant) host
galaxy extinction.  We compare the observed light curve properties to
those of SNe Ibc compiled from the literature and engine-driven SNe
within the same volume, and we derive the physical parameters for the
extended sample using a systematic modeling approach.  The parameters
are compared for SNe Ib, Ic and SNe Ic-BL with the goal of shedding
light on the progenitor channels.  Finally, we provide our SNe Ibc
light curves and light curve templates to the community to facilitate
on-going and future efforts to distinguish between SNe Ia and SN Ibc
at larger distances and/or without spectroscopic diagnostics.

\section{Observations and Photometry}
\label{sec:OR}

We used the robotic Palomar 60-inch telescope (P60;
\citealt{cfm+06}) to obtain multi-epoch, multi-band optical observations for
our SN follow-up campaign.  P60 is a queue-scheduled facility
dedicated to rapid-response and monitoring observations of optical
transients.  We selected targets for our study from the sample of
local ($d\lesssim 150$ Mpc) SN discoveries publicly announced via the
on-line circulars, including the International Astronomical Union
Circulars (IAUC), Central Bureau for Electronic Telegrams (CBET), and
the Astronomers Telegram (ATEL) between Fall 2004 and Spring 2007.
We note that the majority of these SNe were discovered by SN searches
that target the most luminous (and likely metal-rich) nearby galaxies (e.g., LOSS).

Here we present the P60 data from our campaign for all 25 SNe
spectroscopically classified and reported in the on-line circulars as
Type Ib (11 events) and Ic (11 events) and Ic-BL (3 events) for which
we were able to obtain at least eight photometric measurements.  We note
that two SNe Ib in our sample subsequently developed hydrogen features
prompting their re-classification as Type IIb and we include them here
given their connection to the Type Ib class (e.g,~\citealt{cs10}).

The majority of the SNe in this sample (15 of the 25) were observed
between Fall 2005 and Spring 2007 (see \citealt{ams07}).  An
additional 10 SNe were observed as part of the Caltech Core-Collapse
supernova Program (CCCP; \citealt{gcf+07}) which carried out
photometric and spectroscopic follow-up of all young ($\lsim 30$ d
since explosion) and nearby ($d\lsim 120$ Mpc) SNe of Types Ibc or II
between July 2004 and Sep 2005.  We note that the CCCP data for the
peculiar, Ca-rich SN\,2005E were reported separately by \citet{pgm+10}
and so we do not include them here.

The properties of each SN in our sample are summarized in
Table~\ref{tab:list}, including spectroscopic classification, host
galaxy distance, and Galactic extinction \citep{sfd98}.  For
SNe with conflicting classification reports, we adopt the latest
classification since He I absorption features often become more
prominent with age (e.g., SN\,1999ex;
\citealt{hmp+02}). We adopt host galaxy distances from the NASA/IPAC
Extragalactic Database of cosmology independent distances
(NED-D\footnote{\tt http://nedwww.ipac.caltech.edu/Library/Distances/}) when available.  For the remaining objects
we adopt the NED distances after correction for Virgo, Great
Attractor, and Shapley Supercluster Infall and assuming $\rm H_0 = 73$ km
s$^{-1}$ Mpc$^{-1}$ \citep{mhf+00}.

\begin{figure}
\plotone{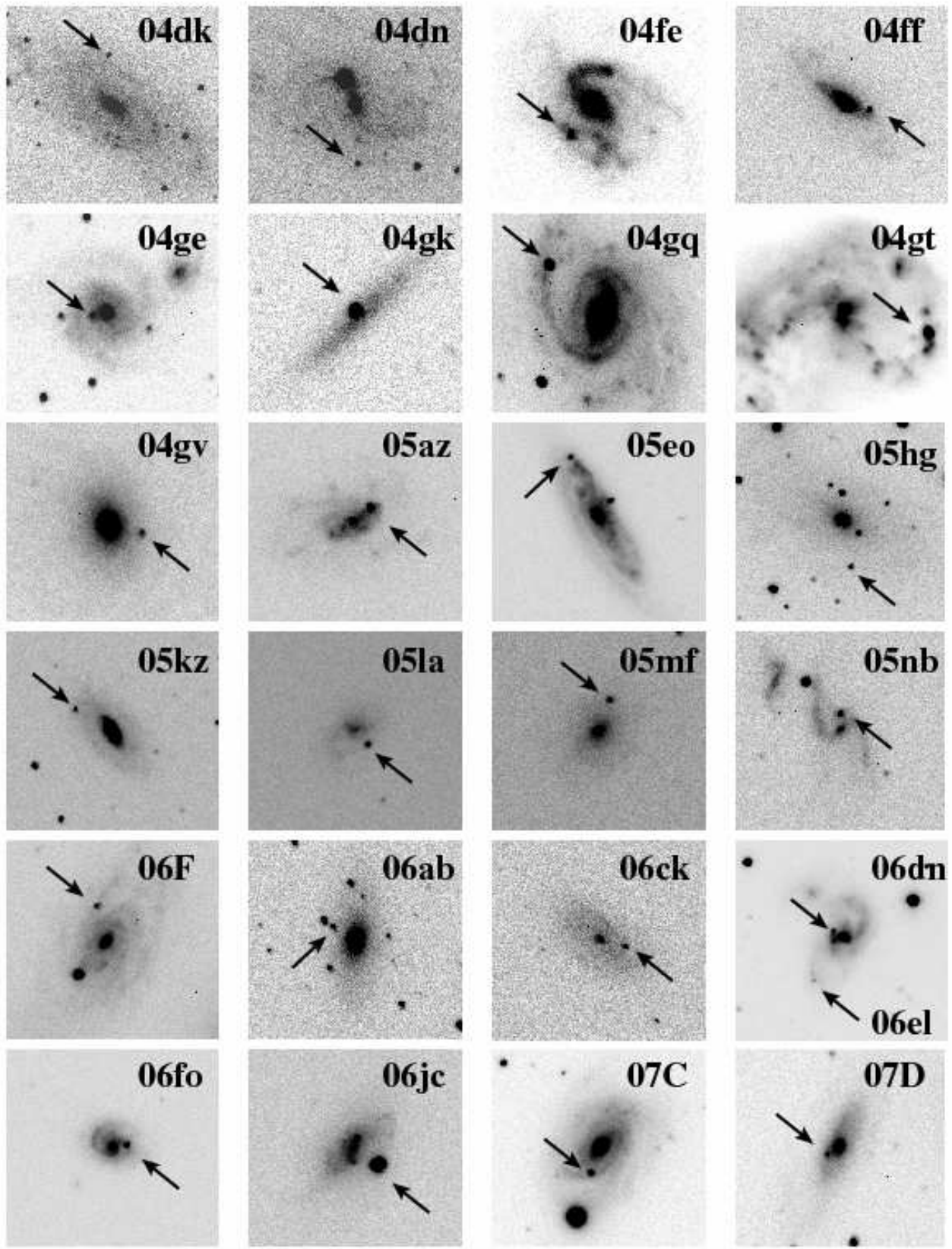}
\caption{P60 images of 25 SNe Ibc in our study. Each stamp is 1.5 x 1.5 arcmin and north is up, east is to the left.  The SNe are marked with arrows.}
\label{fig:stamps}
\end{figure}

For each newly discovered SN accessible to P60, we promptly initiated
nightly photometric follow-up in the Johnson $V-$ and $R-$bands.  The
sensitivity functions of these filters are described and displayed in
\citet{cfm+06}. Our monitoring campaign was optimized for a cadence of
$\sim 1$ day, with observations typically extending for several months
after discovery.  All P60 images were reduced in IRAF\footnote{IRAF is
distributed by the National Optical Astronomy Observatory, which is
operated by the Association for Research in Astronomy, Inc., under
cooperative agreement with the National Science Foundation.} using a
custom real-time reduction pipeline \citep{cfm+06}.  In
Figure~\ref{fig:stamps} we display a montage of the $1.5'\times 1.5'$
region around each SN as observed with P60.

We used PSF  photometry to extract the relative  magnitudes of the SNe
with  respect to  several stars  within the  full $13'\times  13'$ P60
field-of-view\footnote{We compare our  PSF photometry light curves for
SNe 2004dk and 2004dn  with those produced after digitally subtracting
the host galaxy emission using the  Common PSF Method and find them to
be comparable (Gal-Yam {\it  et al.}, in prep)}.  Absolute calibration
of the relative light curves was  performed using a combination of P60
observations of  Landolt standard fields and SDSS  photometry of field
stars.  We used the \citet{stk+02} transformations to convert the SDSS
{\it ugriz} photometry of field stars to the $V-$ and $R-$bands.

The P60 photometry for all 25 SNe is available in Table~\ref{tab:allphot} (machine-readable) and are characterized by typical uncertainties of $\sim
0.02$ to 0.05 mag per epoch.  We supplemented our photometry with
measurements from the literature, primarily drawn from the initial
discovery circulars, thereby providing early-time and often
pre-discovery flux measurements including upper limits (see
Table~\ref{tab:list} for references).  In this compilation, we have
assumed that unfiltered magnitudes from the circulars are roughly
equivalent to $R-$band measurements and assign to the detections an
uncertainty of $\pm 0.2$ mag.  In cases where extensive photometry is
available from the literature, we find that our photometry is fully
consistent (i.e., SN\,2006jc, \citealt{psm+07}; SN\,2005la, \citealt{pqs+08})

\subsection{\label{LC}Sample Definitions}

\begin{figure}
\plotone{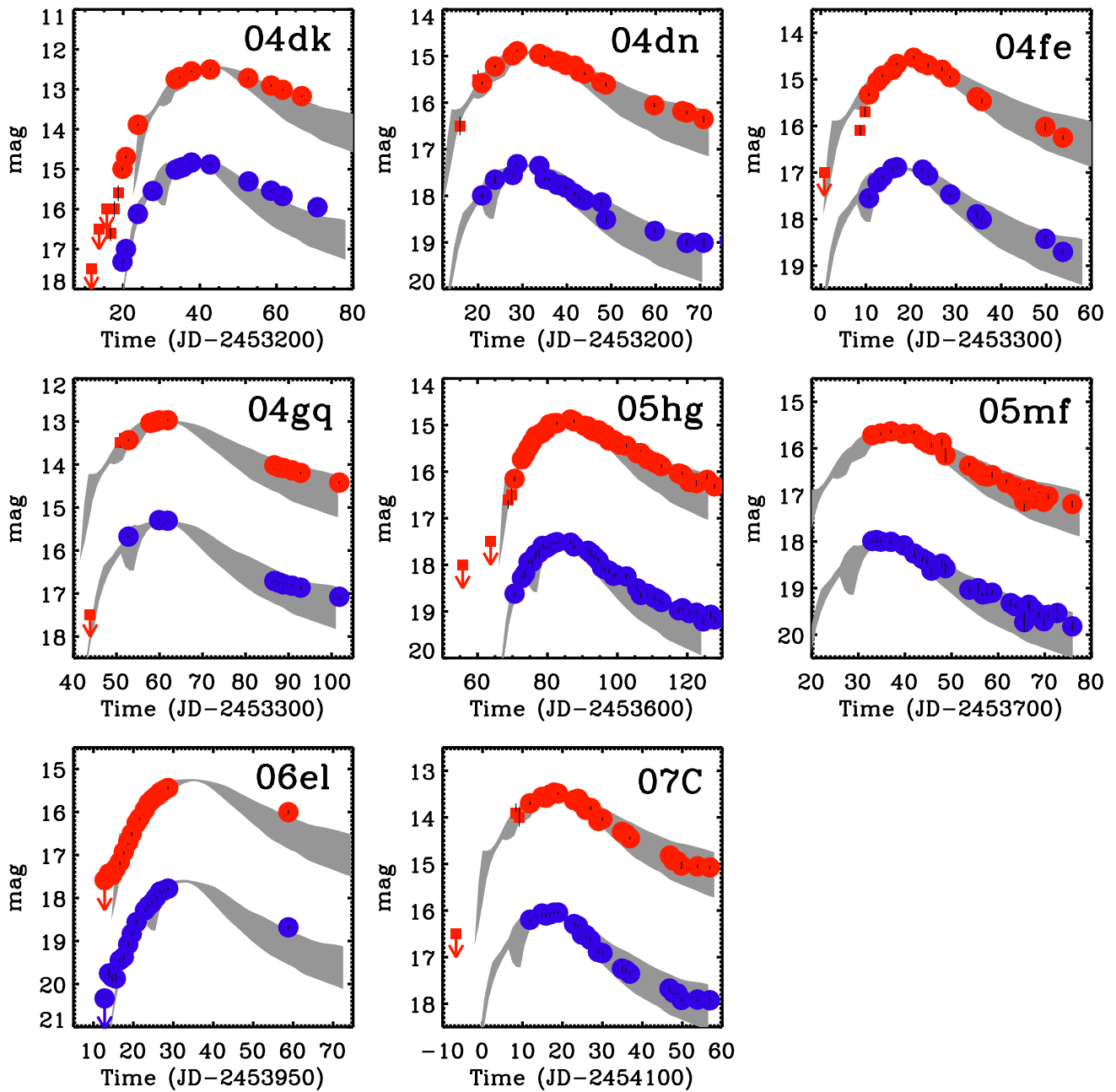}
\caption{The apparent $V-$band (blue) and $R-$band (red) light curves for our 
Gold sample of SNe Ibc.  This sample includes only those SNe for which
our P60 data reveal the epoch and flux of maximum light in both bands.
Circles designate P60 points and squares represent measurements drawn
from the discovery circulars (see Table~\ref{tab:list} for
references).  The $R-$band points have been brightened by 2 mag for
clarity.  We have made no corrections for extinction (Galactic or host
galaxy).  The grey curves are the template light curves constructed
from the sample of literature SNe Ibc and overplotted for comparison
(see \S\ref{sec:templates} and Figure~\ref{fig:VR_curves}).}

\label{fig:postage1}
\end{figure}

\begin{figure}
\plotone{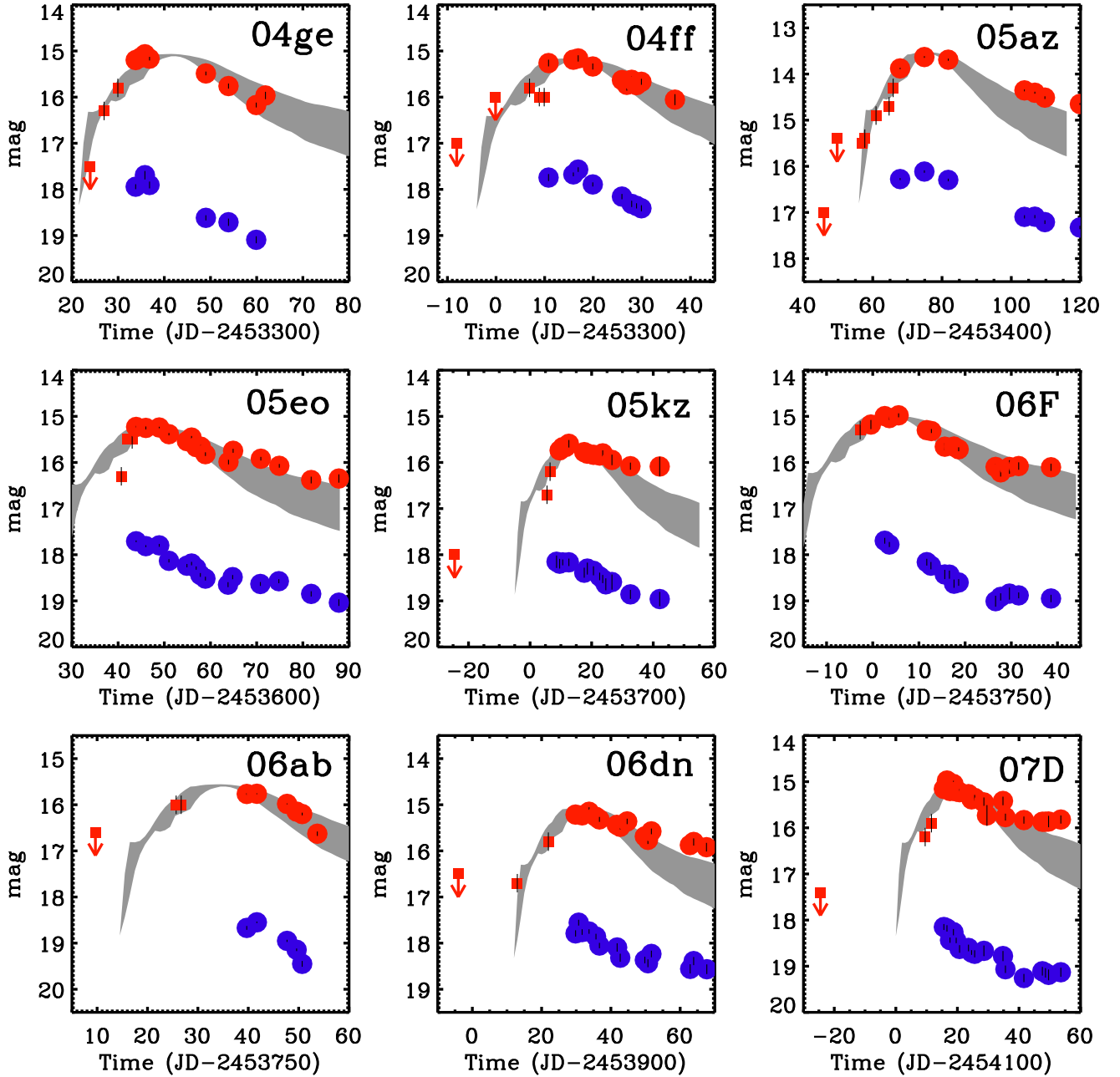}
\caption{The apparent $V-$band (blue) and $R-$band (red) light curves
for our Silver sample of SNe Ibc which includes SNe for which the
epoch and flux of maximum light in the $R-$band is well constrained by
supplementing our P60 measurements with photometry from the
literature.  Circles designate P60 points and squares represent
measurements adopted from the discovery circulars (see
Table~\ref{tab:list} for references).  The $R-$band points have been
brightened by 2 mag for clarity.  We have made no corrections for
extinction (Galactic or host galaxy).  The grey curves are the
template light curves constructed from the sample of literature SNe
Ibc and overplotted for comparison (see \S\ref{sec:templates} and
Figure~\ref{fig:VR_curves})}.
\label{fig:postage2}
\end{figure}

\begin{figure}
\plotone{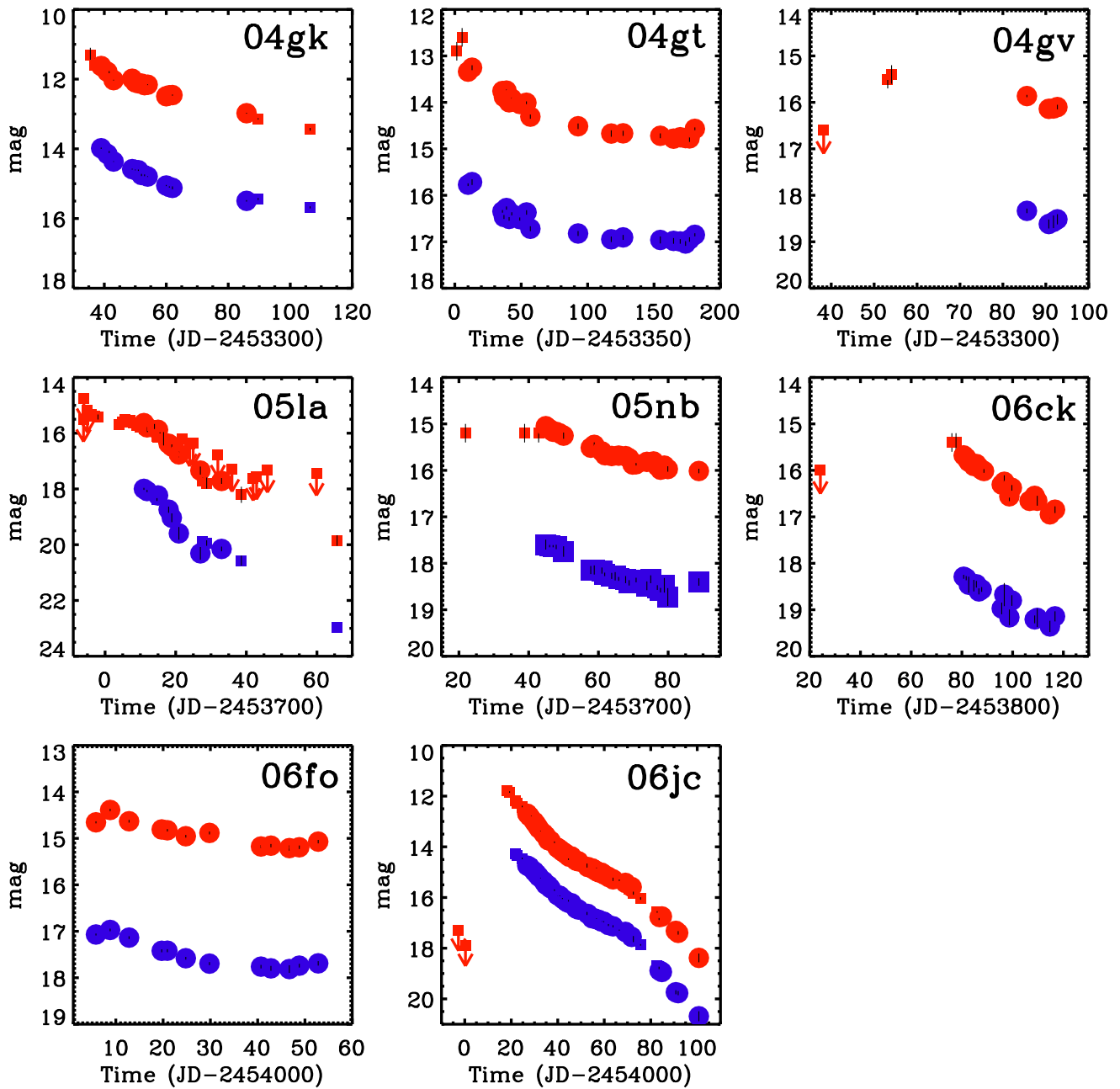}
\caption{The apparent $V-$band (blue) and $R-$band (red) light curves
for our Bronze sample of SNe Ibc which includes SNe for which the
epoch and flux of maximum light are poorly constrained.  Circles
designate P60 points and squares represent measurements drawn from the
literature including discovery circulars (see Table~\ref{tab:list})
and in the case of SNe 2005la and 2006jc, measurements were also drawn
from \citet{pqs+08} and \citet{psm+07}, respectively.  The $R-$band
points have been brightened by 2 mag for clarity. We have made no
corrections for extinction (Galactic or host galaxy).}
\label{fig:postage3}
\end{figure}

Our SN Ibc light curves can be divided into three groups, based on
constraints on the epoch and apparent magnitude of the light curve
maxima.  For the eight SNe in the Gold group, our P60 observations
covered the light curve maximum in both $V-$ and $R-$band
(Figure~\ref{fig:postage1}).  These data alone nearly double the
existing sample of well-studied SNe Ibc light curves available in the
literature.  The Silver group is composed of nine SNe for which the
$R-$band light curve maximum is observationally constrained through a
combination of P60 monitoring and literature measurements
(Figure~\ref{fig:postage2}).  Finally, there are eight SNe in our
Bronze group, for which the epoch and magnitude of the light curve
maximum is poorly constrained in both bands
(Figure~\ref{fig:postage3}).  As shown in
Figures~\ref{fig:postage1}-\ref{fig:postage3}, the light curve shapes
are diverse.

\section{Observed Light Curve Properties}
\label{sec:diag}

The shape of the SN Ibc light curves is fully determined by the three
explosion parameters, $E_K$, $M_{\rm ej}$, and $M_{\rm Ni}$
(e.g.,~\citealt{arn82}).  Specifically, the peak optical luminosity,
$L_{\rm peak}$, scales with the mass of $^{56}$Ni as $L_{\rm
peak}\propto M_{\rm Ni} \tau_c^{-1}$.  Here, $\tau_c\approx 8~M_{\rm ej,{\odot}}^{3/4}E_{K,51}^{-1/4}~\rm days$ is the characteristic
time or width of the light curve and our notation refers to a
normalization of $E_K$ by $10^{51}$ erg and of $M_{\rm ej}$ by
$M_{\odot}$ \citep{inm+03}.  light curves with larger $\tau_c$ values
have correspondingly slower post-maximum decay rates.  Therefore,
while $L_{\rm peak}$ is primarily determined by the fractional mass
of $^{56}$Ni, the width of the light curve (and, in turn, the decay
rate) is determined by $M_{\rm ej}$ and $E_K$.  In our study of SNe
Ibc light curves we therefore focus on their two main observable
properties, namely their peak absolute magnitudes and post-maximum
decay rates.  To enable a systematic comparison of these observables,
we first define a procedure and apply it to an existing set of
well-studied SNe Ibc available in the literature.

\subsection{The Sample of Literature SNe}
\label{sec:templates}

We compiled $V-$ and $R-$band photometry for the existing sample of
well-studied SNe Ibc within a similar distance of $d\lsim 150$ Mpc
that are available in the literature.\footnote{We note that with this
selection criteria, we exclude the rare, overluminous SNe Ibc
detected in magnitude-limited surveys, e.g., SN\,2007bi
\citep{gmo+09} and SN\,2005ap \citep{qkm+09}.}  Our selection criteria
required the light curve maxima to be well-sampled and the total
extinction to have been estimated using the equivalent width of the Na
I D lines and/or fitting of the broadband photometric spectrum.  After
excluding the SNe associated with GRBs (SN\,1998bw, \citealt{gvv+98};
SN\,2006aj, \citealt{pmm+06}) and the notably peculiar events (i.e.,
the Ca-rich SN\,2005E; \citealt{pgm+10}, the double-peaked SN\,2005bf;
\citealt{fcp+06}, and the SNe Ibn 2006jc and 2005la; \citealt{fsg+07,psm+07,pqs+08}), we
construct a sample of 10 SNe Ibc: 1994I, 1999ex, 2002ap, 2003jd,
2004aw, 2007Y, 2007gr, 2007ru, 2008D, and 2008ax.  In
Table~\ref{tab:templates} we compile the basic properties of this
literature sample including the SN Type, the Galactic extinction (from
\citealt{sfd98}), and the host galaxy name, distance estimate, and
inclination.  In line with the P60 sample, we adopt up to date NED-D
host galaxy distances for the literature sample when available.  In
Table~\ref{tab:templates_lc} we further list the epoch and apparent
magnitude of the $V-$ and $R-$band light curve maxima.  For three of
the literature SNe, the published light curves were obtained in the
SDSS $r$-band filter, and we converted these to $R$-band using the
transformations of \citet{stk+02}.  These are noted in
Table~\ref{tab:templates}.  We note that more detailed transformations
(e.g., S corrections; \citealt{shs+02}) are outside the scope of this
paper, and thus the derived $R-$band photometry for these three SNe
may include a small systematic uncertainty.

After correcting the light curves for the typically low Galactic
extinction ($\langle E(B-V)_{\rm Gal}\rangle\approx 0.06$ mag), we
derive peak absolute magnitudes in the $V-$ and $R-$bands and find
that they span $\sim 3-4$ magnitudes.  This has been noted previously
based on similar samples of SNe Ibc drawn from the literature
\citep{rbb06,graph}.  In the $R-$band, the range extends
from $M_{R,\rm peak}\approx -15.6$ to $-18.9$ mag.  Dividing the sample
into SNe Ib, Ic, and Ic-BL (4, 3, and 3 SNe, respectively) we
calculate mean values of $\langle M_{R,\rm peak}\rangle= -16.3\pm 0.6$
mag (Ib), $\langle M_{R,\rm peak}\rangle = -17.1\pm 0.4$ mag (Ic), and
$\langle M_{R,\rm peak}\rangle =-18.4\pm 0.8$ mag (Ic-BL).

\begin{figure}
\plotone{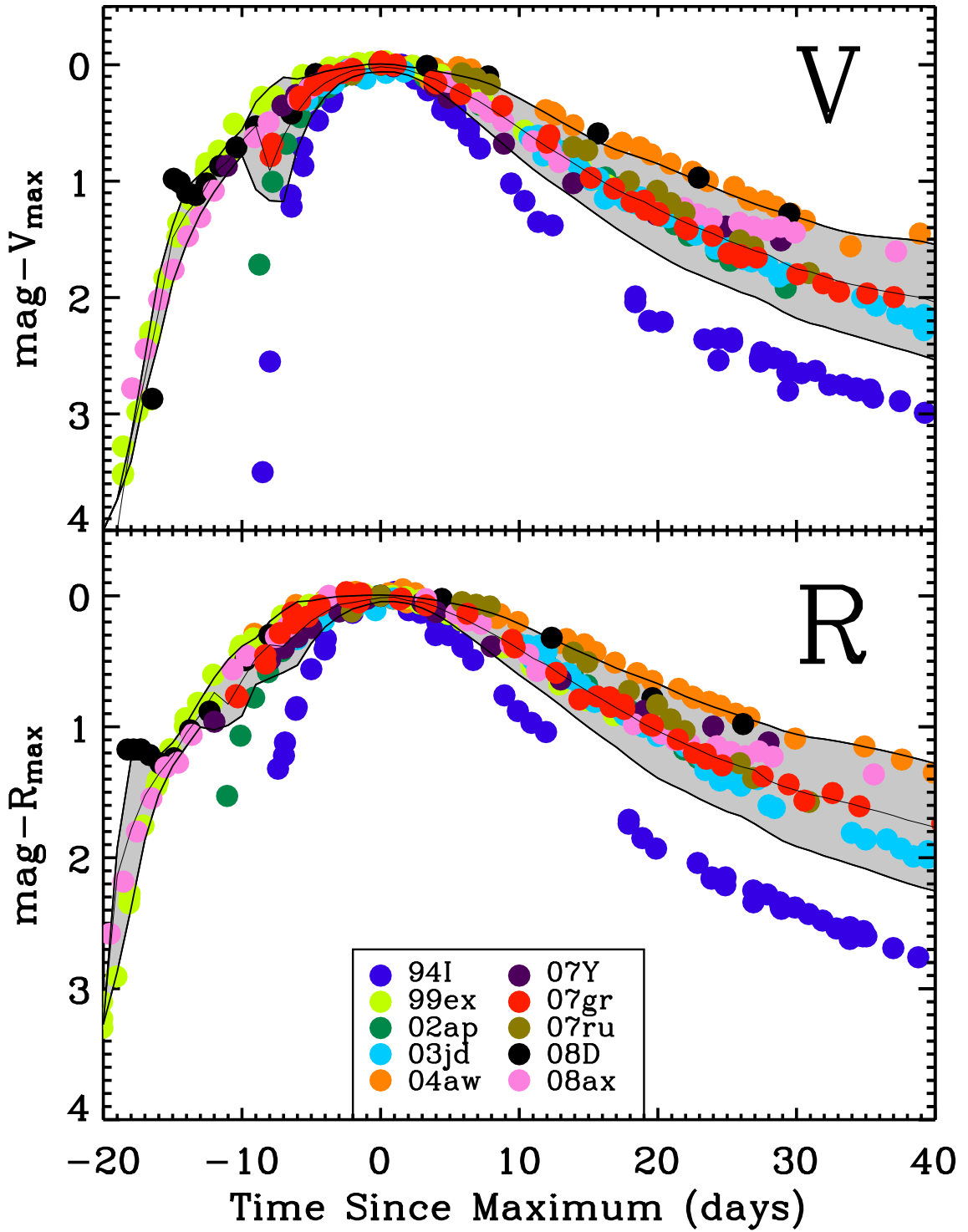}
\caption{We compare the $V-$ and $R-$band light curves for the sample of 
10 well-studied SNe Ibc currently available in the literature (see Table~\ref{tab:templates}).  A compilation of their $V-$band (top panel) and $R-$band (bottom panel) light curves are scaled to their respective epoch of $V_{\rm max}$($R_{\rm max}$) and used to construct a mean template light curve (grey curve). The width of the curve is derived from the $1\sigma$ deviation from the mean at each epoch.}   
\label{fig:VR_curves}
\end{figure}

We next compare the light curve shapes of the literature sample after
shifting each curve relative to the epoch and magnitude of maximum
light (Figure~\ref{fig:VR_curves}).  This figure highlights the
significant dispersion in the light curve widths (see also
\citealt{cw97a},\citealt{cw97b}).  We build template light curves in
the $V-$ and $R-$bands by interpolating over these normalized light
curves and extracting the weighted mean flux density for time
intervals spanning $-20$ to $+40$ days with respect to the epoch of
maximum light.  The uncertainties in the resulting templates are
derived from the standard deviation in the interpolated curves at each
time interval and represent the intrinsic dispersion in light curve
widths.  Our resulting template curves are shown in
Figure~\ref{fig:VR_curves}.

\begin{figure}
\plotone{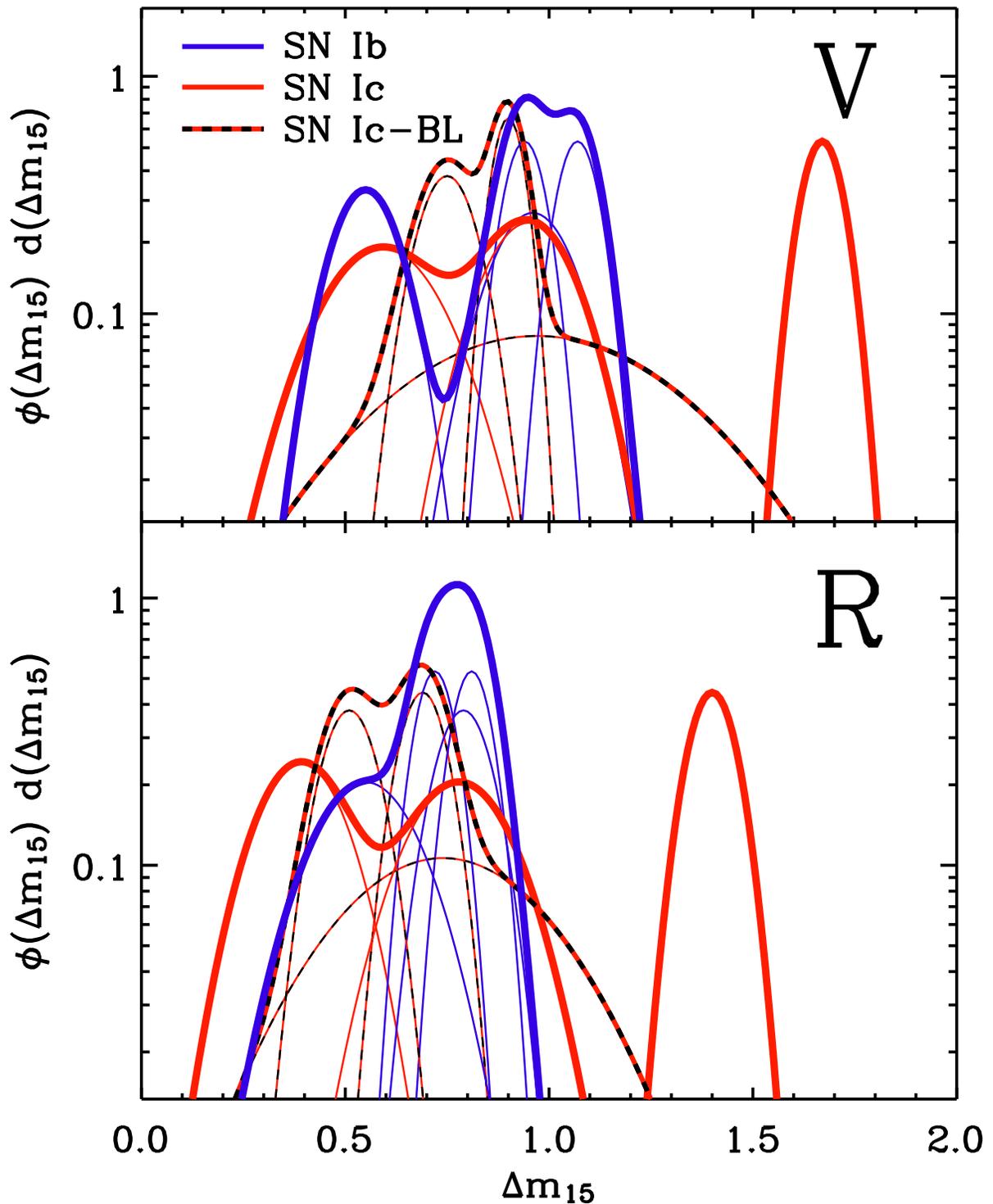}
\caption{We extract the $\Delta m_{15}$ values for the sample of literature SNe in the $V-band$ (top) and $R-$band (bottom).  Within each band, we compare the
$\Delta m_{15}$ values for each of the SNe Ib (blue thin lines), Ic
(red thin lines), and Ic-BL (red/black dashed lines) assuming Gaussian
errors for each of the estimates.  We normalize the summed
differential distributions to unity producing probability density
distributions for SNe Ib, Ic, and Ic-BL (thick lines).  A
statistical test shows no evidence that the typical $\Delta m_{15}$
values vary across the sub-classes.}
\label{fig:template_delta_m15}
\end{figure}

We measure the post-maximum decay rate for each of the literature
SNe using the $\Delta m_{15}$ diagnostic which represents the decay
in magnitudes during the 15 days following the light curve maximum
\citep{phi93}.  By interpolating over the well-sampled $V-$ and
$R-$band light curves, we derive $\Delta m_{15,V}$ and $\Delta
m_{15,R}$ estimates, respectively (Table~\ref{tab:templates_lc}). We
note that this is a variation on the standard definition for $\Delta
m_{15}$ which assumes a time baseline relative to the $B-$band
light curve maximum.  As shown in Figure~\ref{fig:template_delta_m15},
the literature sample reveals a broad dispersion in $\Delta m_{15}$
with mean values of $\langle\Delta m_{15,V}\rangle=0.94\pm 0.31$ mag and
$\langle\Delta m_{15,R}\rangle =0.74\pm 0.27$ mag.  Thus, the $R-$band
light curves decay more slowly than those in the $V-$band.  

After dividing the literature sample into the SN sub-classes, we find
that the $\Delta m_{15,R}$ values are statistically consistent:
$\langle \Delta m_{15,R}\rangle =0.7\pm 0.1$ mag (Ib), $\langle \Delta
m_{15,R}\rangle =0.9\pm 0.5$ mag (Ic), and $\langle \Delta
m_{15,R}\rangle =0.7\pm 0.1$ mag (Ic-BL).  We also note that the
fast-fading Type Ic SN\,1994I is a $2\sigma$ outlier of the overall SN
Ibc distribution.

\subsection{The Sample of P60 SNe}
\label{sec:lc_params}

Since the light curves for some of the SNe in our P60 sample are not
as densely sampled as those of the literature sample, we use the $V-$
and $R-$band template light curves described above to fit the epoch
and magnitude of the light curve maxima for all Gold and Silver P60
SNe.  We use a $\chi^2$ minimization technique for the template
fitting analysis and extract typical uncertainties on the epoch and
apparent magnitude at maximum of about $\pm 0.5$ ($\pm 1.0$) d and
$\pm 0.1$ ($\pm 0.2$) mag, respectively for the Gold (Silver) SNe.
For the Silver SNe, we fit only the $R-$band light curves since the
$V-$band data are insufficient to estimate the parameters of the
light curve maxima.  The Gold and Silver P60 light curves are compared
with the individual template fits in Figures~\ref{fig:postage1} and
\ref{fig:postage2}, respectively, and the estimated epoch and
magnitude of maximum light for our sample are listed in
Table~\ref{tab:p60_lc}.  The Bronze sample do not have sufficient
early data to reasonably constrain the epoch and magnitude of the
light curve peak through template fitting. 

\begin{figure}
\plotone{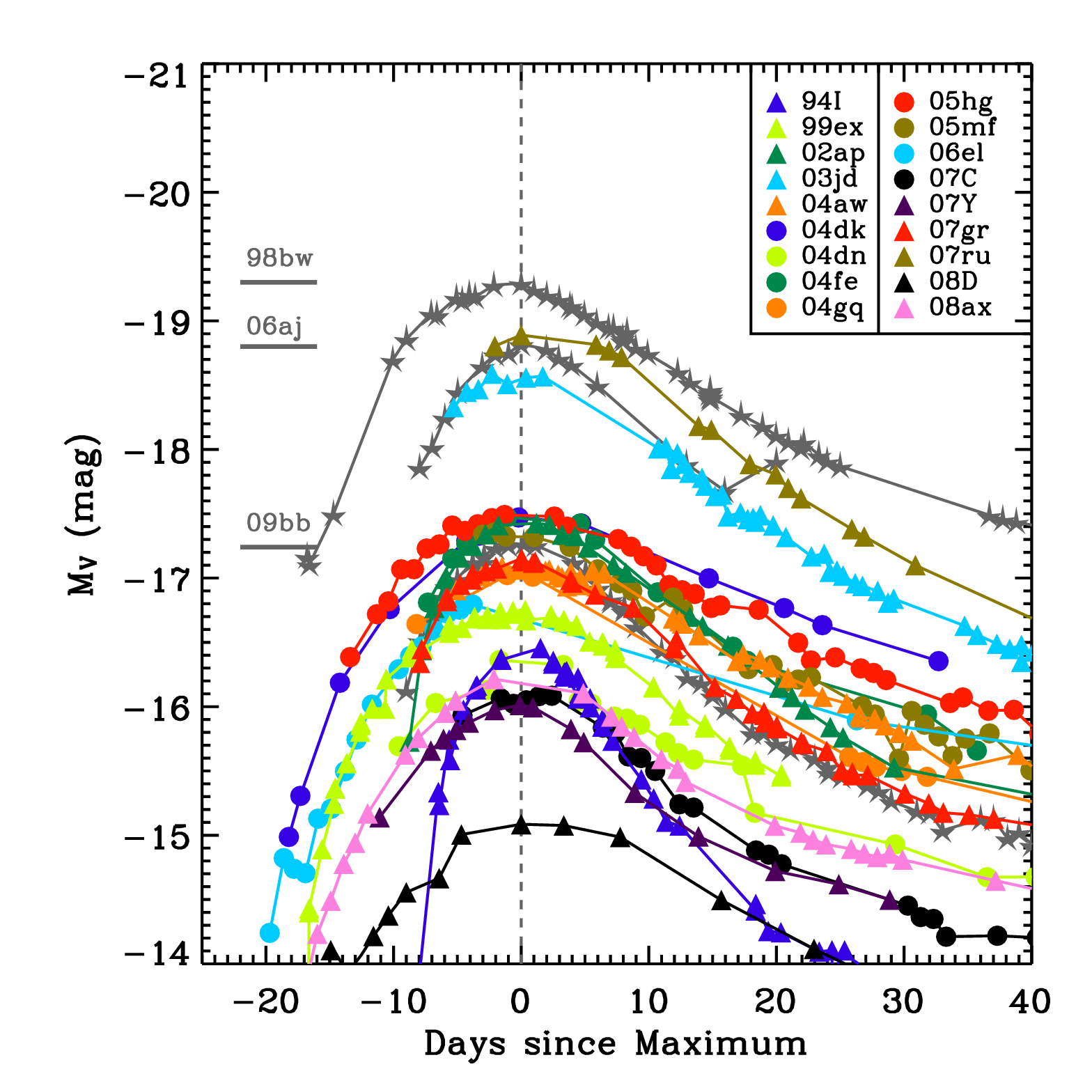}
\caption{The absolute V-band light curves for our Gold SNe Ibc and literature sample, {\it before} correcting for host galaxy extinction. 
The light curves for engine-driven SNe 1998bw, 2006aj, and 2009bb are indicated
by grey star symbols and the peak magnitudes are labeled with grey bars.}
\label{fig:moal_V}
\end{figure}

\begin{figure}
\plotone{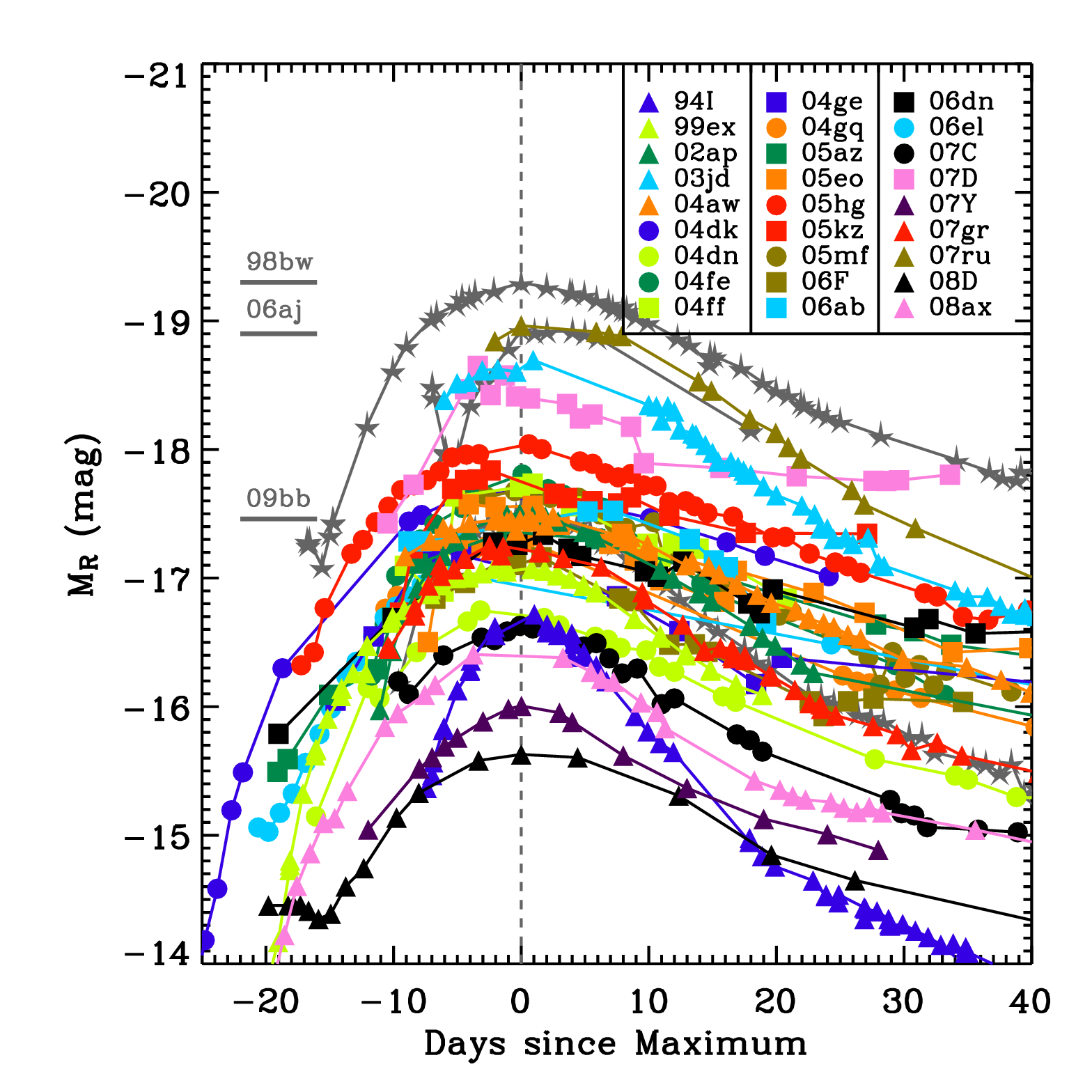}
\caption{The absolute R-band light curves for our Gold and Silver SNe Ibc and literature sample, before correcting for host galaxy extinction. 
The light curves for engine-driven SNe 1998bw, 2006aj, and 2009bb are
indicated by grey star symbols and the peak magnitudes are labeled
with grey bars.}
\label{fig:moal_R}
\end{figure}

In Figures~\ref{fig:moal_V} and \ref{fig:moal_R} we display the
absolute magnitude light curves for the P60 Gold and Silver samples
after shifting them to the epoch of maximum light and correcting for
Galactic extinction ($\langle E(B-V)_{\rm Gal}\rangle\approx 0.09$ mag
for the P60 sample).  Similar to the case of the literature sample, we
find a broad distribution in $M_{V,\rm peak}$ and $M_{R,\rm peak}$
values that span $\sim 2-3$ magnitudes.  Dividing the Gold and Silver
samples into the SNe Ib, Ic, and Ic-BL (8, 7, and 2 SNe, respectively),
we report mean values of $\langle M_{R,\rm peak}\rangle =-17.4\pm 0.5$
mag (Ib), $\langle M_{R,\rm peak}\rangle =-17.5\pm 0.4$ mag (Ic), and
$\langle M_{R,\rm peak}\rangle =-18.2\pm 0.5$ mag (Ic-BL).  These
values are overall consistent with those of the literature sample.

\begin{figure}
\plotone{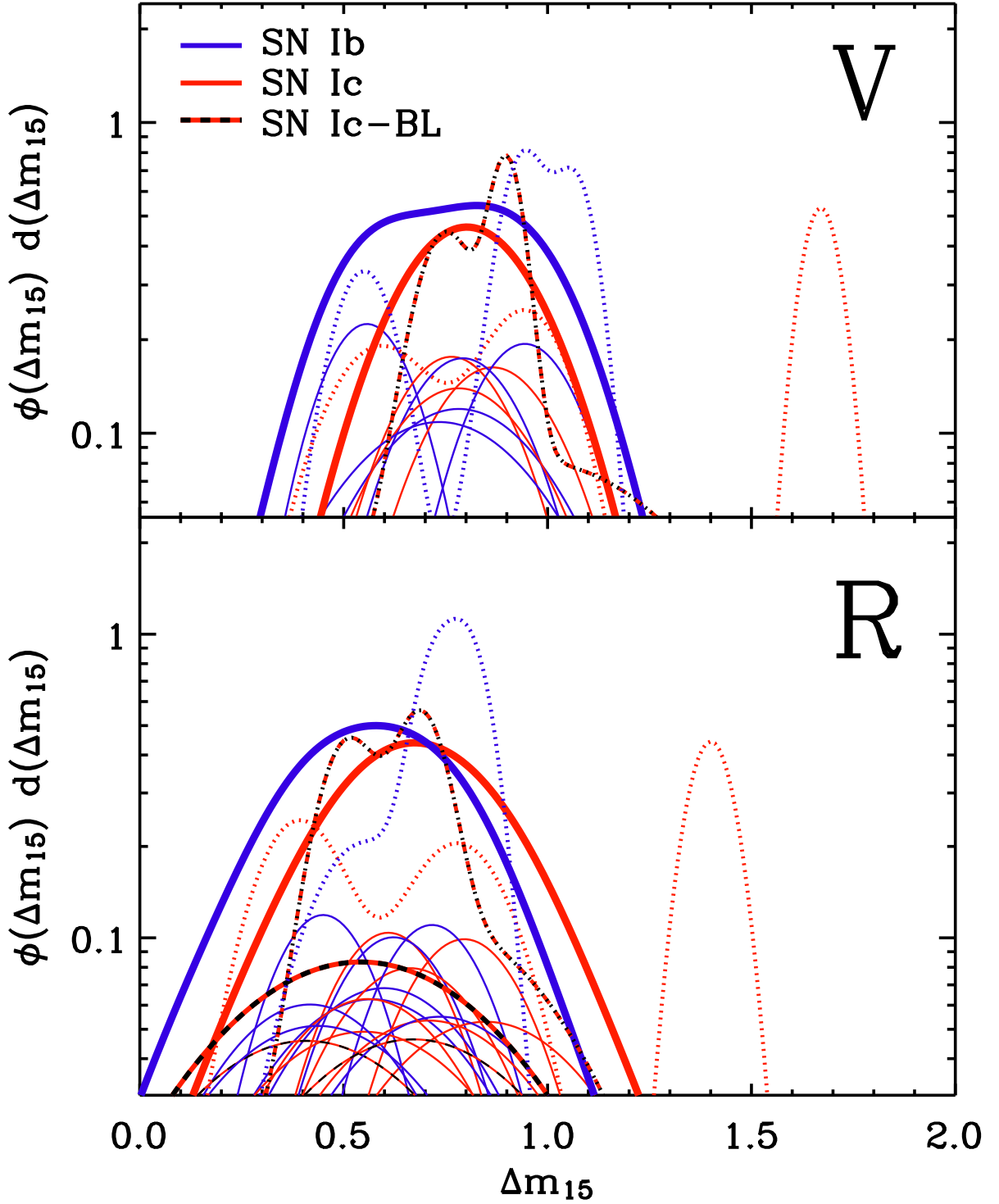}
\caption{We extracted the $\Delta m_{15}$ values for our Gold SNe in the 
$V-$ and $R-$bands and in the $R-$band for the Silver SNe.  In each band, 
we compare the $\Delta m_{15}$ estimates
for SNe Ib (thin blue lines), Ic (thin red lines), and Ic-BL
(red/black dashed lines) assuming Gaussian errors.  We normalize the
summed differential distributions to unity (thick lines).  For
comparison, we overplot the summed differential distributions from the
literature sample (Figure~\ref{fig:template_delta_m15}; dotted lines).
The $\Delta m_{15}$ distributions are similar for the literature and
P60 samples, across the sub-classes.}
\label{fig:delta_m15}
\end{figure}

Adopting the fitted parameters for the light curve maxima, we next
extract $\Delta m_{15,V}$ and $\Delta m_{15,R}$ values for our Gold
and Silver SNe following the same procedure adopted for the literature
sample (\S\ref{sec:templates}).  In Table~\ref{tab:p60_lc} we list the
resulting $\Delta m_{15}$ values for the $V-$ and $R-$bands, and in
Figure~\ref{fig:delta_m15} we compare their distributions with those
from the literature sample.  We find mean values of $\langle\Delta
m_{15,V}\rangle=0.78\pm 0.11$ mag and $\langle\Delta
m_{15,R}\rangle=0.61\pm 0.13$ mag, somewhat lower than those derived for
the literature sample but consistent within the $1\sigma$
uncertainties. Similar to the case of the literature sample, we find
no evidence for a statistically significant difference between the
decay rates of the SNe sub-classes.  

Finally, we comment on the rise time of the SNe Ibc in this sample,
i.e. the time interval between the epoch of first detection  
and the epoch of maximum light. As shown in Figures~\ref{fig:moal_V} and
\ref{fig:moal_R}, SNe Ibc with broader light curves tend to have
correspondingly longer rise times.  We find that several SNe in this
P60 sample have $R-$band rise times exceeding 20 days and thus are
comparable (even longer) than those observed for SNe Ia
\citep{chh+06,hgk+10}.  This study emphasizes the dispersion in rise times
of SNe Ibc light curves through dense, multi-band photometry of young
explosions.

\subsection{The Combined Sample}
\label{sec:comb}

By combining the Gold and Silver P60 samples with the literature
sample, we derive mean values for the peak absolute magnitudes and
light curve decline rates for a larger sample of SNe Ibc to search for
statistical differences.  Correcting for Galactic extinction only, we
find $M_{V,\rm peak}=-17.0\pm 0.9$ mag and $M_{R,\rm peak}=-17.4\pm
0.7$ mag.  For comparison, \citet{llc+10} recently reported that the
typical unfiltered peak absolute magnitudes for SNe Ibc within
$d\approx 60$ Mpc are $M_{\rm peak}=-16.1\pm 1.1$ mag.  Restricting
our sample to this same distance, we find a slightly higher value of
$M_{R,\rm peak}=-16.9\pm 0.6$ mag in the $R-$band, although we note
that these values are consistent within the $1\sigma$ uncertainties.

\begin{figure}
\plotone{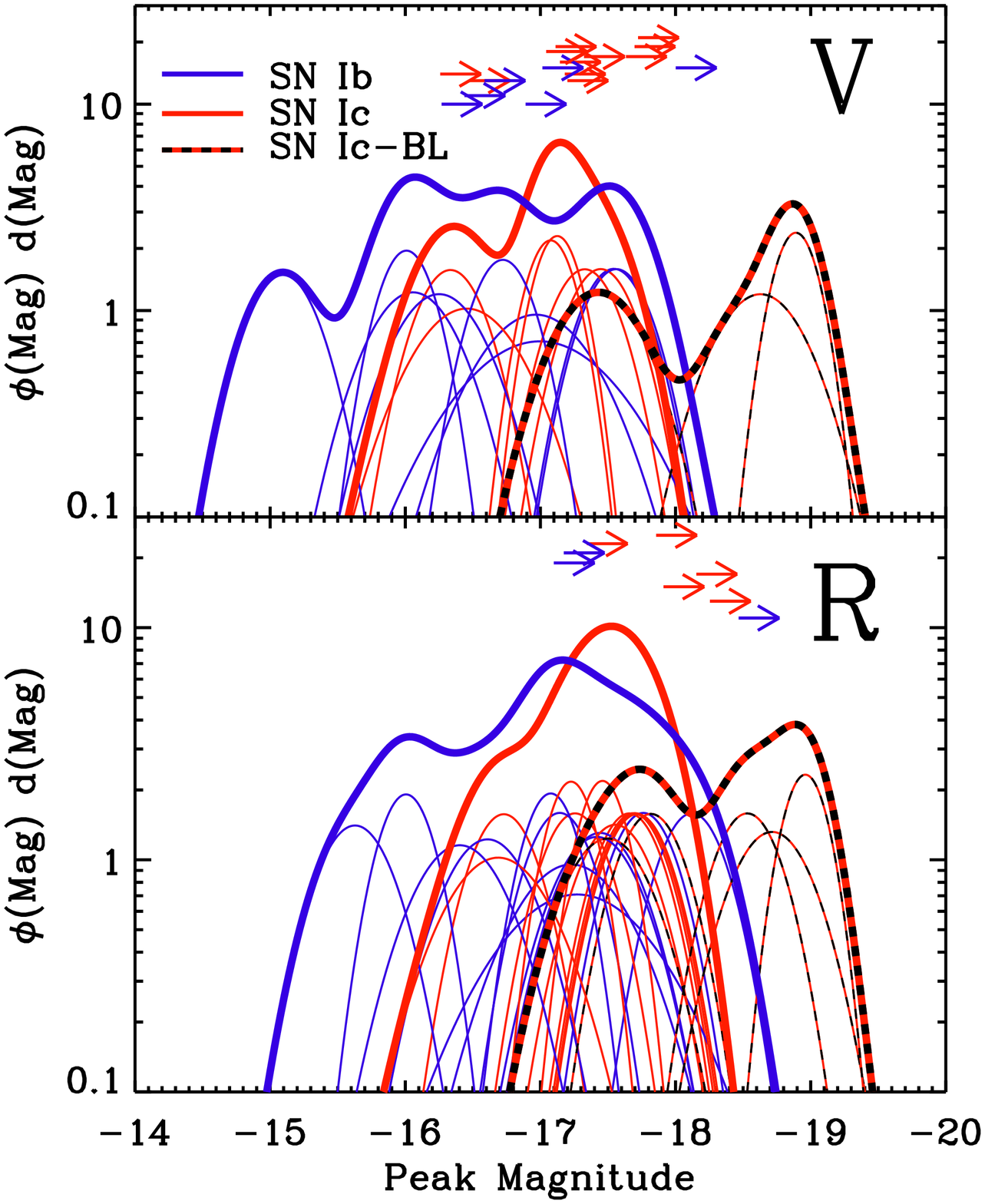}
\caption{We compare the peak absolute magnitudes in the $V-$ and $R-$bands
for SNe Ib (blue thin lines), Ic (red thin lines), and Ic-BL
(red/black dashed lines) from the Gold and Silver P60 and literature
samples {\it before} correcting for host galaxy extinction.  We assume
the errors are Gaussian and that the area under each thin curve is
normalized to one.  The thick lines are the summed differential
distributions for SNe Ib, Ic, and SNe Ic-BL (thick lines).  Lower limits
on the peak absolute magnitudes are shown as arrows for the Bronze P60
sample.}
\label{fig:absmags_g}
\end{figure}

Dividing our extended sample into the three spectroscopic sub-classes
(12 Ib, 10 Ic, and 5 Ic-BL), we report mean peak magnitudes with
higher statistical confidence: $\langle M_{R,\rm peak}\rangle
=-17.0\pm 0.7$ mag (Ib), $\langle M_{R,\rm peak}\rangle=-17.4\pm 0.4$
mag (Ic), and $\langle M_{R,\rm peak}\rangle =-18.3\pm 0.6$ mag
(Ic-BL).  This is in contrast to the report of \citet{llc+10} that SNe
Ib are marginally brighter than SNe Ic; we attribute this
apparent difference to small number statistics. While the light curve maxima
for our sample of Bronze SNe are not well constrained, we are able to
place strict lower limits on the peak absolute magnitudes by adopting
the brightest data point the $V-$ and $R-$bands and correcting for
Galactic extinction.  These lower limits range from $M_{V,\rm
peak}=-16.3$ to $-18.0$ mag and $M_{R,\rm peak}=-17.1$ to $-18.5$
mag. In Figure~\ref{fig:absmags_g} we present the differential
distribution of peak absolute magnitudes for our combined sample,
including the lower limits derived for the Bronze P60 SNe.

We next calculate mean decay rates for the combined sample of $\langle
\Delta m_{15,V}\rangle=0.87\pm 0.25$ mag and $\langle\Delta
m_{15,R}\rangle=0.66\pm 0.20$ mag.  Dividing into the sub-classes, we
find $\langle\Delta m_{15,R}\rangle=0.62\pm 0.14$ mag (Ib),
$\langle\Delta m_{15,R}\rangle=0.73\pm 0.27$ mag (Ic), and
$\langle\Delta m_{15,R}\rangle=0.60\pm 0.14$ mag (Ic-BL).  A
Kolmogorov-Smirnov (K-S) test on these distributions reveals an 89\%
probability that SNe Ib and Ic are drawn from the same population, and
the probability that SNe Ic and Ic-BL are the same is 86\%.  Summing
the SNe Ib and Ic distributions together, a K-S test reveals a 91\%
probability that SNe Ic-BL are drawn from the ordinary SNe Ibc sample.
This study demonstrates that the post-maximum decay rate of SN Ibc
optical light curves cannot be used to distinguish between SNe Ib, Ic,
and Ic-BL.  

\section{\label{sec:EC} Host Galaxy Extinction Corrections}

In \S\ref{sec:comb} we report the peak absolute magnitudes for our
combined sample of SNe Ibc light curves after a typically small
correction for Galactic extinction.  However, SNe Ibc are known to
reside in dusty star-forming regions \citep{vhf96,kkp08}, and thus
extinction within the host galaxy likely dominates the total
line-of-sight extinction.

\begin{figure}
\plotone{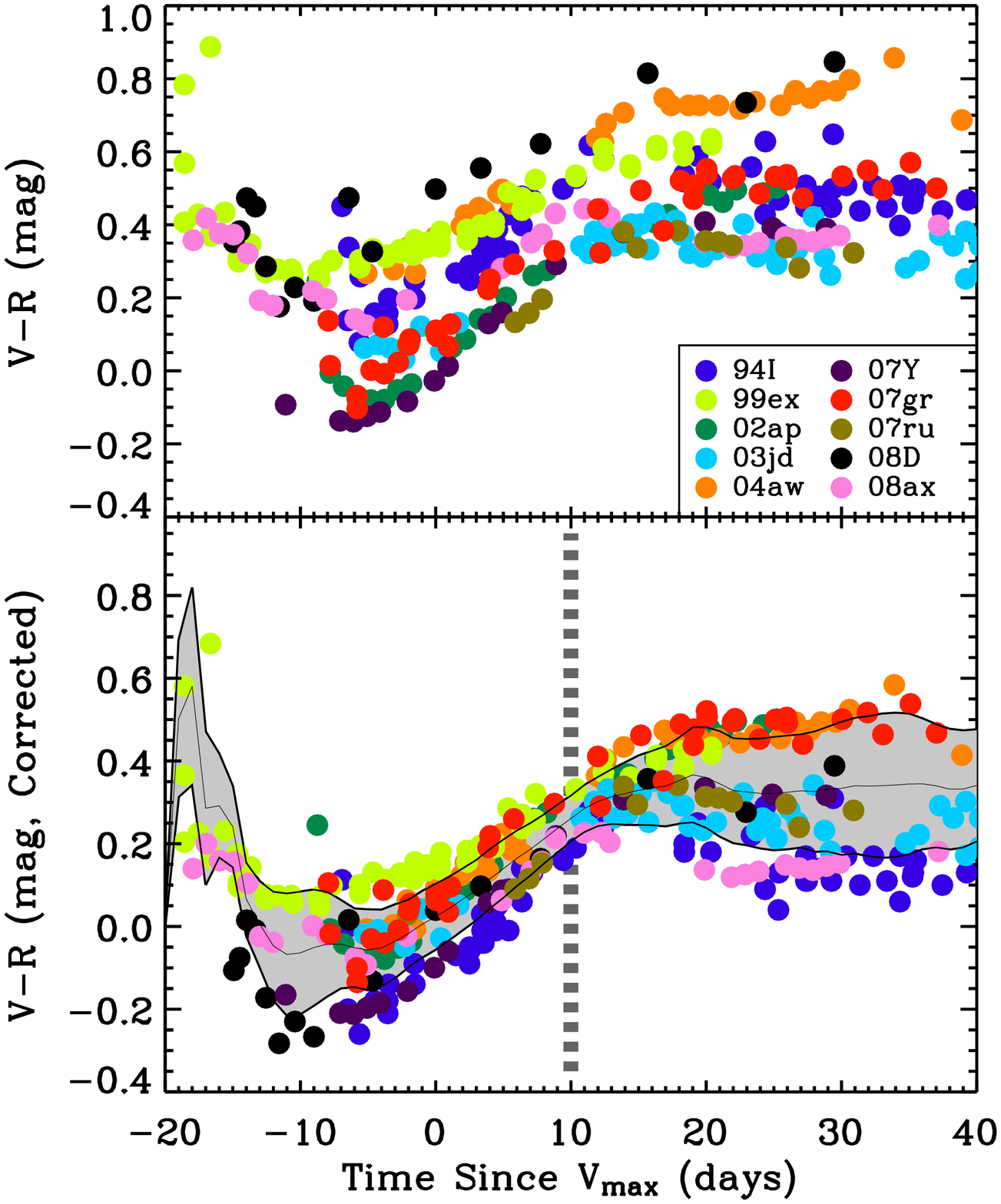}
\caption{We compare the $(V-R)$ color evolution in days since $V-$band maximum light for the sample of 10 well-studied SNe Ibc 
currently available in the literature (see Table~\ref{tab:templates}).  Top: $(V-R)$ color
evolution after correcting for Galactic extinction only.  Bottom: $(V-R)$ color
evolution after correcting for Galactic and host galaxy extinction.
The dispersion in the color evolution is decreased by accounting for the
(often large) host galaxy extinction with a minimum dispersion observed
on a timescale of roughly 10 days after the observed epoch of $M_{V,\rm peak}$ (dashed grey line) with mean value of $0.26\pm 0.06$ mag (grey curve).}
\label{fig:VR_templates_V_max}
\end{figure}

\begin{figure}
\plotone{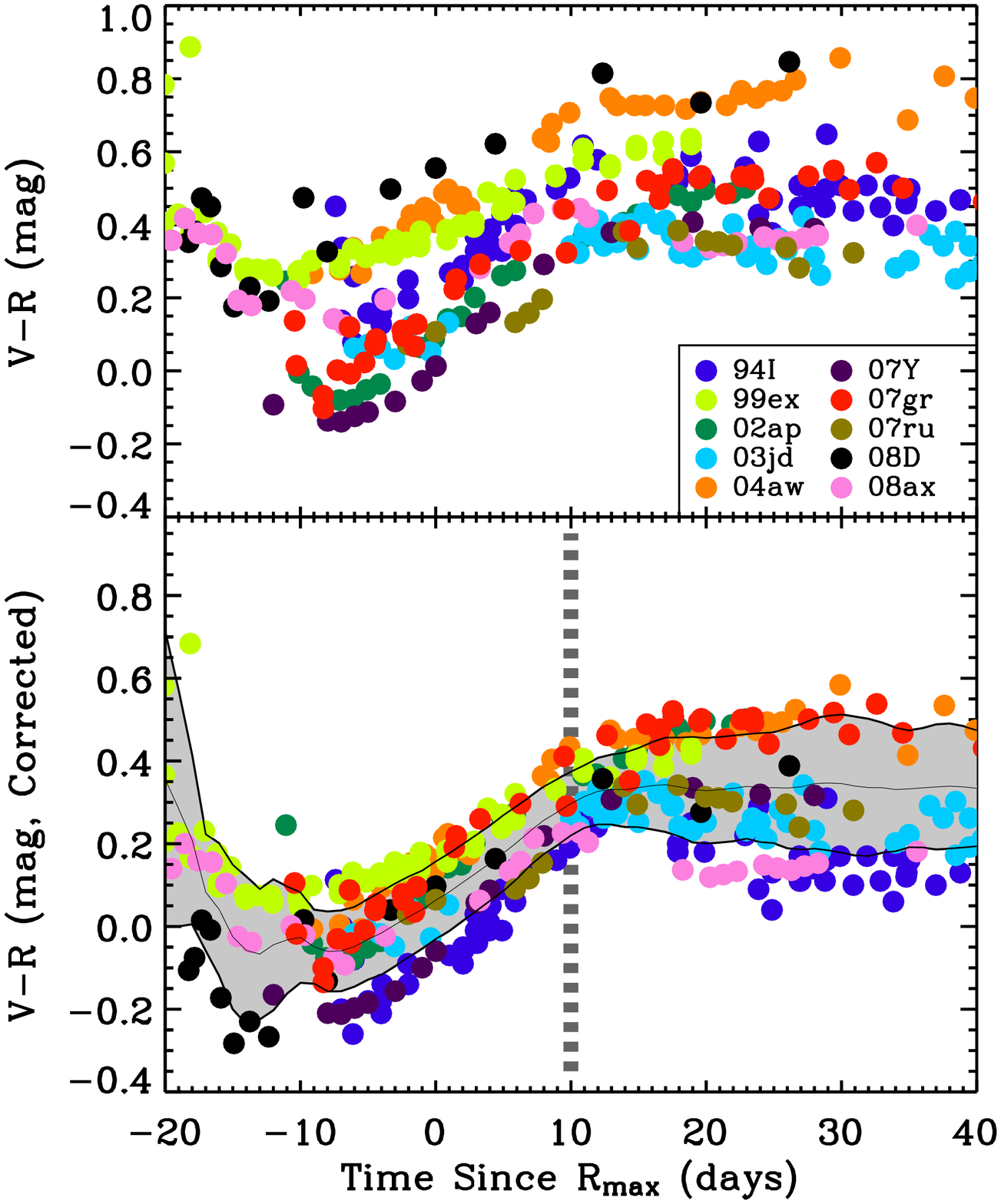}
\caption{
We compare the $(V-R)$ color evolution in days since $R-$band maximum light for the sample of 10 well-studied SNe Ibc 
currently available in the literature (see Table~\ref{tab:templates}).  Top: $(V-R)$ color
evolution after correcting for Galactic extinction only.  Bottom: $(V-R)$ color
evolution after correcting for Galactic and host galaxy extinction.
The dispersion in the color evolution is decreased by accounting for the
(often large) host galaxy extinction with a minimum dispersion observed
on a timescale of roughly 10 days after the observed epoch of $M_{R,\rm peak}$ (dashed grey line) with mean value of $0.29\pm 0.08$ mag (grey curve).
}
\label{fig:VR_templates_R_max}
\end{figure}

Motivated by the well-studied Lira relation observed for SNe Ia
\citep{pls+99}, we investigate whether the color evolution of SN Ibc
may be used to infer the total line-of-sight extinction.  To this end,
we construct $(V-R)$ color curves for the literature sample and
compare their temporal evolution with respect to the epoch of $V-$ and
$R-$band light curve maxima.  In Figures~\ref{fig:VR_templates_V_max}
and \ref{fig:VR_templates_R_max} (top panels), we display the $(V-R)$
color curves after correcting {\it only} for Galactic extinction.  All
of these SNe show a clear trend of increasing $(V-R)$ color beginning
$\sim 1$ week prior to maximum light, however, there is a broad
dispersion in the $(V-R)$ colors at any single epoch.

For each of these literature sample SNe, host galaxy extinction
estimates, $E(B-V)_{\rm host}$, are available based on the equivalent
width of the Na I D lines and/or fitting of the spectral energy
distribution (Table~\ref{tab:templates_lc}).  The respective authors
assume an $R_V=3.1$ Milky Way extinction law for the host galaxy in
every case.  A comparison of $E(B-V)_{\rm Gal}$ and $E(B-V)_{\rm
host}$ values for each SN reveals that host galaxy extinction is
(nearly) always dominant.  Indeed, we find a mean host galaxy
extinction of $\langle E(B-V)_{\rm host}\rangle=0.21\pm 0.20$ mag for
the literature sample, a factor of $\sim 4$ higher than the average
Galactic extinction for these same SNe.

By summing the Galactic contribution with the host galaxy extinction
estimates, we correct the color curves for the total line-of-sight
extinction\footnote{We do not include a $k-$correction to the
$E(B-V)_{\rm host}$ values since (i) the SNe all reside in the local
Universe, and (ii) the uncertainties in the host galaxy extinction
values are often significant.}  As shown in
Figures~\ref{fig:VR_templates_V_max} and \ref{fig:VR_templates_R_max}
(bottom panels), the dispersion in the resulting $(V-R)$ color curves
is significantly decreased after correcting for host galaxy
extinction.  We construct extinction-corrected template color curves
over timescales spanning $-20$ to $+40$ days since light curve maxima
in both the $V-$ and $R-$bands by extracting the mean and standard
deviation of the $(V-R)$ color at each epoch.  As shown in the
Figures, the dispersion in the template curves is minimized at roughly
$\Delta t\approx 10$ d after the $V-$ and $R-$band light curve maxima.

Interpolating the color curves to $\Delta t=10$ d after maximum, we
estimate mean values of $\langle (V-R)_{V10}\rangle=0.26\pm 0.06$ mag
$\langle(V-R)_{R10}\rangle=0.29\pm 0.08$ mag and a full range
extending from $(V-R)_{V10}\approx 0.18$ to 0.34 mag and
$(V-R)_{R10}\approx 0.20$ to 0.44 mag.  We conclude that the $(V-R)$
color evolution observed for SNe Ibc can be exploited as a useful
diagnostic for estimating their host galaxy extinction.

We adopted these extinction-corrected mean colors and their associated
dispersions at $\Delta t\approx 10$ d as template values and used them
to infer the total host galaxy extinction for our P60 sample.  To this
end, we first constructed $(V-R)$ color curves for the P60 sample and
corrected each for Galactic extinction (see Table~\ref{tab:list}). We
interpolated over the P60 color curves to estimate the respective
values of $(V-R)_{V10}$ and $(V-R)_{R10}$ for all Gold SNe and
$(V-R)_{R10}$ for all Silver SNe.  By attributing any residual color
excess over the template values to host galaxy extinction, we derived
$E(B-V)_{\rm host}$ estimates for each SN.

\begin{figure}
\plotone{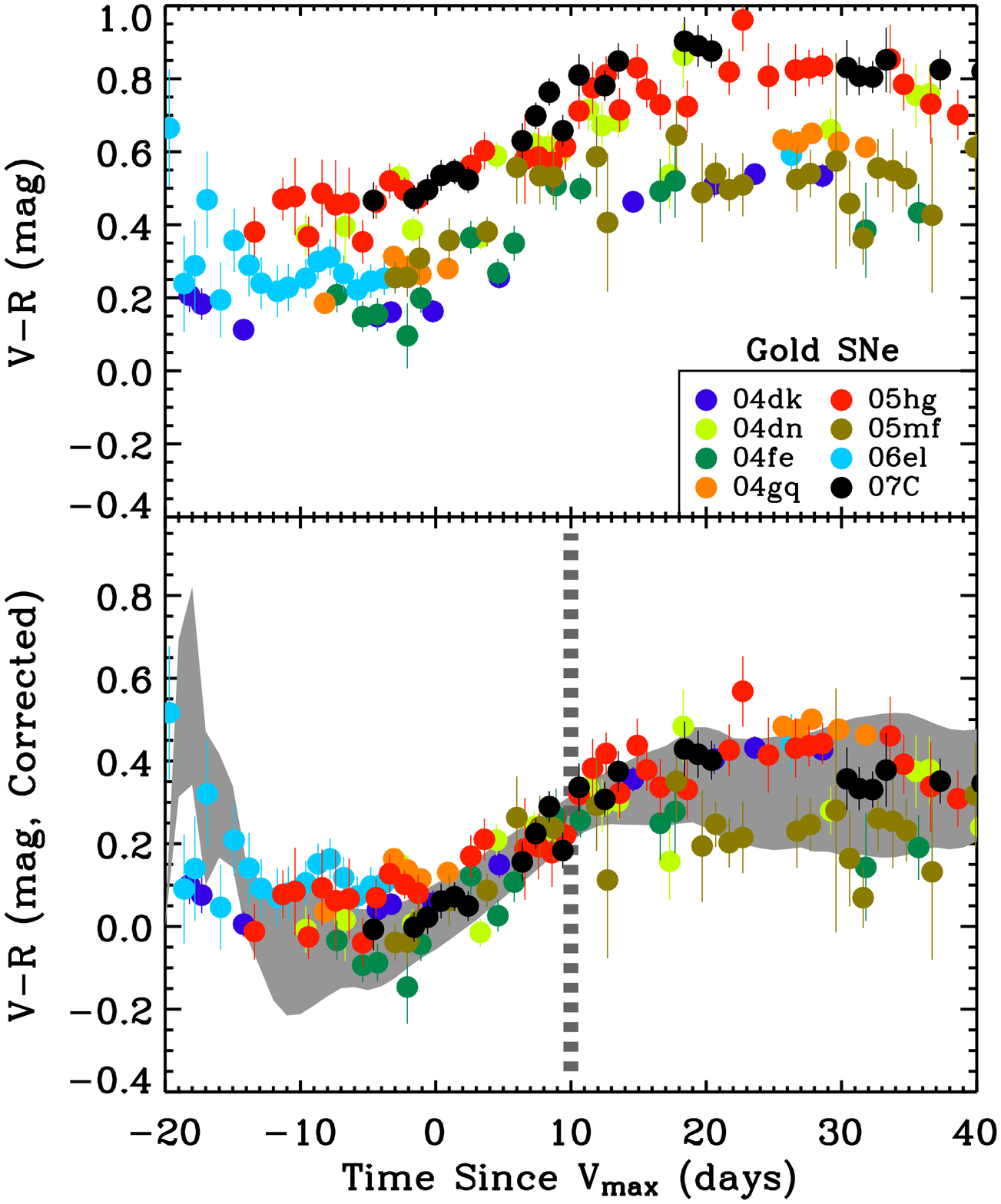}
\caption{We compare the $(V-R)$ color evolution in days since $V-$band maximum light for our sample of Gold SNe Ibc after correcting for Galactic extinction only (top) and the total extinction (bottom) as estimated using the
intrinsic $(V-R)_{V10}$ colors for well-studied SNe Ibc from the literature (grey band).
}
\label{fig:VRsne1_V_max}
\end{figure}

\begin{figure}
\plotone{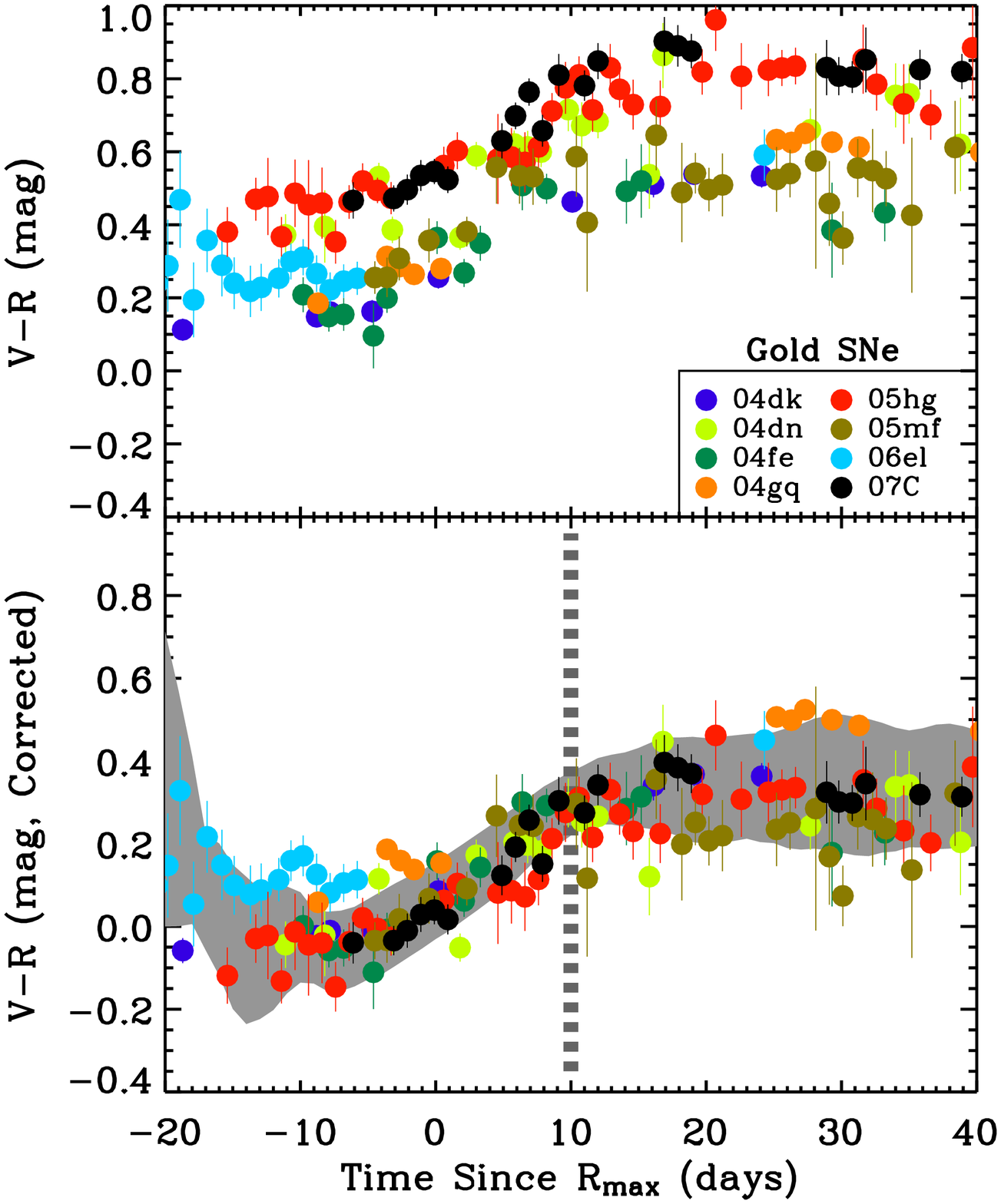}
\caption{We compare the $(V-R)$ color evolution in days since $R-$band maximum light for our sample of Gold SNe Ibc after correcting for Galactic extinction only (top) and the total extinction (bottom) as estimated using the
intrinsic $(V-R)_{R10}$ colors for well-studied SNe Ibc from the literature (grey band).
}
\label{fig:VRsne1_R_max}
\end{figure}

\begin{figure}
\plotone{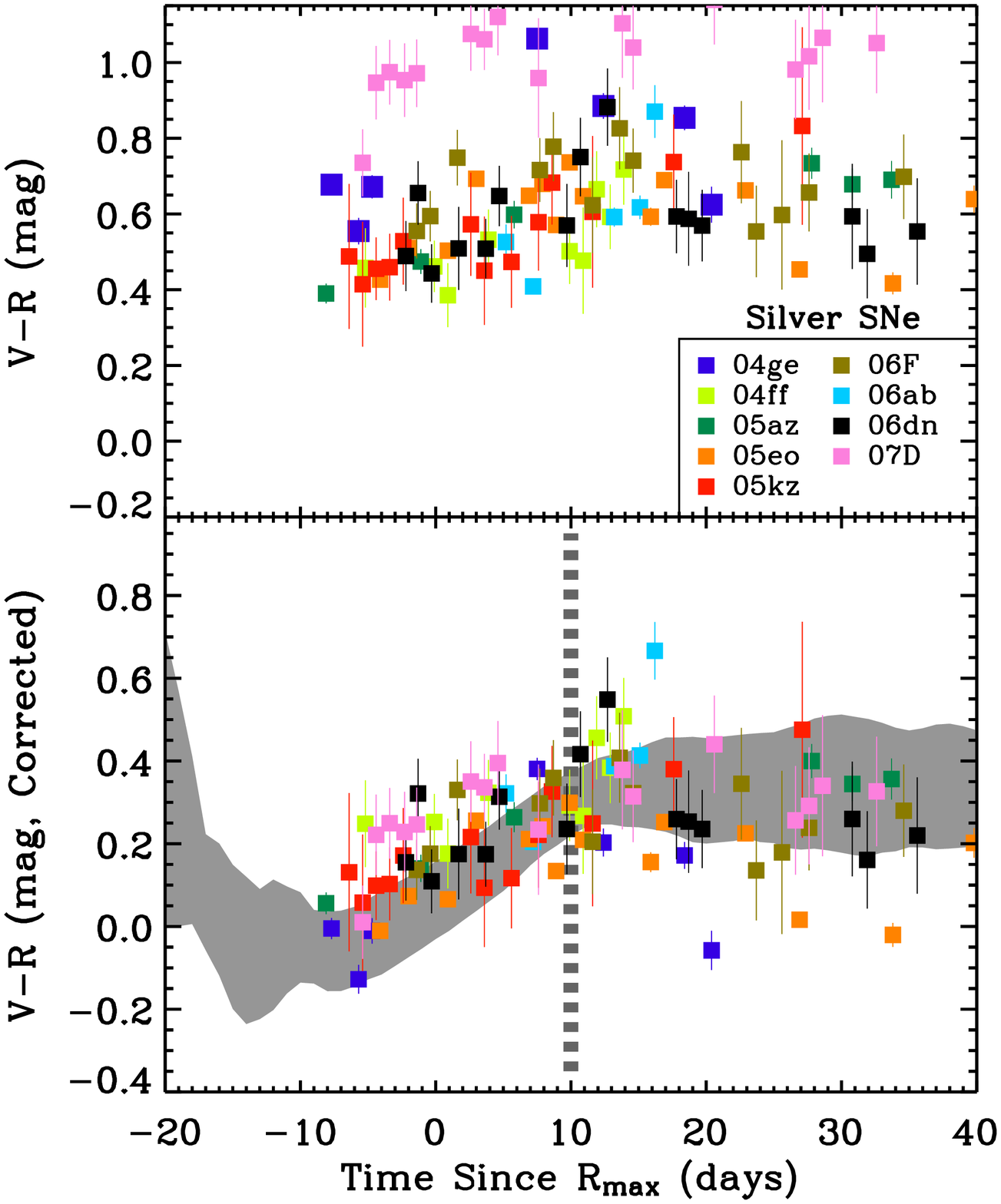}
\caption{We compare the $(V-R)$ color evolution in days since $R-$band maximum light for our sample of Silver SNe Ibc after correcting for Galactic extinction only (top) and the total extinction (bottom) as estimated using the
intrinsic $(V-R)_{R10}$ colors for well-studied SNe Ibc from the literature (grey band).
}
\label{fig:VRsne2_R_max}
\end{figure}

As shown in Figures~\ref{fig:VRsne1_V_max}-\ref{fig:VRsne2_R_max}, the
extinction-corrected color curves for our P60 sample are generally
consistent with the templates curves constructed from the literature
sample.  The host galaxy extinction values are listed in
Table~\ref{tab:p60_lc} where the associated uncertainties are
dominated by the standard deviation of the template values at 10 days
after $V-$ and $R-$band maxima.  To ensure consistency with the
literature sample extinction estimates, we similarly adopted a Milky
Way extinction law for the host galaxies.

We combine the $E(B-V)_{\rm host}$ values for the literature and P60
samples and report a mean extinction of $E(B-V)_{\rm host}=0.36\pm
0.24$ mag.  Dividing the sample into spectroscopic sub-classes, we
find no evidence for a statistically significant difference in their
host galaxy extinction values.  A K-S test reveals that the
probability that SNe Ib, Ic and Ic-BL are drawn from the same
population of events is $\gtrsim 17\%$.  Therefore, based on our
compilation of $V-$ and $R-$band light curves from the literature and
our P60 study, we conclude that most SNe Ibc show evidence for
significant host galaxy extinction.  This is slightly higher than the
extinction estimates for SNe IIP ($\langle A_V\rangle \approx 0.9$ mag;
\citealt{sec+09}) suggesting that SNe Ibc tend to be more embedded in
the star-forming regions of their host galaxies than SNe IIP and
consistent with the results of \citep{kkp08}.  Finally we note that
this result is broadly consistent with the extinction estimates
derived systematically from optical and near-IR afterglow studies of a
sample of long-duration GRBs \citep{pcb+09,gkk+10}.

\subsection{Consistency with Independent Extinction Estimates}
\label{sec:NaID}

For the SNe from our sample with high $E(B-V)_{\rm host}$ estimates,
we searched for an indication of high equivalent width Na I D
absorption features which have been shown to correlate with total
line-of-sight extinction (e.g.,
\citealt{mz97}). Indeed, in the case of our highest extinction
SNe, ($E(B-V)\gtrsim 0.9$ mag), strong Na I D features ($EQ\gtrsim
0.1$ nm) at the redshift of the host galaxies were detected in the
initial classification spectra: SNe 2004ge \citep{i8543} and 2007D
(Foley \& Gal-Yam; private communication)\footnote{We caution,
however, that the uncertainty in the extinction correction likely
increases for the more highly extinguished SNe
(e.g.,~\citealt{bpp+09})}. These independent extinction diagnostics
support our photometry-based technique.

\subsection{Host Galaxy Inclination}
\label{sec:inc}

\begin{figure}
\plotone{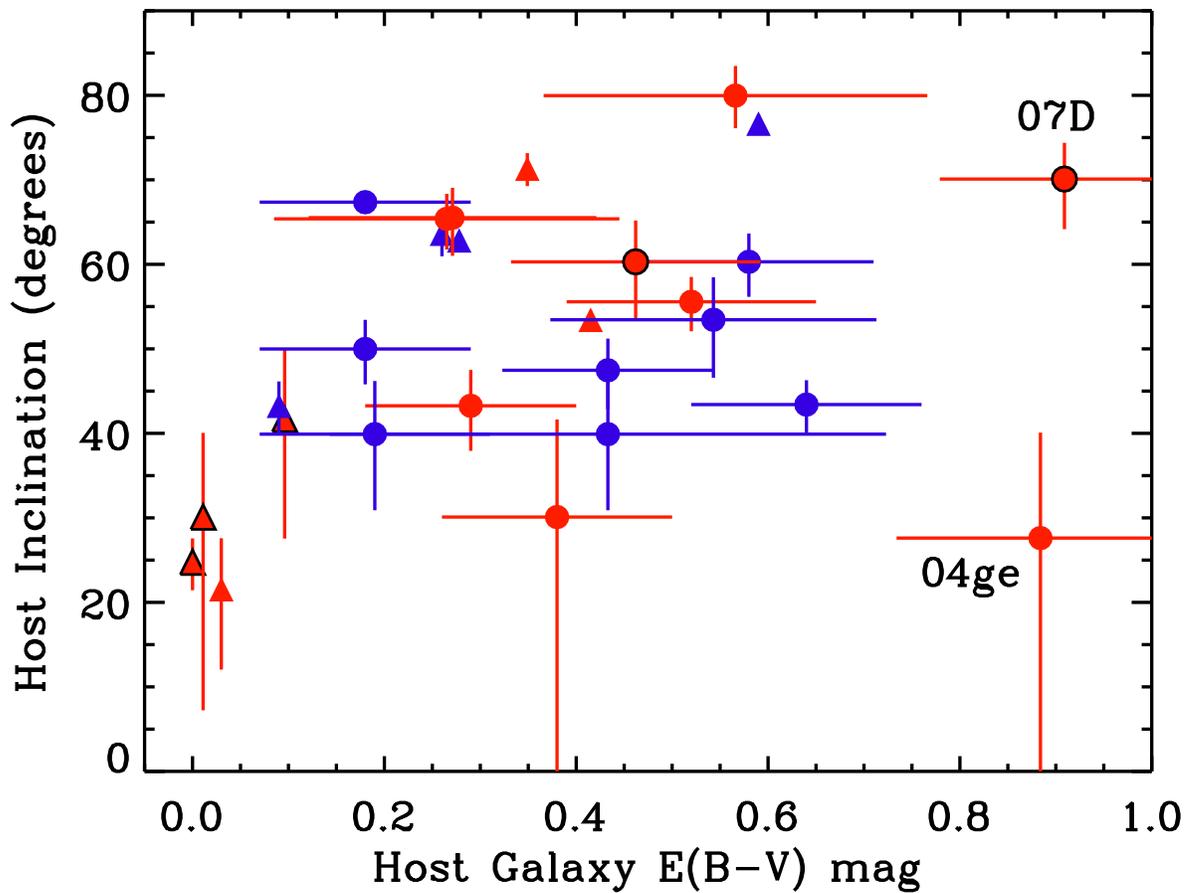}

\caption{The inferred total line-of-sight extinction, $E(B-V)$, for SNe Ib 
(blue), Ic (red), and Ic-BL (red encircled black) are compared with
their host galaxy inclinations. A general trend is seen with the more
heavily extinguished SNe residing in more highly inclined host
galaxies with a 94\% confidence level. The two SNe with the highest
inferred extinctions, SNe 2004ge and 2007D (labeled), show evidence
for unusually strong Na I D lines in their optical spectra, thus
supporting our photometrically derived extinction estimates.}
\label{fig:inc}
\end{figure}

To investigate the proximity of the dust to the explosion site, we
compare the host galaxy inclination and the $E(B-V)_{\rm
host}$ values.  We estimate the inclination for each host galaxy in
the literature and P60 samples using the traditional formalism of
\citet{hub26}, adopting the major and minor host galaxy axes from 
\citet{rc3}.  
In Figure~\ref{fig:inc} we compare the host inclinations with the
derived $E(B-V)_{\rm host}$ values.  A general trend is seen whereby
the more heavily extinguished SNe reside within more highly inclined
host galaxies.  A similar result was recently reported by \citet{llc+10} based
on the distribution of peak absolute magnitudes prior to host galaxy
extinction corrections.  We find evidence for a significant correlation (94\%
confidence level) between host inclination and extinction with a
Spearman correlation coefficient of $\rho\approx 0.37$.  As the Figure
reveals, however, there is (at least) one notable exception to this
trend, the highly extinguished Type Ic SN\,2004ge and its nearly
face-on galaxy, UGC 3555.  There is also evidence that some SNe (e.g.,
Type Ibn) form significant dust through the dynamical interaction of
the SN blastwave with the local environment \citep{sff08}.  We
conclude that dust within the host galaxy dominates the derived
$E(B-V)_{\rm host}$ values for SNe Ibc, but local dust ($\lsim 1$ kpc)
may contribute significantly in some cases.

\section{Extinction Corrected Peak Luminosities}
\label{sec:lum}

\begin{figure}
\plotone{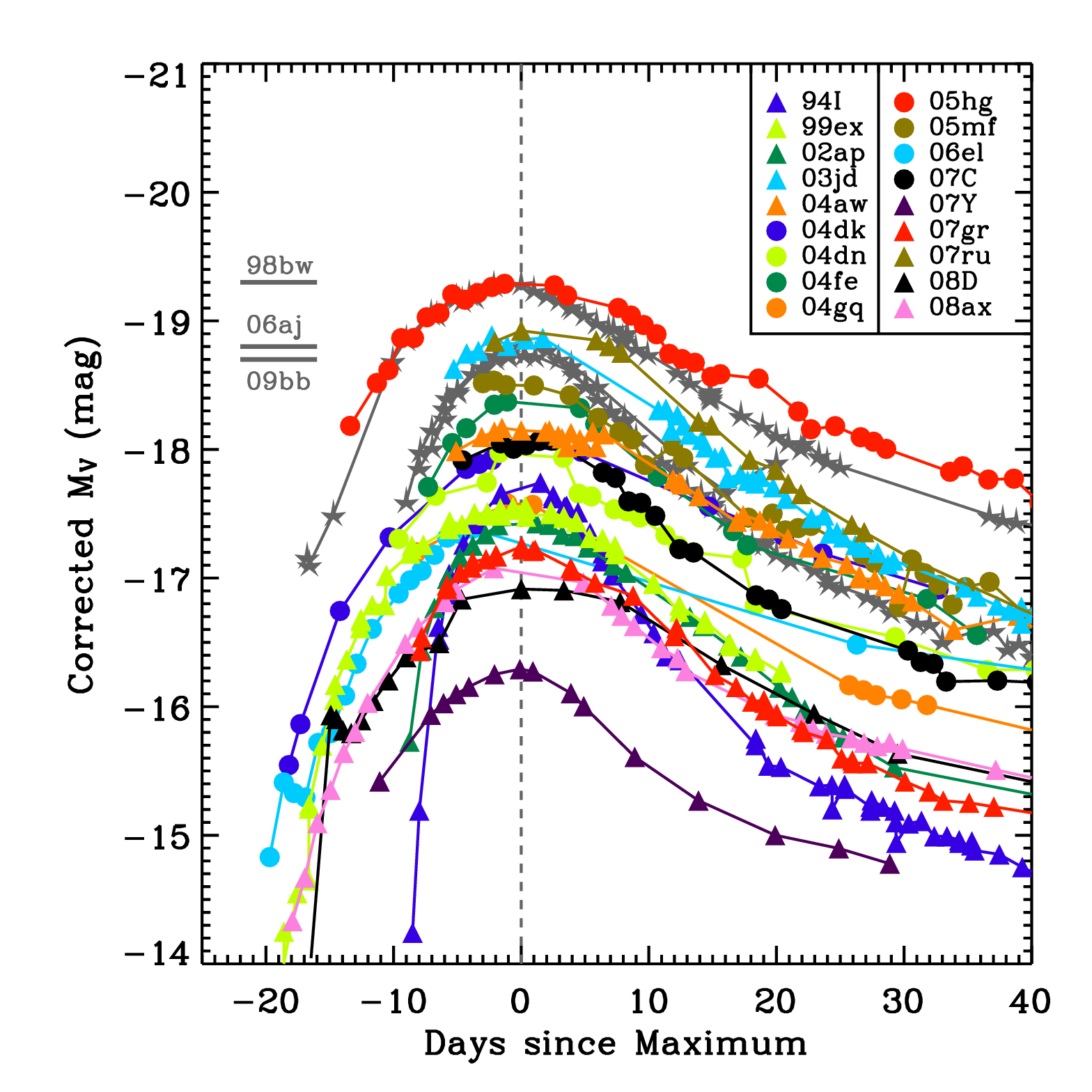}
\caption{The absolute V-band lightcurves for our Gold SNe Ibc and literature 
sample, {\it after} correcting for host galaxy extinction.  The
symbols are the same as those in Figure~\ref{fig:moal_V}.  The
extinction corrected light curves for engine-driven SNe 1998bw,
2006aj, and 2009bb are indicated by grey star symbols and the peak
magnitudes are labeled with grey bars.}
\label{fig:moal_Vcor}
\end{figure}

\begin{figure}
\plotone{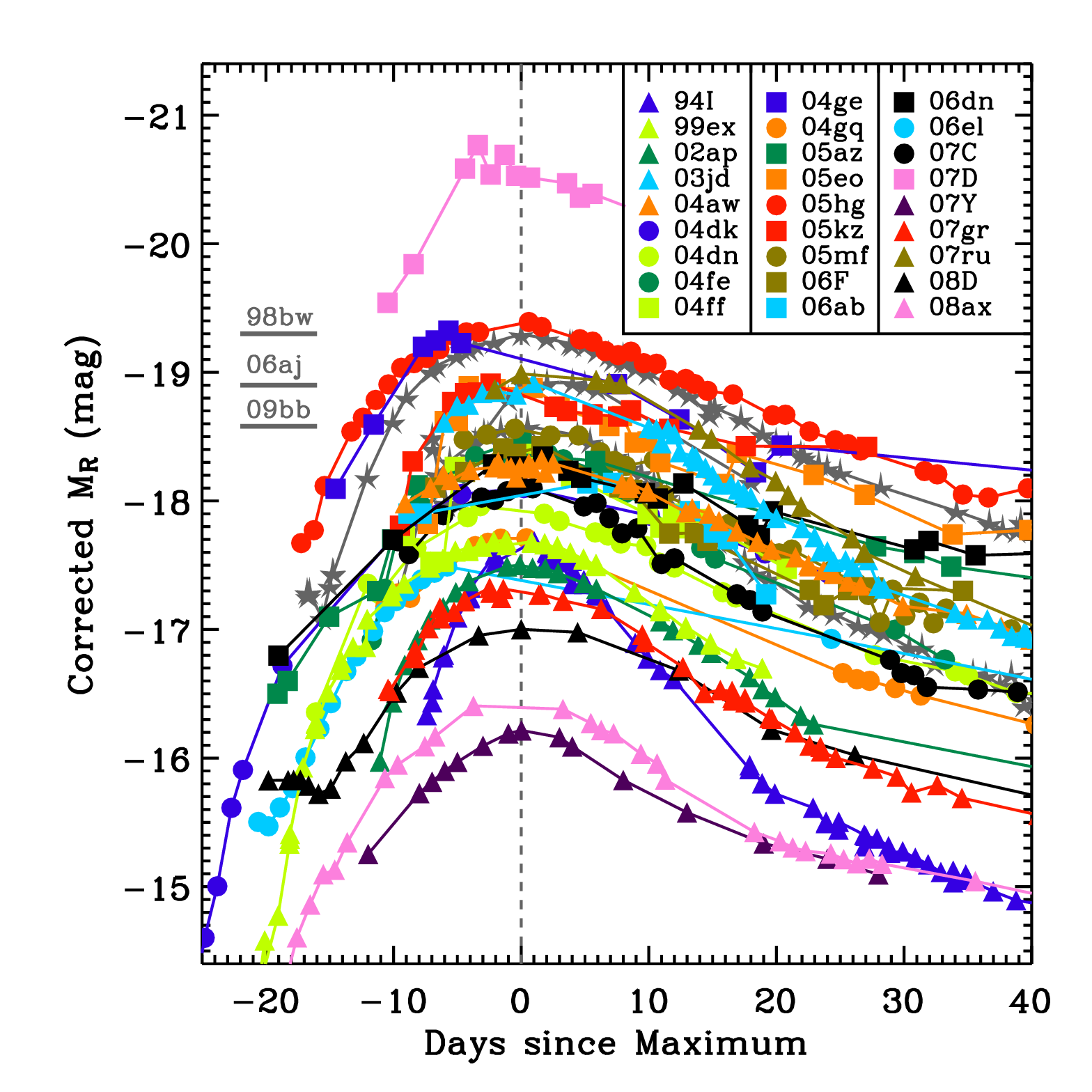}
\caption{The absolute R-band lightcurves for our Gold and Silver SNe Ibc and literature sample, after correcting for host galaxy extinction.    The
symbols are the same as those in Figure~\ref{fig:moal_R}.  The
extinction corrected light curves for engine-driven SNe 1998bw,
2006aj, and 2009bb are indicated by grey star symbols and the peak
magnitudes are labeled with grey bars.}
\label{fig:moal_Rcor}
\end{figure}

We next construct a sample of {\it total} extinction-corrected
light curves by correcting for the $E(B-V)_{\rm host}$ values compiled
in Tables~\ref{tab:templates_lc} and \ref{tab:p60_lc}.  The corrected
$V-$ and $R-$band light curves are shown collectively in
Figures~\ref{fig:moal_Vcor} and ~\ref{fig:moal_Rcor}.  From this
combined sample, we derive the mean, total extinction-corrected peak
R-band magnitude of all SNe Ibc to be $\langle M_{R\rm,peak}\rangle
=-18.2\pm 0.9$ mag.  Focusing on the SNe Ibc within $d\approx 60$ Mpc,
we find a value of $\langle M_{R,\rm peak}\rangle=-17.6 \pm 0.6$ mag.
A direct comparison with the values discussed in \S\ref{sec:comb}
reveals the importance of including host galaxy extinction corrections in
deriving the luminosity function for SNe Ibc (c.f.,~\citealt{llc+10}).

\begin{figure}
\plotone{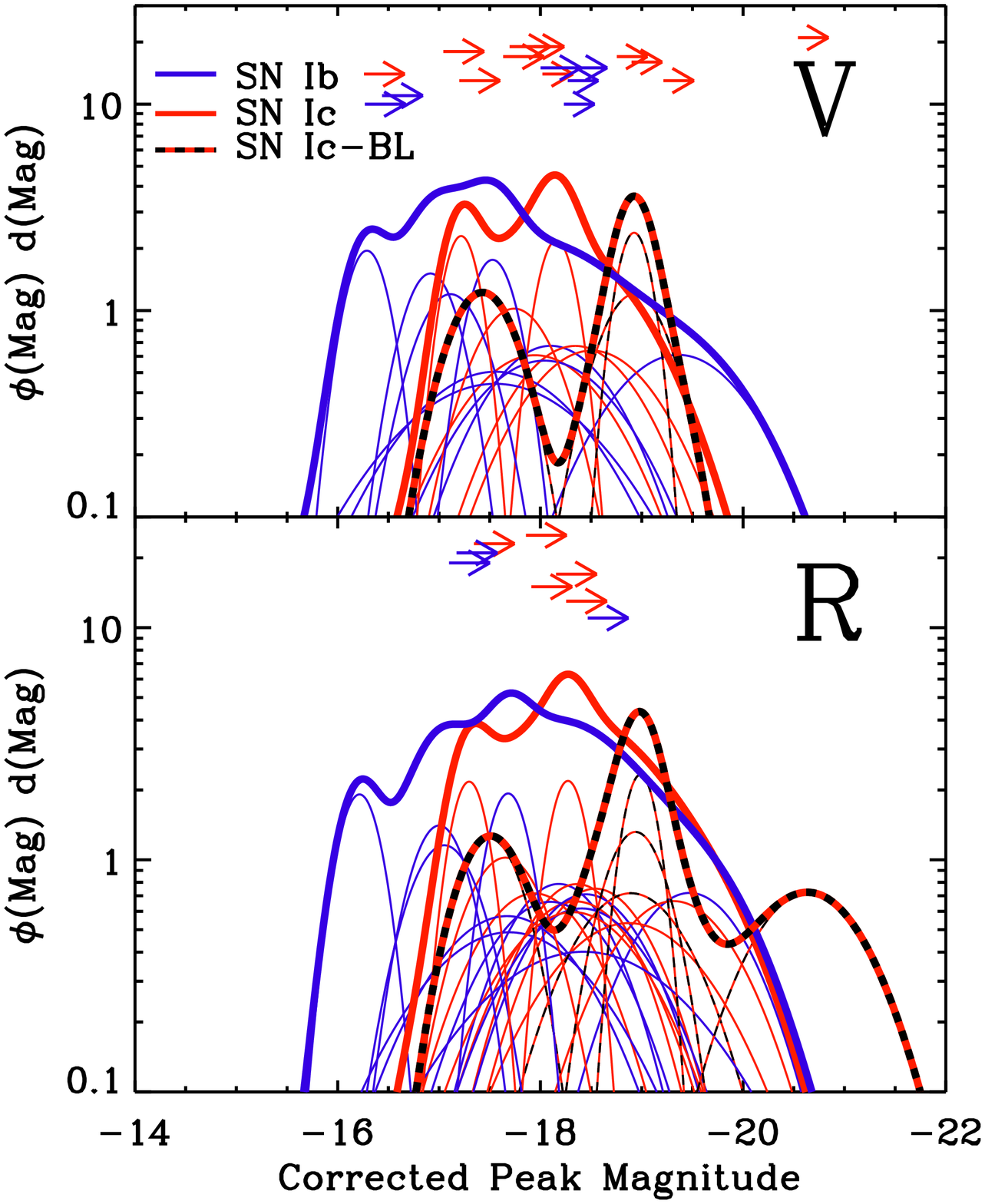}
\caption{We compare the peak absolute magnitudes in the $V-$ and $R-$bands
for SNe Ib (blue thin lines), Ic (red thin lines), and Ic-BL
(red/black dashed lines) from the Gold and Silver P60 and literature
samples {\it after} correcting for host galaxy extinction.  We assume
the errors are Gaussian and that the area under each thin curve is
normalized to one.  The thick lines are the summed differential
distributions for SNe Ib, Ic, and SNe Ic-BL (thick lines).  Lower limits
on the peak absolute magnitudes are shown as arrows for the Bronze P60
sample and do not include a correction for host galaxy extinction.}
\label{fig:absmags_gcor}
\end{figure}

The differential distribution of corrected peak absolute magnitudes
for the combined sample are shown in Figure~\ref{fig:absmags_gcor}.
We note that the errors on these estimates are fairly large
(e.g.,~$\pm 0.5$ mag) since we have properly accounted for the
uncertainties in host galaxy distance, host galaxy extinction correction,
and photometric measurement error.  Dividing the
sample into SNe Ib and ordinary SNe Ic, we find $\langle M_{R,\rm peak}\rangle
=-17.9\pm 0.9$ mag (Ib) and $-18.5\pm 0.8$ mag (Ic).  

\begin{figure}
\plotone{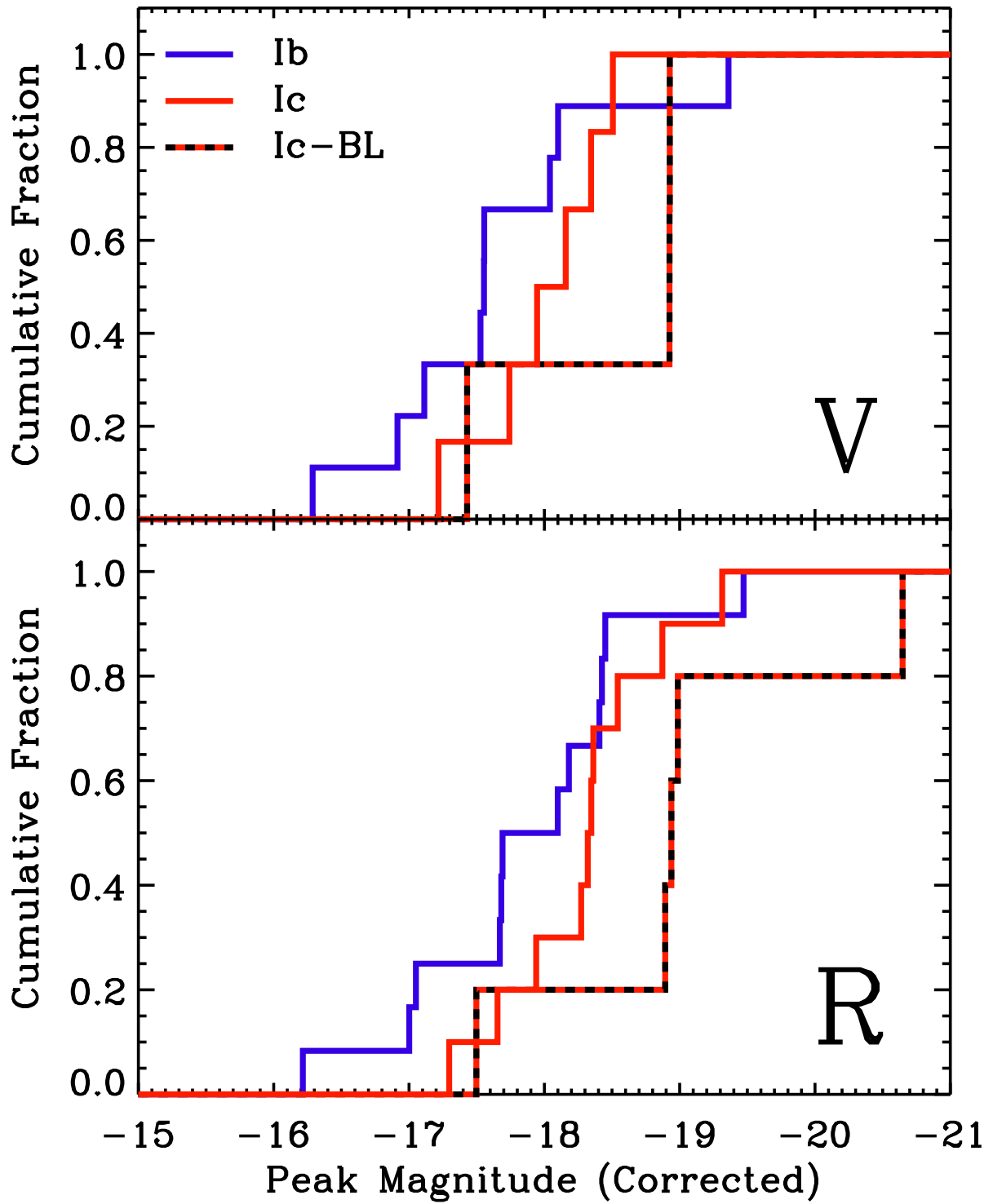}
\caption{We compare the cumulative distributions of peak absolute magnitude 
in the $V-$ and $R-$bands for SNe Ib (blue), Ic (red), and Ic-BL
(red/black) after correcting for Galactic and host galaxy extinction.
A K-S test reveals a 36\% probability that SNe Ib and Ic are drawn
from the same population of explosions while the probability that the
SN Ic-BL distribution is consistent with ordinary SNe Ibc is just 1.6\%.}
\label{fig:cum_hist}
\end{figure}

As shown in Figure~\ref{fig:cum_hist}, the cumulative distributions of
total extinction-corrected $R_{\rm max}$ values for SNe Ib and Ic
indicate that they are similar; a K-S test reveals a 36\% probability
that these events are from the same parent population of explosions.
Thus, we find no statistical evidence for a difference in the peak
luminosities of He-rich and He-poor explosions.  As discussed in
\S\ref{sec:disc}, this suggests that SNe Ib and Ic synthesize a
similar mass of $^{56}$Ni.

Separating out the broad-lined SNe Ic, we next compare their peak
luminosities with ordinary SNe Ic and Ib.  While the sample is small,
a K-S test reveals only a 1.6\% probability that they are drawn from
the same population of explosions.  We report a mean, total extinction
corrected, peak absolute magnitude for SNe Ic-BL of $\langle M_{R,\rm
peak}\rangle =-19.0\pm 1.1$ mag, comparable to those of SNe Ia
\citep{rfl+99}. We further note that none of these SNe show evidence for a
central engine, which may be revealed through the detection of an
associated GRB and/or strong non-thermal synchrotron emission (an
afterglow) from a relativistic blastwave (see
\citealt{snb+06} and references therein).

Our peak magnitudes are roughly consistent with the broad range of
values reported by \citet{rbb06} based on their compilation of SN Ibc
light curves available in the literature. However, we do not confirm
evidence for a bimodal distribution of peak absolute
magnitudes and suggest that this may be due to a
disproportionate representation of overluminous SNe in their
literature compilation.

\begin{figure}
\plotone{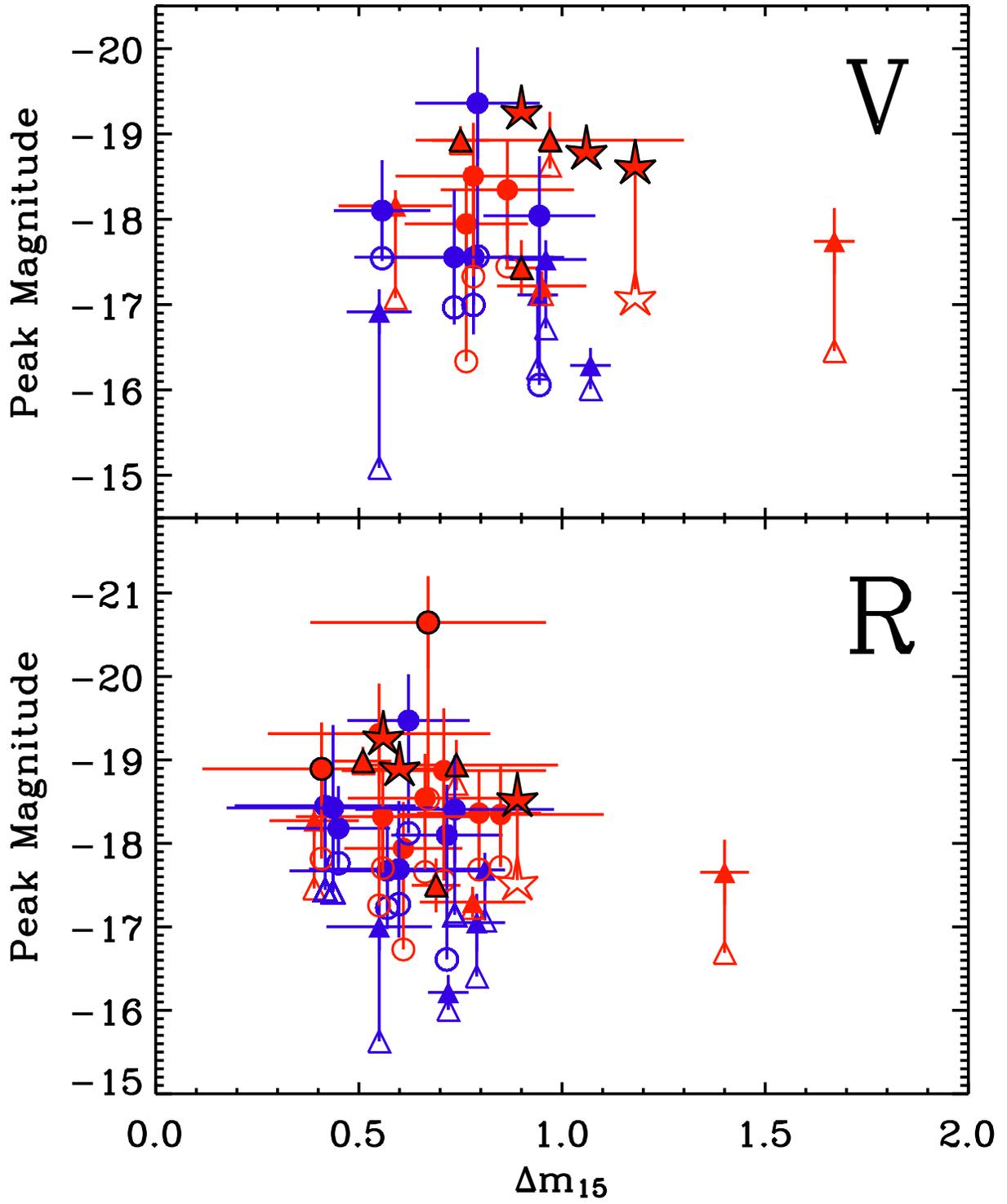}
\caption{We compare the peak absolute magnitudes in the
$V-$ (top) and $R-$bands (bottom) for SNe Ib (blue) and Ic (red) with
their respective $\Delta m_{15}$ values.  SNe Ic-BL are emphasized with outlined
symbols and engine-driven SNe are indicated as red stars.  There is
no evidence for a statistically significant correlation between
the observed parameters as is seen for SNe Ia.}  
\label{fig:absmags_m15}
\end{figure}

Finally, in Figure~\ref{fig:absmags_m15}, we compare the peak absolute
magnitudes in the $V-$ and $R-$bands with their associated $\Delta m_{15,V}$ 
and $\Delta m_{15,R}$ estimates.  In contrast to SNe Ia where $\Delta m_{15}$ is seen to
correlate tightly with $M_{V,\rm peak}$, we find no evidence for a
``Phillips relation'' in our sample of SNe Ibc.  We calculate a
Spearman correlation coefficient of $\rho \approx 0.058$ with a
confidence level of 23\%.  We therefore conclude that the decline rate
of SNe Ibc light curves cannot be used as a reliable proxy for the
peak absolute magnitude and discuss this point further in the context
of analytical models in \S\ref{sec:disc}.  

\subsection{The Bronze Sample}
\label{sec:bronze}

We note that none of the SNe in our sample show unusual light curve
shapes with the exception of two Bronze SNe, 2005la and 2006jc, whose
unusual photometric and spectroscopic properties have already been
discussed in detail in the literature
\citep{fsg+07,psm+07,pqs+08}.  We consider the various
effects plaguing the Bronze sample and leading to their poor
light curve coverage.  The Bronze SNe, on average, are {\it not}
characterized by larger distances, higher $E(B-V)_{\rm Gal}$ values,
or higher host galaxy inclinations than the other SNe in our sample.
We also note that the Bronze group is not populated by intrinsically
faint SNe since the lower limits on the peak absolute magnitudes imply
$M_{R,\rm peak}\lsim -17$ mag (Figure~\ref{fig:absmags_gcor}).  We
conclude that the lack of early-time data for the Bronze sample is
related to the insufficient cadence of many SN discovery surveys, which
determines the epoch of first detection.

\subsection{A Comparison to Engine-Driven SNe}
\label{sec:grbs}

We next compared the optical luminosities of our SN Ibc sample to
those of GRB-SNe discovered within a similar volume, $d\lsim 150$ Mpc.
While additional GRB-SNe are available in the literature we chose a
distance-limited approach in an effort to minimize observational
biases associated with the $\gamma-$ray discovery of GRB-SNe. Our
comparison GRB-SNe sample includes just two events: SN\,1998bw
\citep{gvv+98} associated with GRB\,980425 \citep{paa+00} at $d\approx
38$ Mpc and SN\,2006aj
\citep{pmm+06} associated with XRF\,060218 at
$d\approx 141$ Mpc.  To this small sample of central engine-drive
explosions, we add the broad-lined Type Ic SN\,2009bb at $d\approx 40$
Mpc which was not associated with a detected gamma-ray burst but was
shown to produce copious relativistic ejecta based on radio
observations of the ``orphaned'' afterglow \citep{scp+10}.

We compiled the optical SN photometry in the $V-$ and $R-$bands for
SNe 1998bw, 2006aj, and 2009bb from \citet{gvv+98}, \citet{mha+06},
and \citet{pig10}, respectively.  After correcting for Galactic
extinction, the peak magnitudes for GRB-SNe 1998bw (2006aj) are
$M_{V,\rm peak}\approx -19.3$ ($-18.8$) mag and $M_{R,\rm peak}\approx
-19.3$ ($-18.9$) mag.  No evidence for host galaxy extinction was
reported for either SN\,1998bw or SN\,2006aj and we test this result
using our ($V-R$) color diagnostic.  Interpolating over the light
curves, we find $(V-R)_{V10}=0.25\pm 0.09$ and $0.34\pm 0.08$ mag for
SNe 1998bw and 2006aj, respectively.  Indeed, these values are
consistent with the {\it total} extinction corrected colors for our
literature sample. This exercise serves as independent confirmation
for negligible host galaxy extinction in these two cases.  The peak
magnitudes for SN\,2009bb are nearly identical to those of SN\,2006aj
after correcting for a host galaxy extinction of $E(B-V)_{\rm
host}\approx 0.48$ mag \citep{pig10}.  Finally, we note that the 
recent photometric study of SN\,2010bh associated with XRF\,100316D 
further supports the utility of our $(V-R)_{10}$ extinction correction 
technique for engine-driven SNe with moderate reddening \citep{cbg+11}.

A comparison of the extinction-corrected peak absolute magnitudes for
engine-driven SNe and ordinary SNe reveals a handful of objects with
luminosities comparable to and/or higher than SNe 1998bw, 2006aj, and
2009bb.  We use a Monte Carlo test to determine whether the peak
absolute luminosities of these three engine-driven and relativistic
SNe are consistent with the $M_{R,\rm peak}$ distribution for SNe Ibc
(Figure~\ref{fig:absmags_gcor}).  We find a $\sim 2.3\%$ probability
that the engine-driven SNe are drawn from the ordinary SNe Ibc
population and a $\sim 22\%$ chance that they are consistent with the
SNe Ic-BL sample.

%Thus, the nearest two GRB-SNe appear to have minimal host
%galaxy extinction, in stark constrast to ordinary SNe Ibc which are
%typically extincted by $E(B-V)\gtrsim 0.3$ mag (see \S\ref{sec:EC}).
%Meanwhile, the relativistic SN\,2009bb shows evidence for significant
%host galaxy extinction.

%We further compare the light curve shapes of GRB-SNe to those of
%optically-selected SNe Ibc in Figure~\ref{fig:absmags_m15})....
%KS test 

\section{Physical Parameters}
\label{sec:disc}

In this paper, we derived the light curve properties (peak luminosity
and decay rate) in a systematic fashion for the 17 Gold and Silver P60
SNe, the ten well-studied SNe Ibc currently available from the
literature, and the three central engine-driven SNe identified to date
within the same volume.  As discussed in \S\ref{sec:diag}, these
observed properties are directly determined by the three explosion
parameters: $M_{\rm Ni}$, $M_{\rm ej}$, and $E_K$.  Here we present a
systematic method to map the observed light curve properties of SNe
Ibc ($M_R$, $\Delta m_{15}$) to the physical parameters of the
explosion.

We adopt the analytic light curve models of \citet{vbc+08} which are
based on original formalism of \citet{arn82} for Type I SNe.  The
models account for the energy deposition of $^{56}$Co in addition to
$^{56}$Ni.  We assume a homogeneous density distribution within the SN
ejecta and a fixed optical opacity ($\kappa=0.05~\rm g~cm^{-2}$) to
produce a family of 1000 model light curves for a wide range of reasonable
explosion parameters spanning $E_K\approx (0.5-50)\times 10^{51}$ erg,
$M_{\rm ej}\approx 0.1-10~\rm M_{\odot}$, and $M_{\rm Ni}=0.05-1.5~\rm
M_{\odot}$.  We made the further requirement that $M_{\rm Ni}\lsim
0.5~M_{\rm ej}$ and converted the resulting bolometric light curves to
the $R-$band assuming a typical near-IR contribution of $\sim 25\%$
at maximum light (e.g.,~\citealt{smp+09}).

\begin{figure}
\plotone{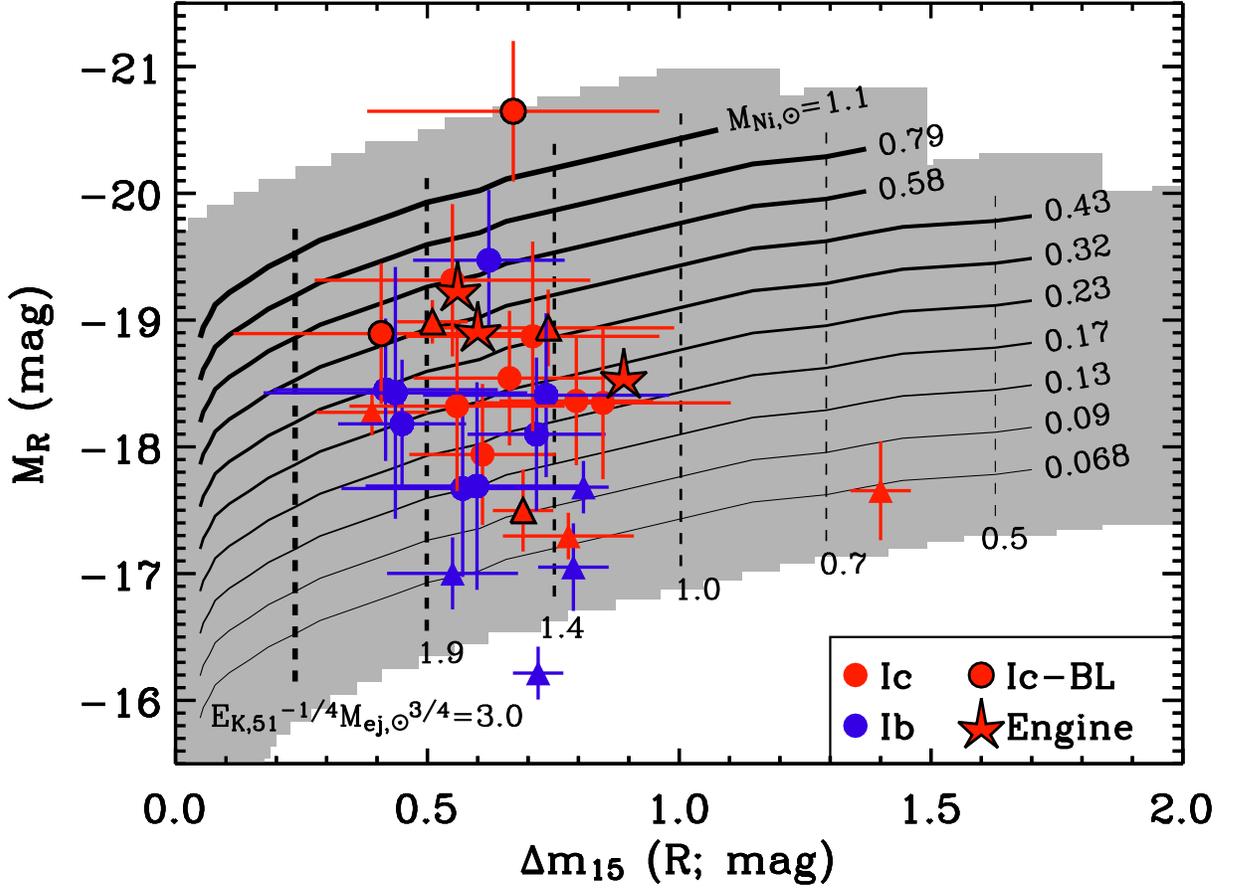}
\caption{We produce a two-dimensional parameter space for $M_R$ and $\Delta
m_{15,R}$ based on a range of reasonable values for $M_{\rm Ni}$,
$M_{\rm ej}$, and $E_K$ (grey region; see \S\ref{sec:disc}).  The
$M_{\rm Ni}$ values are shown with increasing thickness from bottom to
top (solid lines).  The quantity, $E_{K,51}^{-1/4}M_{\rm
ej,\odot}^{3/4}$ scales with $\tau_c$ and $\Delta m_{15,R}$ and
shown in decreasing thickness from left to right (dashed lines).  The
utility of this Figure is that it enables the explosion parameters to
be reasonably and systematically estimated without detailed modeling
of the light curves and spectra.}
\label{fig:tc_ni}
\end{figure}

For each model light curve (associated with specific values of $M_{\rm
Ni}$, $M_{\rm ej}$, and $E_K$), we {\it measure} the $\Delta m_{15,R}$
and $M_{R,\rm peak}$ values.  These two measurable quantities define a
two-dimension region in the $M_{\rm Ni}$ and $\tau_c$ parameter space
since a normalization constant is all that distinguishes $\tau_c$ from
the quantity, $M_{\rm ej,{\odot}}^{3/4}E_{K,51}^{-1/4}$.  Therefore,
within this parameter space, the two observable light curve properties
are mapped to the three physical explosion parameters
(Figure~\ref{fig:tc_ni}). A photospheric velocity measurement is
additionally required to break the degeneracy between $M_{\rm ej}$ and
$E_K$.

In Figure~\ref{fig:tc_ni} we compare the $\Delta m_{15,R}$ and
$M_{R,\rm peak}$ measurements for our sample of SNe Ibc with the grid
of synthetic values measured for our family of model light curves.
The utility of this Figure is that it enables the explosion parameters
to be reasonably and systematically estimated using only the two
observed quantities and sparing detailed modeling of the light curves
and spectra.  Extrapolating over the grid of synthetic parameters
($M_{R,\rm peak}$, $\Delta m_{15,R}$) for the family of models, we
convert our measurements to estimates for $M_{\rm Ni}$, $\tau_c$, and
$M_{\rm ej,\odot}^{3/4}E_{K,51}^{-1/4}$ for the SNe in our Gold and
Silver P60 samples (Table~\ref{tab:params}).  For a few of the
literature SNe and engine-driven SNe, detailed modeling of the light
curves and spectra have resulted in direct estimates for the $M_{\rm
Ni}$, and $M_{\rm ej}$, and $E_K$ values (e.g., 1994I;
\citealt{nmm+01}, 2002ap;
\citealt{mkm+07}, 1998bw; \citealt{imn+98}, 2006aj; \citealt{mdn+06},
and 2008D; \citealt{sbp+08}).  In these cases, our inferred physical
parameter estimates from Figure~\ref{fig:tc_ni} are reasonably
consistent with the values derived from detailed modeling thereby
supporting our method.

\subsection{Nickel Masses}

We report mean $^{56}$Ni masses for SNe Ib of $\langle M_{\rm
Ni}\rangle=0.20\pm 0.16~\rm M_{\odot}$ and for SNe Ic of $\langle
M_{\rm Ni}\rangle=0.24\pm 0.15~\rm M_{\odot}$ while for SNe Ic-BL we
find $\langle M_{\rm Ni}\rangle=0.58\pm 0.55~\rm M_{\odot}$.  For
comparison, we note that the typical $^{56}$Ni masses measured for SNe
Ia and IIP are $M_{\rm Ni}\approx 0.5~M_{\odot}$
\citep{mrb+07} and $M_{\rm Ni}\approx 0.07~M_{\odot}$, respectively
\citep{ntu+06}.

A K-S test on the $M_{\rm Ni}$ estimates reveals a 31\% probability
that SNe Ib and ordinary Ic are drawn from the same group of
explosions.  Meanwhile a K-S test on SNe Ic-BL and ordinary SNe Ic
reveals only a 3.9\% probability of originating from the same
population.  After grouping SNe Ib and ordinary Ic together, their
difference from SNe Ic-BL is accentuated; a K-S test shows only a 1.6\%
probability that they represent the same explosions.  

Focusing on the nearby engine-driven SNe, we find an average $^{56}$Ni
mass of $M_{\rm Ni}=0.40\pm 0.18~M_{\odot}$ for the three local
events.  Based on these $M_{\rm Ni}$ values, we calculate a 10\%
probability that engine-driven SNe are drawn from the overall SNe Ibc
population and a 40\% probability that they are drawn from the SNe
Ic-BL population.  We conclude that SNe Ib and Ic are characterized by
similar $^{56}$Ni mass values while SNe Ic-BL appear more extreme,
bearing more similarity to engine-driven explosions.  In this context,
we note that the especially bright Type Ic-BL SN\,2007D synthesized
an estimated $^{56}$Ni mass of $M_{\rm Ni}\approx 1.5~\rm M_{\odot}$,
and the peak luminosity is comparable to the pair-instability
candidate, SN\,2007bi \citep{gmo+09}.

\begin{figure}
\plotone{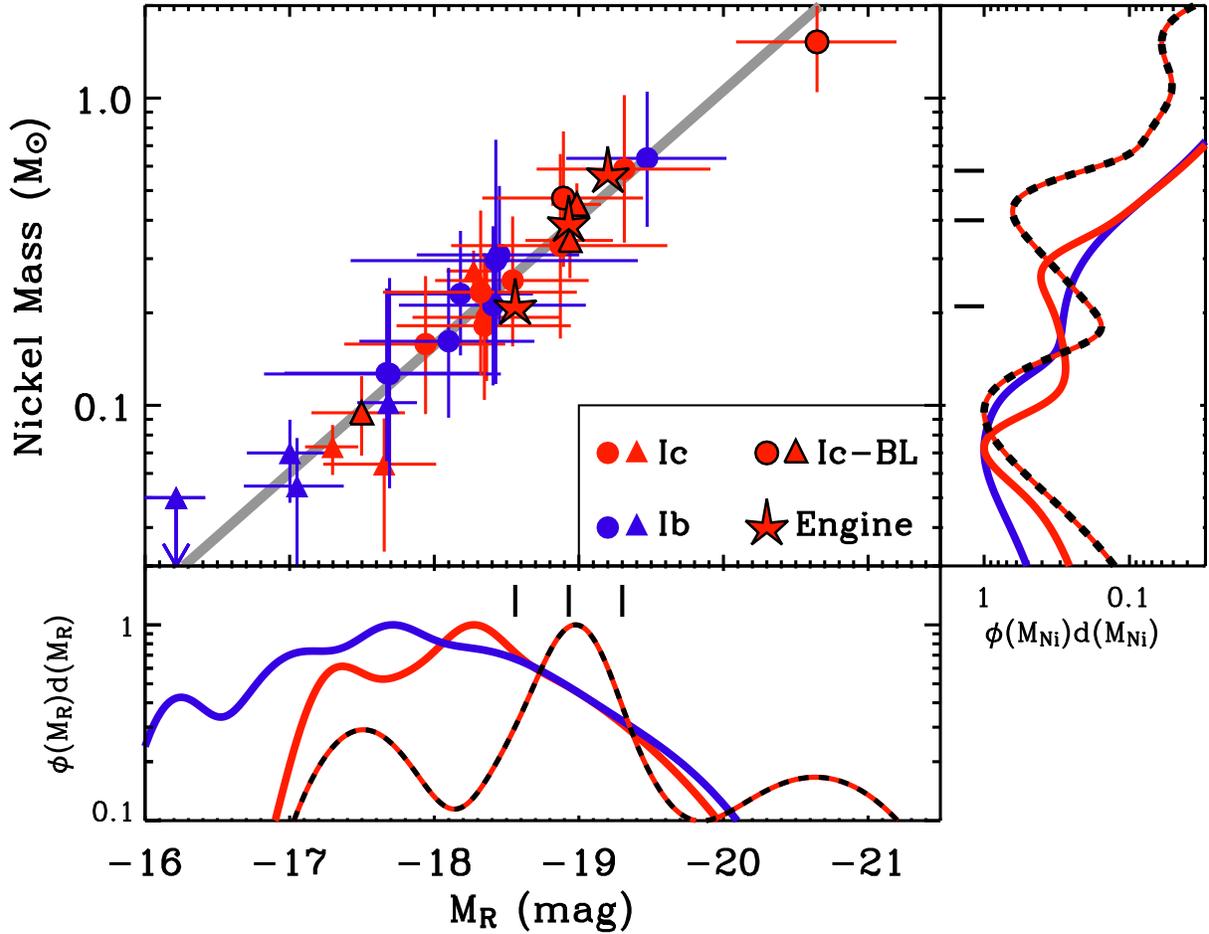}
\caption{The model-derived Nickel-56 mass, $M_{\rm Ni}$, and peak absolute magnitude, $M_R$, are compared for SNe Ib (blue), Ic (red) and broad-lined Ic (red encircled black) for the literature (triangles) and P60 (circles) samples.  Engine-driven SNe Ic-BL are highlighted as stars. A clear trend
is seen with an observed scatter due to the diversity in light curve
widths. We find a best fit for the function, $\rm log(M_{Ni})\approx
-0.41\,M_R-8.3$ (grey solid line).  The distributions of $M_R$ and
$M_{\rm Ni}$ are shown on the bottom and side panels, respectively,
with engine-driven explosions shown as tick marks.  SNe Ic-BL are more
luminous and synthesize more Nickel-56 than SNe Ib and ordinary SNe
Ic.  Engine-driven explosions are statistically consistent with the SNe Ic-BL
population.}
\label{fig:absmag_ni}
\end{figure}

In Figure~\ref{fig:absmag_ni}, we compare the derived estimates for
$M_{\rm Ni}$ with their observed $M_{R,\rm peak}$ values.  A clear
trend is seen with an intrinsic scatter due to the observed dispersion
in light curve width. The relation can be fit by the function, $\rm
log(M_{\rm Ni,\odot})\approx -0.41\,M_{R,\rm peak}-8.3$. This function
has roughly the same slope as the fit to a literature sample of
core-collapse SNe by \citet{pgm+10} but it is somewhat brighter.  For
future studies of SNe Ibc in which only an extinction-corrected
$M_{R,\rm peak}$ value is known, this relation can be used to estimate
$M_{\rm Ni}$ in a systematic way.

\subsection{Light Curve Widths}

We next extract $\tau_c$ estimates for each SN in our extended sample
and find similar average values for the sub-types of $\langle \tau_c
\rangle=13\pm 3$ d (Type Ib), $\langle \tau_c \rangle=12\pm 4$ d (Type
Ic), and $\langle \tau_c \rangle=14\pm 3$ d (Type Ic-BL; see
Table~\ref{tab:peak}).  A KS test on the sample $\tau_c$ values
suggests a 90\% probability that SNe Ib, Ic, and Ic-BL are drawn from
the same parent population of explosions.  For the engine-driven
explosions, we find a similar average value of $\langle \tau_c
\rangle=12\pm 3$ d.  Thus, light curve width and/or early decline rate
cannot be used to distinguish He-rich from He-poor events, nor does it
serve as a proxy for which SNe Ic-BL harbor central engines.  At the
same time, the large dispersion in light curve widths points to a
significant variation in the $M_{\rm ej}$ values and, in turn, the
progenitor mass.

\subsection{Estimates for $M_{\rm ej}$ and $E_K$}

With reasonable assumptions for the average photospheric velocities of
SNe Ib, Ic and Ic-BL, we may break the degeneracy between $M_{\rm ej}$
and $E_K$ implied by our mean $\tau_c$ estimates derived from our
$\Delta m_{15}$ measurements.  Within the framework of the analytic
light curve models, these two physical parameters depend on $v_{\rm
ph}$ as follows:

\begin{eqnarray}
M_{\rm ej} & \approx & 0.8~\left(\frac{\tau_c}{8~{\rm d}}\right)^2 \left(\frac{v_{\rm ph}}{10,000~\rm km~s^{-1}}\right) ~M_{\odot} \\
E_{K,51} & \approx & 0.5~\left(\frac{\tau_c}{8~{\rm d}}\right)^2 \left(\frac{v_{\rm ph}}{10,000~\rm km~s^{-1}}\right)^3~{\rm erg}.
\end{eqnarray}

\begin{figure}
\plotone{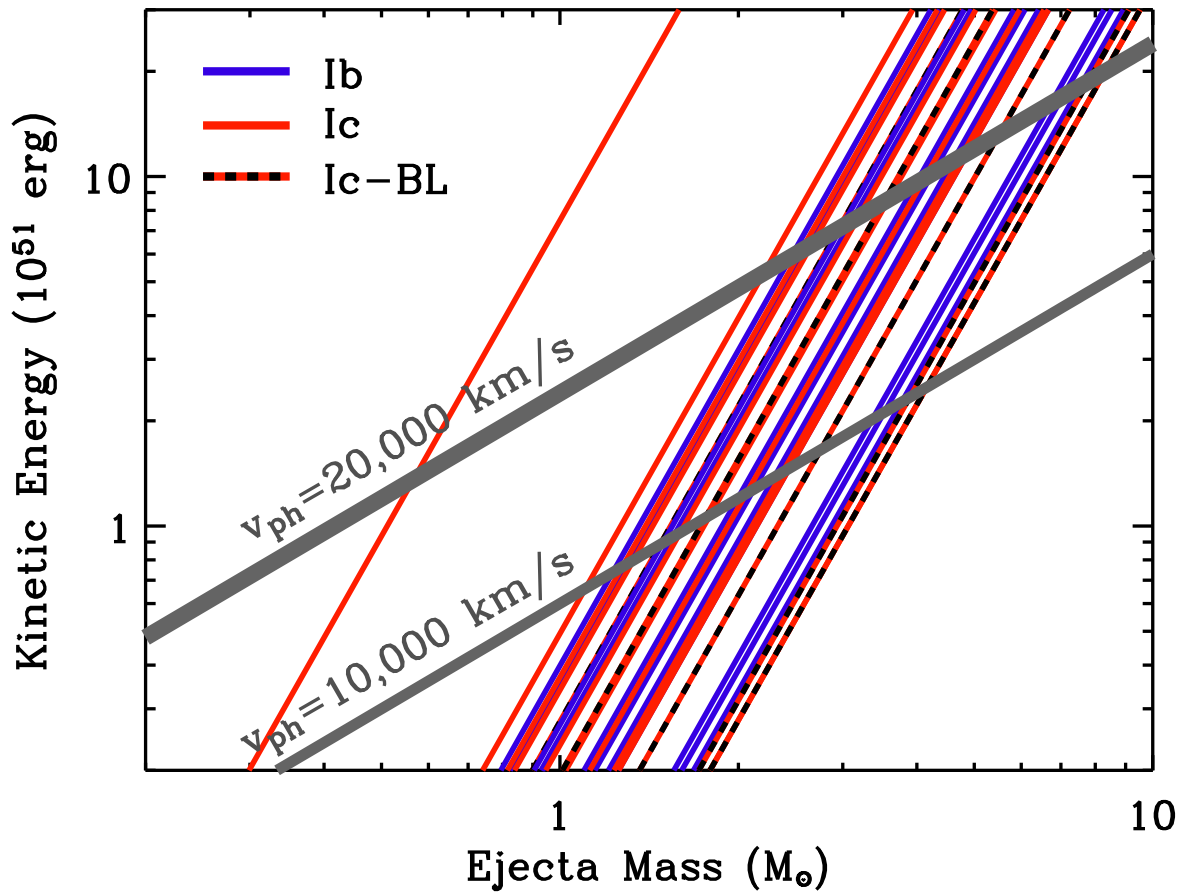}
\caption{Our light curve modeling for each of the SNe Ib (blue), Ic (red), and
Ic-BL (red/black) in our extended sample results in a degenerate
solution for $M_{\rm ej}$ and $E_K$.  Making reasonable assumptions
for the photospheric velocities, $v_{\rm ph}$, breaks this degeneracy.
We assume SNe Ibc have $v_{\rm ph}=10,000~\rm km~s^{-1}$ and SNe Ic-BL
have $v_{\rm ph}=20,000~\rm km~s^{-1}$ at the epoch of maximum light
(grey lines).  The intersection point of the $v_{\rm ph}$ lines with
each of the degenerate modeling solutions represents the implied
physical parameters.  We find average values of $M_{\rm ej}\approx
2~M_{\odot}$ and $E_{K,51}\approx 1~\rm erg$ for SNe Ib and Ic while
SNe Ic-BL show more extreme parameters: $M_{\rm ej}\approx
5~M_{\odot}$ and $E_{K,51}\approx 10~\rm erg$.}
\label{fig:me}
\end{figure}

\noindent
For SNe Ib and Ic we adopt $v_{\rm ph}\approx 10,000~\rm km~s^{-1}$ at
maximum light, in line with spectroscopic observations \citep{mfl+01}.
For SNe Ic-BL and engine-driven SNe, the observed range of $v_{\rm
ph}$ values is broader; we adopt $v_{\rm ph}=20,000~\rm km~s^{-1}$ for
the SNe Ic-BL at maximum light (e.g.,~\citealt{pmm+06}).  Combining
these assumptions with our average $\tau_c$ estimates for the combined
sample (Table~\ref{tab:params}), we find typical values of $M_{\rm
ej}\approx 2~M_{\odot}$ and $E_{K,51}\approx 1$ erg for SNe Ib and Ic
while for SNe Ic-BL we find higher values, $M_{\rm ej}\approx
5~M_{\odot}$ and $E_{K,51}\approx 10$ erg (Figure~\ref{fig:me}).  For
the three local engine-driven explosions we estimate $M_{\rm
ej}\approx 4~M_{\odot}$ and $E_{K,51}\approx 9$ erg
(Table~\ref{tab:peak}).  Thus, based on these reasonable estimates of $v_{\rm ph}$, we find no evidence for different
explosion parameters among SNe Ib and Ic, while those of SNe Ic-BL are
distinct and more closely resemble the values inferred for
engine-driven explosions.

\section{Implications for SNe Ibc Progenitors}
\label{sec:progenitors}

Here we consider the implications for the progenitors of SNe Ibc based
on their derived explosion parameters together with host galaxy and/or
explosion site diagnostics.  While we derive {\it similar} explosion
properties for SNe Ib and Ic, their explosion site properties imply
statistically significant {\it dissimilarities}. SNe Ic favor the more
luminous and metal-rich regions of their host galaxies than SNe Ib
(\citealt{kkp08,mbf+10} but see \citealt{acj+10}).  These explosion
site diagnostics suggest that SNe Ic progenitors are more massive
and/or younger and characterized by a slightly higher metal content
than the progenitors of SNe Ib.  More massive progenitors may produce
explosions with larger $M_{\rm ej}$ values unless they are able to
lose mass more efficiently throughout.  Since the
line-driven winds of Wolf-Rayet stars are enhanced at higher
metallicity ($\dot{M}\propto Z^{0.8}$;
\citealt{vd05}), we speculate that SNe Ic progenitors are initially more
massive than those of SNe Ib but lose mass more efficiently, resulting
in explosion parameters that appear indistinguishable from their helium-rich
cousins.

Next we consider the more extreme explosion parameters we infer for
SNe Ic-BL in comparison to those of SNe Ibc.  Explosion site
metallicities for SNe Ic-BL do not reveal strong dissimilarities from
those of SNe Ib and Ic (\citealt{acj+10,mbf+10} but see
\citealt{agk+10}).  However, a study of the explosion site 
luminosities indicate that SNe Ic-BL trace the star-forming light of
their host galaxies at least as tightly as SNe Ic and perhaps as
tightly as GRB-SNe \citep{kkp08}.  At the same time, the light curves of
SNe Ic-BL show evidence for more powerful explosions, with somewhat
larger $M_{\rm ej}$ values than ordinary SNe Ibc
(Table~\ref{tab:peak}).  We therefore speculate that SNe Ic-BL may
represent the explosion of more massive progenitor stars than other
SNe Ibc across a broad range of metallicities.  This may also explain
the existence of broad-lined SNe of Types Ib and IIb that stem
from metal-poor progenitors (e.g., SN\,2003bg;
\citealt{hdm+09}).

Finally we consider the relation of engine-driven SN progenitors to
those of SNe Ic-BL.  Based on their optical properties (light curves
and spectra), SNe Ic-BL and engine-driven SNe appear statistically
indistinguishable, however radio observations reveal relativistic
outflows in only a small fraction \citep{scp+10}.  Meanwhile,
explosion site metallicity measurements indicate that engine-driven
explosions tend to populate the low end of the metallicity distribution
for SNe Ibc \citep{mbf+10}, although there are some notable exceptions (see
\citealt{lsf+10}). In these cases, an additional key parameter may be
at play and possibly related to the rotation, binarity, or the nature
of the compact remnant of the progenitor system (black hole, neutron
star or magnetar;
\citealt{ywl10}).

%Additional clues may be derived from radio observations since they
%probe the dynamical interaction of the fastest ejecta (``blastwave'')
%with the local circumstellar medium \citep{che98}.  Radio observations
%to date indicate that SNe Ib, Ic, and Ic-BL are not dissimilar
%\citep{ams07}.  However, they are clearly distinguished from the
%exceptionally strong radio emission seen from engine-driven
%explosions, and attributed to a relativistic and energetic blastwave
%\citep{scp+10}.

\section{Conclusions}
\label{sec:future}

We present the first uniform and statistical sample of SNe Ibc
multi-band light curves available to date.  We find a significant
dispersion among the light curves both in peak luminosity and decay
rate. Through a comparison with the existing small sample of
well-observed SNe Ibc light curves from the literature, we find that a
significant fraction of SNe Ibc are heavily extinguished with
$E(B-V)_{\rm host}\approx 0.4$ mag.  After correcting our light curves for host
galaxy extinction, we compare differential and cumulative
distributions for the peak absolute magnitudes of SNe Ib, Ic, and
Ic-BL.  We find that the peak luminosity distributions for SNe Ib and
Ic are statistically indistinguishable, with $\langle M_{R,\rm
peak}\rangle \approx -18$ mag and are therefore not inconsistent with
a single progenitor channel for helium-poor and helium-rich events.  A
comparison of their early decline rates supports this hypothesis and
we note there is no evidence for a correlation with peak absolute
magnitude as is seen for SNe Ia.  However, we find that SNe Ic-BL are
typically more luminous with $\langle M_R\rangle \approx -19$ mag.
The probability that they belong to the same set of progenitors and/or
explosions as ordinary SNe Ib and Ic is just 2\%.

We compare these observed light curve properties with those of the
three nearby engine-driven SNe discovered within the same volume and
find a significant overlap with the SNe Ic-BL population.  The
probability that engine-driven SNe are drawn from the SNe Ic-BL
population is 25\%, based on these optical diagnostics alone.  This
result underscores the point that neither high luminosity nor 
fast photospheric velocity can be used as a robust indicator for an
engine-driven explosion, and thus radio searches for a relativistic
outflow are required to distinguish (e.g., \citealt{scp+10}).  

We model the optical light curves in a systematic fashion to extract
estimates of the three physical parameters, $M_{\rm Ni}$, $M_{\rm
ej}$, and $E_K$ and find that the results reiterate those reviewed
above.  Namely, there is no significant difference between SNe Ib and
Ic and we estimate their physical parameters to be $M_{\rm Ni}\approx
0.2~M_{\odot}$, $M_{\rm ej}\approx 2~M_{\odot}$, and $E_{K,51}\approx
1$ erg.  Meanwhile, SNe Ic-BL and engine-driven events are similar to
one another and distinct from ordinary SNe Ib and Ic; for them we
estimate more extreme physical parameters of $M_{\rm Ni}\approx
0.5~M_{\odot}$, $M_{\rm ej}\approx 5~M_{\odot}$, and $E_{K,51}\approx
10$ erg.  The statistically significant difference between ordinary
SNe Ibc and SNe Ic-BL (including engine-driven events) suggests that
the latter share a distinct $^{56}$Ni production mechanism and/or
ejecta distribution (e.g., bipolar outflow).

Looking forward, the inclusion of spectroscopic diagnostics (e.g.,
$v_{\rm ph}$ measurements, He I feature intensities) would enable an
extension of this systematic study to search for additional
correlations and to properly break the model degeneracies between
$M_{\rm ej}$ and $E_K$.  Along this line, we reiterate that ten of the
SNe in our study were observed as part of the CCCP and thus
spectroscopic observations (often extensive) exist for this
sub-sample.  This project will be the focus of a separate study.
Furthermore, a direct comparison of host galaxy diagnostics with
optical light curve properties (e.g., peak luminosity) may reveal
further clues on the nature of SN Ibc progenitors and their relation
to those of engine-driven explosions.  Since our sample is drawn from
targeted surveys, it is likely biased toward metal-rich host galaxies.
Fortunately, new wide-field transient surveys such as Pan-STARRS
\citep{kab+02} and the Palomar Transient Factory \citep{lkd+09} are
discovering SNe in a broad range of host galaxy environments and will
enable such comparisons to be made in an unbiased fashion.  Moreover,
thanks to the higher cadence of these optical surveys, SNe Ibc will be
discovered earlier with respect to the explosion date, reducing the
fraction of Bronze SNe in future follow-up samples.  Finally we note
that the significant $E(B-V)_{\rm host}$ values inferred here
motivates the regular use of near-IR facilities in future studies of
SNe Ibc.

\medskip

M.R.D. and A.M.S. acknowledge support by the National Science Foundation
Research Experiences for Undergraduates (REU) and Department of
Defense Awards to Stimulate and Support Undergraduate Research
Experiences (ASSURE) programs under Grant no. 0754568 and by the
Smithsonian Institution.  The work of A.G. is supported by grants from
the Israeli Science Foundation (ISF), an EU/FP7 Marie Curie IRG
Fellowship and a research grant from the Gruber Awards.  S.B.C
acknowledges generous support from Gary and Cynthia Bengier and the
Richard and Rhoda Goldman Foundation.  The work of D.C.L. is supported
by National Science Foundation (NSF) grant AST-1009571.  D.C.L. is
grateful for an NSF Astronomy and Astrophysics Postdoctoral Fellowship
under award AST-0401479, during which part of this work was completed.

\bibliographystyle{apj1b}
%\bibliography{journals_apj,refs}

\begin{thebibliography}{}

\bibitem[{Aldering} {\it et al.}\ (2005)]{A451}
{Aldering}, G. {\it et al.}\  2005, The Astronomer's Telegram, 451, 1.

\bibitem[{Anderson} {\it et al.}\ (2010)]{acj+10}
{Anderson}, J.~P., {Covarrubias}, R.~A., {James}, P.~A., {Hamuy}, M., and
  {Habergham}, S.~M. 2010, \mnras, 407, 2660.

\bibitem[{Antilogus} {\it et al.}\ (2006)]{A854}
{Antilogus}, P. {\it et al.}\  2006, The Astronomer's Telegram, 854, 1.

\bibitem[{Arcavi} {\it et al.}\ (2010)]{agk+10}
{Arcavi}, I. {\it et al.}\  2010, \apj, 721, 777.

\bibitem[{Arnett}(1982)]{arn82}
{Arnett}, W.~D. 1982, \apj, 253, 785.

\bibitem[{Baek} {\it et al.}\ (2006)]{i8660}
{Baek}, M., {Li}, W., {Puckett}, T., {Sostero}, G., and {Garzia}, S. 2006,
  \iaucirc, 8660, 1.

\bibitem[{Begelman} \& {Sarazin}(1986)]{bs86}
{Begelman}, M.~C. and {Sarazin}, C.~L. 1986, \apjl, 302, L59.

\bibitem[{Berger} {\it et al.}\ (2003)]{bkf+03}
{Berger}, E., {Kulkarni}, S.~R., {Frail}, D.~A., and {Soderberg}, A.~M. 2003,
  \apj, 599, 408.

\bibitem[{Blondin} {\it et al.}\ (2006)]{c626}
{Blondin}, S., {Modjaz}, M., {Kirshner}, R., {Challis}, P., and {Calkins}, M.
  2006, Central Bureau Electronic Telegrams, 626, 1.

\bibitem[{Blondin} {\it et al.}\ (2009)]{bpp+09}
{Blondin}, S. {\it et al.}\ 2009, \apj, 693, 207.

\bibitem[{Cano} {\it et al.}\ (2011)]{cbg+11}
{Cano}, Z. {\it et al.}\ 2011, arxiv:1104.5141.

\bibitem[{Cenko} {\it et al.}\ (2006)]{cfm+06}
{Cenko}, S.~B. {\it et al.}\  2006, \pasp, 118, 1396.

\bibitem[{Chevalier}(1998)]{che98}
{Chevalier}, R.~A. 1998, \apj, 499, 810.

\bibitem[{Chevalier} \& {Fransson}(2006)]{cf06}
{Chevalier}, R.~A. and {Fransson}, C. 2006, \apj, 651, 381.

\bibitem[{Chevalier} \& {Soderberg}(2010)]{cs10}
{Chevalier}, R.~A. and {Soderberg}, A.~M. 2010, \apjl, 711, L40.

\bibitem[{Clocchiatti} {\it et al.}\ (1996)]{cwb+96}
{Clocchiatti}, A., {Wheeler}, J.~C., {Benetti}, S. and {Frueh}, M. 1996, \apj, 459, 547.

\bibitem[{Clocchiatti} \& Wheeler (1997a)]{cw97a}
{Clocchiatti}, A. and {Wheeler}, J.~C. 1997a, \apj, 491, 375.

\bibitem[{Clocchiatti} \& Wheeler (1997b)]{cw97b}
{Clocchiatti}, A. and {Wheeler}, J.~C. 1997b, astro-ph/9601023.

\bibitem[{Colesanti} {\it et al.}\ (2006)a]{i8658}
{Colesanti}, C. {\it et al.}\  2006a, \iaucirc, 8658, 2.

\bibitem[{Colesanti} {\it et al.}\ (2006)b]{i8713}
{Colesanti}, C. {\it et al.}\  2006b, \iaucirc, 8713, 1.

\bibitem[{Conley} {\it et al.}\ (2006)]{chh+06}
{Conley}, A. {\it et al.}\  2006, \aj, 132, 1707.

\bibitem[{de Vaucouleurs} {\it et al.}\ (1991)]{rc3}
{de Vaucouleurs}, G., {de Vaucouleurs}, A., {Corwin}, Jr., H.~G., {Buta},
  R.~J., {Paturel}, G., and {Fouque}, P. 1991, { {Third Reference Catalogue of
  Bright Galaxies}}, ).

\bibitem[{Elias} {\it et al.}\ (1985)]{emn+85}
{Elias}, J.~H., {Matthews}, K., {Neugebauer}, G., and {Persson}, S.~E. 1985,
  \apj, 296, 379.

\bibitem[{Filippenko}(1997)]{fil97}
{Filippenko}, A.~V. 1997, \araa, 35, 309.

\bibitem[{Filippenko} \& {Foley}(2004)a]{i8543}
{Filippenko}, A.~V. and {Foley}, R.~J. 2004a, \iaucirc, 8453, 3.

\bibitem[{Filippenko} \& {Foley}(2004)b]{i8453}
{Filippenko}, A.~V. and {Foley}, R.~J. 2004b, \iaucirc, 8453, 3.

\bibitem[{Filippenko}, {Foley} \& {Matheson}(2005)]{i8639b}
{Filippenko}, A.~V., {Foley}, R.~J., and {Matheson}, T. 2005, \iaucirc, 8639,
  2.

\bibitem[{Filippenko} {\it et al.}\ (2004)]{i8404}
{Filippenko}, A.~V., {Ganeshalingam}, M., {Serduke}, F.~J.~D., and {Hoffman},
  J.~L. 2004, \iaucirc, 8404, 1.

\bibitem[{Filippenko} \& {Sargent}(1985)]{fs85}
{Filippenko}, A.~V. and {Sargent}, W.~L.~W. 1985, \nat, 316, 407.

\bibitem[{Folatelli} {\it et al.}\ (2006)]{fcp+06}
{Folatelli}, G. {\it et al.}\  2006, \apj, 641, 1039.

\bibitem[{Foley} \& {Gal-Yam}(2007)]{c805}
{Foley}, R.~J. and {Gal-Yam}, A. 2007, Central Bureau Electronic Telegrams,
  805, 1.

\bibitem[{Foley} {\it et al.}\ (2003)]{fps+03}
{Foley}, R.~J. {\it et al.}\  2003, \pasp, 115, 1220.

\bibitem[{Foley} {\it et al.}\ (2007)]{fsg+07}
{Foley}, R.~J., {Smith}, N., {Ganeshalingam}, M., {Li}, W., {Chornock}, R., and
  {Filippenko}, A.~V. 2007, \apjl, 657, L105.

\bibitem[{Gal-Yam} {\it et al.}\ (2007)]{gcf+07}
{Gal-Yam}, A., {Cenko}, S.~B., {Fox}, D.~B., {Leonard}, D.~C., {Moon}, D.-S.,
  {Sand}, D.~J., and {Soderberg}, A.~M. 2007, in { The Multicolored Landscape
  of Compact Objects and Their Explosive Origins}, ed.\ T. {di Salvo}, G.~L.
  {Israel}, L. {Piersant}, L. {Burderi}, G. {Matt}, A. {Tornambe}, and M.~T.
  {Menna}, volume 924 of { American Institute of Physics Conference Series},
  297.

\bibitem[{Gal-Yam} {\it et al.}\ (2005)]{gfk+05}
{Gal-Yam}, A. {\it et al.}\  2005, \apjl, 630, L29.

\bibitem[{Gal-Yam} {\it et al.}\ (2009)]{gmo+09}
{Gal-Yam}, A. {\it et al.}\  2009, \nat, 462, 624.

\bibitem[{Galama} {\it et al.}\ (1998)]{gvv+98}
{Galama}, T.~J. {\it et al.}\  1998, \nat, 395, 670.

\bibitem[{Ganeshalingam} {\it et al.}\ (2004)]{i8456}
{Ganeshalingam}, M., {Swift}, B.~J., {Serduke}, F.~J.~D., and {Filippenko},
  A.~V. 2004, \iaucirc, 8456, 4.

\bibitem[{Graham} \& {Li}(2004)a]{i8377}
{Graham}, J. and {Li}, W. 2004a, \iaucirc, 8377, 2.

\bibitem[{Graham} \& {Li}(2004)b]{i8381}
{Graham}, J. and {Li}, W. 2004b, \iaucirc, 8381, 1.

\bibitem[{Greiner} {\it et al.}\ (2010)]{gkk+10}
{Greiner}, J. {\it et al.}\  2010, in { American Institute of Physics
  Conference Series}, ed.\ {N.~Kawai \& S.~Nagataki}, volume 1279 of { American
  Institute of Physics Conference Series}, 144.

\bibitem[{Hamuy} {\it et al.}\ (2009)]{hdm+09}
{Hamuy}, M. {\it et al.}\  2009, \apj, 703, 1612.

\bibitem[{Hamuy} {\it et al.}\ (2002)]{hmp+02}
{Hamuy}, M. {\it et al.}\  2002, \aj, 124, 417.

\bibitem[{Hayden} {\it et al.}\ (2010)]{hgk+10}
{Hayden}, B.~T. {\it et al.}\  2010, \apj, 712, 350.

\bibitem[{Hubble}(1926)]{hub26}
{Hubble}, E.~P. 1926, \apj, 64, 321.

\bibitem[{Hunter} {\it et al.}\ (2009)]{hvk+09}
{Hunter}, D.~J. {\it et al.}\  2009, \aap, 508, 371.

\bibitem[{Itagaki} {\it et al.}\ (2006)]{i8762}
{Itagaki}, K. {\it et al.}\  2006, \iaucirc, 8762, 1.

\bibitem[{Iwamoto} {\it et al.}\ (1998)]{imn+98}
{Iwamoto}, K. {\it et al.}\  1998, \nat, 395, 672.

\bibitem[{Iwamoto} {\it et al.}\ (2003)]{inm+03}
{Iwamoto}, K., {Nomoto}, K., {Mazzali}, P.~A., and {et al.} 2003, in {
  Supernovae and Gamma-Ray Bursters}, ed.\ K. {Weiler}, volume 598 of { Lecture
  Notes in Physics, Berlin Springer Verlag}, 243.

\bibitem[{Joubert} {\it et al.}\ (2007)]{i8794}
{Joubert}, N., {Li}, W., {Puckett}, T., {Orff}, T., {Gagliano}, R., and
  {Sehgal}, A. 2007, \iaucirc, 8794, 3.

\bibitem[{Kaiser} {\it et al.}\ (2002)]{kab+02}
{Kaiser}, N. {\it et al.}\  2002, in { Society of Photo-Optical Instrumentation
  Engineers (SPIE) Conferen\ ce Series}, ed.\ J.~A. {Tyson} and S. {Wolff},
  volume 4836 of { Society of Photo-Optical Instrumentation Engineers (SPIE)
  Conferen\ ce Series}, 154.

\bibitem[{Kelly}, {Kirshner} \& {Pahre}(2008)]{kkp08}
{Kelly}, P.~L., {Kirshner}, R.~P., and {Pahre}, M. 2008, \apj, 687, 1201.

\bibitem[{Law} {\it et al.}\ (2009)]{lkd+09}
{Law}, N.~M. {\it et al.}\  2009, ArXiv e-prints.

\bibitem[{Lee}, {Baek} \& {Li}(2006)]{i8669}
{Lee}, E., {Baek}, M., and {Li}, W. 2006, \iaucirc, 8669, 1.

\bibitem[{Levesque} {\it et al.}\ (2010)]{lsf+10}
{Levesque}, E.~M. {\it et al.}\  2010, \apjl, 709, L26.

\bibitem[{Li} {\it et al.}\ (2010)]{llc+10}
{Li}, W. {\it et al.}\  2010, ArXiv e-prints.

\bibitem[{MacFadyen}, {Woosley} \& {Heger}(2001)]{mwh01}
{MacFadyen}, A.~I., {Woosley}, S.~E., and {Heger}, A. 2001, \apj, 550, 410.

\bibitem[{Matheson} {\it et al.}\ (2001)]{mfl+01}
{Matheson}, T., {Filippenko}, A.~V., {Li}, W., {Leonard}, D.~C., and {Shields},
  J.~C. 2001, \aj, 121, 1648.

\bibitem[{Mazzali} {\it et al.}\ (2006)]{mdn+06}
{Mazzali}, P.~A. {\it et al.}\  2006, \nat, 442, 1018.

\bibitem[{Mazzali} {\it et al.}\ (2007)a]{mkm+07}
{Mazzali}, P.~A. {\it et al.}\  2007a, \apj, 670, 592.

\bibitem[{Mazzali} {\it et al.}\ (2007)b]{mrb+07}
{Mazzali}, P.~A., {R{\"o}pke}, F.~K., {Benetti}, S., and {Hillebrandt}, W.
  2007b, Science, 315, 825.

\bibitem[{Mirabal} {\it et al.}\ (2006)]{mha+06}
{Mirabal}, N., {Halpern}, J.~P., {An}, D., {Thorstensen}, J.~R., and
  {Terndrup}, D.~M. 2006, \apjl, 643, L99.

\bibitem[{Modjaz} {\it et al.}\ (2006)]{c677}
{Modjaz}, M., {Blondin}, S., {Kirshner}, R., {Challis}, P., {Matheson}, T., and
  {Mamajek}, E. 2006, Central Bureau Electronic Telegrams, 677, 1.

\bibitem[{Modjaz} {\it et al.}\ (2010)]{mbf+10}
{Modjaz}, M., {Bloom}, J.~S., {Filippenko}, A.~V., {Kewley}, L., {Perley}, D.,
  and {Silverman}, J.~M. 2010, ArXiv e-prints.

\bibitem[{Modjaz} {\it et al.}\ (2004)a]{i8428}
{Modjaz}, M., {Challis}, P., {Kirshner}, R., {Garg}, A., {Stubbs}, C., and
  {Matheson}, T. 2004a, \iaucirc, 8428, 3.

\bibitem[{Modjaz} {\it et al.}\ (2005)a]{i8605b}
{Modjaz}, M., {Challis}, P., {Kirshner}, R., {Matheson}, T., and {Berlind}, P.
  2005a, \iaucirc, 8605, 2.

\bibitem[{Modjaz} {\it et al.}\ (2004)b]{i8426}
{Modjaz}, M., {Challis}, P., {Kirshner}, R., {Matheson}, T., {Garg}, A., and
  {Stubbs}, C. 2004b, \iaucirc, 8426, 3.

\bibitem[{Modjaz} {\it et al.}\ (2005)b]{i8650}
{Modjaz}, M., {Kirshner}, R., {Challis}, P., {Blondin}, S., and {Berlind}, P.
  2005b, \iaucirc, 8650, 2.

\bibitem[{Modjaz} {\it et al.}\ (2005)c]{c271}
{Modjaz}, M., {Kirshner}, R., {Challis}, P., {Matheson}, T., and {Berlind}, P.
  2005c, Central Bureau Electronic Telegrams, 271, 1.

\bibitem[{Modjaz} {\it et al.}\ (2005)d]{i8461}
{Modjaz}, M., {Kirshner}, R., {Challis}, P., {Matheson}, T., {Falco}, E., and
  {Berlind}, P. 2005d, \iaucirc, 8461, 2.

\bibitem[{Monard} {\it et al.}\ (2004)]{i8454}
{Monard}, L.~A.~G., {Quimby}, R., {Gerardy}, C., {Hoeflich}, P., {Wheeler},
  J.~C., {Chen}, Y., {Smith}, H.~J., and {Bauer}, A. 2004, \iaucirc, 8454, 1.

\bibitem[{Moore}, {Shimasaki} \& {Li}(2004)]{i8443}
{Moore}, M., {Shimasaki}, K., and {Li}, W. 2004, \iaucirc, 8443, 3.

\bibitem[{Mould} {\it et al.}\ (2000)]{mhf+00}
{Mould}, J.~R. {\it et al.}\  2000, \apj, 529, 786.

\bibitem[{Munari} \& {Zwitter}(1997)]{mz97}
{Munari}, U. and {Zwitter}, T. 1997, \aap, 318, 269.

\bibitem[{Newton} \& {Puckett}(2005)]{i8648}
{Newton}, J. and {Puckett}, T. 2005, \iaucirc, 8648, 2.

\bibitem[{Nomoto} {\it et al.}\ (2001)]{nmm+01}
{Nomoto}, K., {Maeda}, K., {Mochizuki}, Y., {Kumagai}, S., {Umeda}, H.,
  {Nakamura}, T., and {Tanihata}, I. 2001, in { Gamma 2001: Gamma-Ray
  Astrophysics}, ed.\ {S.~Ritz, N.~Gehrels, \& C.~R.~Shrader}, volume 587 of {
  American Institute of Physics Conference Series}, 487.

\bibitem[{Nomoto} {\it et al.}\ (2006)]{ntu+06}
{Nomoto}, K., {Tominaga}, N., {Umeda}, H., {Kobayashi}, C., and {Maeda}, K.
  2006, Nuclear Physics A, 777, 424.

\bibitem[{Pastorello} {\it et al.}\ (2008)a]{pkc+08}
{Pastorello}, A. {\it et al.}\  2008a, \mnras, 389, 955.

\bibitem[{Pastorello} {\it et al.}\ (2008)b]{pqs+08}
{Pastorello}, A. {\it et al.}\  2008b, \mnras, 389, 131.

\bibitem[{Pastorello} {\it et al.}\ (2007)]{psm+07}
{Pastorello}, A. {\it et al.}\  2007, \nat, 447, 829.

\bibitem[{Perets} {\it et al.}\ (2010)]{pgm+10}
{Perets}, H.~B. {\it et al.}\  2010, \nat, 465, 322.

\bibitem[{Perley} {\it et al.}\ (2009)]{pcb+09}
{Perley}, D.~A. {\it et al.}\  2009, ArXiv e-prints.

\bibitem[{Phillips}(1993)]{phi93}
{Phillips}, M.~M. 1993, \apjl, 413, L105.

\bibitem[{Phillips} {\it et al.}\ (1999)]{pls+99}
{Phillips}, M.~M., {Lira}, P., {Suntzeff}, N.~B., {Schommer}, R.~A., {Hamuy},
  M., and {Maza}, J. 1999, \aj, 118, 1766.

\bibitem[{Pian} {\it et al.}\ (2000)]{paa+00}
{Pian}, E. {\it et al.}\  2000, \apj, 536, 778.

\bibitem[{Pian} {\it et al.}\ (2006)]{pmm+06}
{Pian}, E. {\it et al.}\  2006, \nat, 442, 1011.

\bibitem[{Pignata} {\it et al.}\ (2011)]{pig10}
{Pignata}, G. {\it et.~al.}\ 2011, \apj, 728, 14.

\bibitem[{Podsiadlowski}, {Joss} \& {Hsu}(1992)]{pjh92}
{Podsiadlowski}, P., {Joss}, P.~C., and {Hsu}, J.~J.~L. 1992, \apj, 391, 246.

\bibitem[{Podsiadlowski} {\it et al.}\ (2004)]{pmn+04}
{Podsiadlowski}, P., {Mazzali}, P.~A., {Nomoto}, K., {Lazzati}, D., and
  {Cappellaro}, E. 2004, \apjl, 607, L17.

\bibitem[{Porter} \& {Filippenko}(1987)]{pf87}
{Porter}, A.~C. and {Filippenko}, A.~V. 1987, \aj, 93, 1372.

\bibitem[{Prasad} {\it et al.}\ (2006)]{i8770}
{Prasad}, R.~R., {Joubert}, N., {Baek}, M., {Li}, W., {Krzeminsky}, W., and
  {Contreras}, C. 2006, \iaucirc, 8770, 1.

\bibitem[{Prieto}, {Stanek} \& {Beacom}(2008)]{psb08}
{Prieto}, J.~L., {Stanek}, K.~Z., and {Beacom}, J.~F. 2008, \apj, 673, 999.

\bibitem[{Puckett} {\it et al.}\ (2006)a]{i8750}
{Puckett}, T., {Gagliano}, R., {Mostardi}, R., {Li}, W., {Frieman}, J., and
  {Grcevich}, J. 2006a, \iaucirc, 8750, 2.

\bibitem[{Puckett} {\it et al.}\ (2007)]{i8792}
{Puckett}, T., {Orff}, T., {Madison}, D., {Li}, W., {Itagaki}, K., {Nakano},
  S., {Newton}, J., and {Kadota}, K. 2007, \iaucirc, 8792, 2.

\bibitem[{Puckett} {\it et al.}\ (2006)b]{i8741}
{Puckett}, T., {Peoples}, M., {Joubert}, N., {Madison}, D.~R., {Mostardi}, R.,
  {Khandrika}, H., {Li}, W., and {Foley}, R.~J. 2006b, \iaucirc, 8741, 1.

\bibitem[{Puckett} {\it et al.}\ (2005)a]{i8605}
{Puckett}, T., {Peoples}, M., {Prasad}, R.~R., {Li}, W., {Lee}, E., {Berlind},
  P., and {Faworski}, S. 2005a, \iaucirc, 8605, 1.

\bibitem[{Puckett} {\it et al.}\ (2005)b]{i8639}
{Puckett}, T., {Sostero}, G., {Quimby}, R., and {Mondol}, P. 2005b, \iaucirc,
  8639, 1.

\bibitem[{Pugh} {\it et al.}\ (2004)]{i8452}
{Pugh}, H., {Li}, W., {Manzini}, F., and {Behrend}, R. 2004, \iaucirc, 8452, 2.

\bibitem[{Pugh}, {Park} \& {Li}(2004)]{i8425}
{Pugh}, H., {Park}, S., and {Li}, W. 2004, \iaucirc, 8425, 1.

\bibitem[{Quimby} {\it et al.}\ (2004)]{i8446}
{Quimby}, R. {\it et al.}\  2004, \iaucirc, 8446, 1.

\bibitem[{Quimby} {\it et al.}\ (2006)]{i8657}
{Quimby}, R., {Mondol}, P., {Castro}, F., {Roman}, B., and {Rostopchin}, S.
  2006, \iaucirc, 8657, 1.

\bibitem[{Quimby} {\it et al.}\ (2005)]{i8503}
{Quimby}, R., {Mondol}, P., {Hoeflich}, P., {Wheeler}, J.~C., and {Gerardy}, C.
  2005, \iaucirc, 8503, 1.

\bibitem[{Quimby} {\it et al.}\ (2009)]{qkm+09}
{Quimby}, R.~M. {\it et al.}\  2009, ArXiv e-prints.

\bibitem[{Richardson}(2009)]{ric09}
{Richardson}, D. 2009, \aj, 137, 347.

\bibitem[{Richardson}, {Branch} \& {Baron}(2006)]{rbb06}
{Richardson}, D., {Branch}, D., and {Baron}, E. 2006, \aj, 131, 2233.

\bibitem[{Richmond} {\it et al.}\ (1996)]{rvh+96}
{Richmond}, M.~W. {\it et al.}\  1996, \aj, 111, 327.

\bibitem[{Riess} {\it et al.}\ (1999)]{rfl+99}
{Riess}, A.~G. {\it et al.}\  1999, \aj, 118, 2675.

\bibitem[{Sahu} {\it et al.}\ (2009)]{sta+09}
{Sahu}, D.~K., {Tanaka}, M., {Anupama}, G.~C., {Gurugubelli}, U.~K., and
  {Nomoto}, K. 2009, \apj, 697, 676.

\bibitem[{Saviane}, {Hibbard} \& {Rich}(2004)]{shr04}
{Saviane}, I., {Hibbard}, J.~E., and {Rich}, R.~M. 2004, \aj, 127, 660.

\bibitem[{Schlegel}, {Finkbeiner} \& {Davis}(1998)a]{sfd98}
{Schlegel}, D.~J., {Finkbeiner}, D.~P., and {Davis}, M. 1998a, \apj, 500, 525.

\bibitem[{Schlegel}, {Finkbeiner} \& {Davis}(1998)b]{sfd+98}
{Schlegel}, D.~J., {Finkbeiner}, D.~P., and {Davis}, M. 1998b, \apj, 500, 525.

\bibitem[{Schwehr} \& {Li}(2006)]{i8728}
{Schwehr}, J. and {Li}, W. 2006, \iaucirc, 8728, 2.

\bibitem[{Shimasaki} \& {Li}(2005)]{i8623}
{Shimasaki}, K. and {Li}, W. 2005, \iaucirc, 8623, 2.

\bibitem[{Silbermann} {\it et al.}\ (1996)]{shm+96}
{Silbermann}, N.~A. {\it et al.}\  1996, \apj, 470, 1.

\bibitem[{Smartt}(2009)]{sma09}
{Smartt}, S.~J. 2009, \araa, 47, 63.

\bibitem[{Smartt} {\it et al.}\ (2009)]{sec+09}
{Smartt}, S.~J., {Eldridge}, J.~J., {Crockett}, R.~M., and {Maund}, J.~R. 2009,
  \mnras, 395, 1409.

\bibitem[{Smith} {\it et al.}\ (2002)]{stk+02}
{Smith}, J.~A. {\it et al.}\  2002, \aj, 123, 2121.

\bibitem[{Smith}, {Foley} \& {Filippenko}(2008)]{sff08}
{Smith}, N., {Foley}, R.~J., and {Filippenko}, A.~V. 2008, \apj, 680, 568.

\bibitem[{Soderberg}(2007)]{ams07}
{Soderberg}, A.~M. 2007, Caltech PhD Thesis.

\bibitem[{Soderberg} {\it et al.}\ (2008)]{sbp+08}
{Soderberg}, A.~M. {\it et al.}\  2008, \nat, 453, 469.

\bibitem[{Soderberg} {\it et al.}\ (2010)]{scp+10}
{Soderberg}, A.~M. {\it et al.}\  2010, \nat, 463, 513.

\bibitem[{Soderberg} {\it et al.}\ (2006)a]{graph}
{Soderberg}, A.~M. {\it et al.}\  2006a, \apj, 636, 391.

\bibitem[{Soderberg} {\it et al.}\ (2006)b]{snb+06}
{Soderberg}, A.~M., {Nakar}, E., {Berger}, E., and {Kulkarni}, S.~R. 2006b,
  \apj, 638, 930.

\bibitem[{Stritzinger} {\it et al.}\ (2002)]{shs+02}
{Stritzinger}, M. {\it et al.}\  2002, \aj, 124, 2100.

\bibitem[{Stritzinger} {\it et al.}\ (2009)]{smp+09}
{Stritzinger}, M. {\it et al.}\  2009, \apj, 696, 713.

\bibitem[{Tak{\'a}ts} \& {Vink{\'o}}(2006)]{tk06}
{Tak{\'a}ts}, K. and {Vink{\'o}}, J. 2006, \mnras, 372, 1735.

\bibitem[{Taubenberger} {\it et al.}\ (2006)]{tpm+06}
{Taubenberger}, S. {\it et al.}\  2006, \mnras, 371, 1459.

\bibitem[{Thompson}, {Chang} \& {Quataert}(2004)]{tcq04}
{Thompson}, T.~A., {Chang}, P., and {Quataert}, E. 2004, \apj, 611, 380.

\bibitem[{Uomoto}, \& {Kirshner}(1985)]{uk85}
{Uomoto}, A. \& {Kirshner}, R.~P. 1985, \aap, 149, L7. 

\bibitem[{Valenti} {\it et al.}\ (2008)]{vbc+08}
{Valenti}, S. {\it et al.}\  2008, \mnras, 383, 1485.

\bibitem[{van Dyk}, {Hamuy} \& {Filippenko}(1996)]{vhf96}
{van Dyk}, S.~D., {Hamuy}, M., and {Filippenko}, A.~V. 1996, \aj, 111, 2017.

\bibitem[{Vink} \& {de Koter}(2005)]{vd05}
{Vink}, J.~S. and {de Koter}, A. 2005, \aap, 442, 587.

\bibitem[{Weiler} {\it et al.}\ (1986)]{wsp+86}
{Weiler}, K.~W., {Sramek}, R.~A., {Panagia}, N., {van der Hulst}, J.~M., and
  {Salvati}, M. 1986, \apj, 301, 790.

\bibitem[{Wheeler} \& {Levreault}(1985)]{wl85}
{Wheeler}, J.~C. and {Levreault}, R. 1985, \apjl, 294, L17.

\bibitem[{Wong} {\it et al.}\ (2006)]{i8677}
{Wong}, D.~S., {Silverman}, J.~M., {Pooley}, D., and {Filippenko}, A.~V. 2006,
  \iaucirc, 8677, 3.

\bibitem[{Woosley} \& {Bloom}(2006)]{Summary}
{Woosley}, S.~E. and {Bloom}, J.~S. 2006, \araa, 44, 507.

\bibitem[{Woosley}, {Heger} \& {Weaver}(2002)]{whw02}
{Woosley}, S.~E., {Heger}, A., and {Weaver}, T.~A. 2002, Reviews of Modern
  Physics, 74, 1015.

\bibitem[{Woosley}, {Langer} \& {Weaver}(1995)]{wlw95}
{Woosley}, S.~E., {Langer}, N., and {Weaver}, T.~A. 1995, \apj, 448, 315.

\bibitem[{Yoon}, {Woosley} \& {Langer}(2010)]{ywl10}
{Yoon}, S., {Woosley}, S.~E., and {Langer}, N. 2010, ArXiv e-prints.

\end{thebibliography}

\clearpage

\clearpage

\clearpage

\clearpage

\clearpage

\clearpage

\clearpage

\clearpage

\clearpage

\clearpage

\clearpage

\clearpage

\clearpage

\clearpage

\clearpage

\clearpage

\clearpage

\clearpage

\clearpage

\clearpage

\clearpage

\clearpage

\clearpage

\begin{deluxetable}{llcccccc}
\setlength{\tabcolsep}{0.07in}
\tablecaption{Sample of SNe Ibc}
%\tablewidth{0pt}
\tablehead{
\colhead{SN}  & 
\colhead{Host Galaxy} & 
\colhead{Distance} &
\colhead{Inclination} & 
\colhead{$E(B-V)$} & 
\colhead{Discovery} &
\colhead{Spectral} & 
\colhead{Classification} \\
\colhead{}  & 
\colhead{} & 
\colhead{(Mpc)} & 
\colhead{(degrees)} & 
\colhead{Galactic (mag)\tablenotemark{1}} &
\colhead{Circular\tablenotemark{2}} &
\colhead{Type} & 
\colhead{Circular\tablenotemark{2}} 
}
\startdata
2004dk\tablenotemark{*} &  NGC 6118      & $23\pm 2$\tablenotemark{\dagger} & $67^{+0.7}_{-0.7}$ & 0.157 & I8377 & Ib & I8404  \\
2004dn\tablenotemark{*} &  UGC 2069      & $51\pm 4$ & $56^{+3}_{-3}$ & 0.048 & I8381 & Ic     & I8381  \\
2004fe\tablenotemark{*} &  NGC 132       & $72\pm 5$ & $43^{+4}_{-5}$ & 0.025 & I8425 & Ic     & I8426  \\
2004ff\tablenotemark{*} &  ESO 552-G40   & $92\pm 7$ & $66^{+4}_{-5}$ & 0.032 & I8425 & Ic     & I8428   \\
2004ge\tablenotemark{*} &  UGC 3555      & $67\pm 5$ & $28^{+12}_{-28}$ & 0.087 & I8443 & Ic     & I8453  \\
2004gk\tablenotemark{*} &  IC 3311       & $17\pm 1$\tablenotemark{\dagger,a} & $\sim 90$ & 0.030 & I8446 & Ic     & I8446    \\ 
2004gq\tablenotemark{*} &  NGC 1832      & $26\pm 6$\tablenotemark{\dagger} & $50^{+3}_{-4}$ & 0.073 & I8452 & Ib     & I8461 \\
2004gt\tablenotemark{*}$^{,3}$ &  NGC 4038      & $23\pm 2$ & $56^{+2}_{-2}$ & 0.046 & I8454 & Ic     & I8456  \\
2004gv\tablenotemark{*} &  NGC 856       & $79\pm 6$ & $46^{+6}_{-9}$ & 0.033 & I8454 & Ib     & I8456  \\
2005az\tablenotemark{*} &  NGC 4961      & $39\pm 4$\tablenotemark{\dagger} & $47^{+4}_{-5}$ & 0.011 & I8503 & Ib     & A451  \\
2005eo &  UGC 4132      & $85\pm 7$\tablenotemark{\dagger} & $80^{+4}_{-4}$ & 0.067 & I8605 & Ic     & I8605  \\
2005hg &  UGC 1394      & $86\pm 6$ & $60^{+3}_{-4}$ & 0.105 & I8623 & Ib     & C271   \\
2005kz &  MCG +08-34-32 & $115\pm 8$ & $60^{+5}_{-7}$ & 0.054 & I8639 & Ic-BL  & I8639   \\
2005la\tablenotemark{4} &  Anon.         & $76\pm 8$\tablenotemark{d} & \nodata & 0.011 & I8639 & IIb    & I8639  \\
2005mf &  UGC 4798      & $113\pm 8$ & $30^{+12}_{-30}$ & 0.018 & I8648 & Ic     & I8650  \\
2005nb &  UGC 7230      & $106\pm 7$ & $55^{+4}_{-5}$ & 0.037 & I8657 & Ic-BL  & I8657  \\
2006F  &  NGC 935       & $55\pm 4$ & $53^{+5}_{-7}$ & 0.190 & I8658 & Ib     & I8660  \\
2006ab &  PGC 10652     & $68\pm 5$ & $65^{+3}_{-4}$ & 0.489 & I8669 & Ic     & I8677  \\
2006ck &  UGC 8238      & $112\pm 11$\tablenotemark{\dagger} & $61^{+5}_{-7}$ & 0.028 & I8713 & Ic     & I8713  \\
2006dn &  UGC 12188     & $70\pm 7$\tablenotemark{d} & $40^{+6}_{-9}$ & 0.113 & I8728 & Ib     & A854  \\
2006el &  UGC 12188     & $70\pm 7$\tablenotemark{d} & $40^{+6}_{-9}$ & 0.113 & I8741 & IIb    & C626  \\
2006fo &  UGC 2019      & $82\pm 6$ & $34^{+12}_{-34}$ & 0.029 & I8570 & Ic     & I8770  \\
2006jc\tablenotemark{5} &  UGC 4904      & $28\pm 2$ & $59^{+4}_{-5}$ & 0.020 & I8762 & Ibn     & C677 \\
2007C  &  NGC 4981      & $25\pm 3$\tablenotemark{\dagger} & $43^{+3}_{-3}$ & 0.042 & I8792 & Ib     & I8792  \\
2007D  &  UGC 2653      & $93\pm 7$ & $70^{+4}_{-6}$ & 0.335 & I8794 & Ic-BL  & C805  \\
\label{tab:list}
\enddata
\tablenotetext{1}{Milky Way extinction estimates adopted from \citet{sfd+98}.}
\tablenotetext{2}{Entries refer to electronic circular numbers
  prefaced by ``I'' (IAUC), ``C'' (CBET), or ``A'' (ATEL) and the full references are listed below.}
\tablenotetext{3}{Previous photometry has been published by \citet{gfk+05}.}
\tablenotetext{4}{Previous photometry has been published by \citet{pqs+08}.}
\tablenotetext{5}{Previous photometry has been published by \citet{fsg+07,psm+07}.}
\tablenotetext{*}{P60 observations conducted under the Caltech Core-Collapse Supernova Program.}
\tablenotetext{\dagger}{Redshift independent distance.}
\tablenotetext{a}{Host galaxy is a member of the Virgo cluster. Distance is adopted from \citet{mhf+00}.}
\tablenotetext{c}{Distance is an average of the value reported in \citet{shr04} and the Tully-Fisher distance referenced by LEDA.}
\tablenotetext{d}{Distance is derived from the host galaxy redshift reported in ATEL 854.}
\noindent
\tablenotetext{---}{References: I8377=\citep{i8377}, I8404=\citep{i8404}, I8381=\citep{i8381}, I8425=\citep{i8425}, I8426=\citep{i8426}, I8428=\citep{i8428}, I8443=\citep{i8443}, I8453=\citep{i8453}, I8446=\citep{i8446}, I8452=\citep{i8452}, I8461=\citep{i8461}, I8454=\citep{i8454}, I8456=\citep{i8456}, I8503=\citep{i8503}, A451=\citep{A451}, I8605=\citep{i8605,i8605b}, I8623=\citep{i8623}, C271=\citep{c271}, I8639=\citep{i8639,i8639b}, I8648=\citep{i8648}, I8650=\citep{i8650}, I8657=\citep{i8657}, I8658=\citep{i8658}, I8660=\citep{i8660}, I8669=\citep{i8669}, I8677=\citep{i8677}, I8713=\citep{i8713}, I8728=\citep{i8728}, A854=\citep{A854}, I8741=\citep{i8741}, C626=\citep{c626}, I8570=\citep{i8750}, I8770=\citep{i8770}, I8762=\citep{i8762}, C677=\citep{c677}, I8792=\citep{i8792}, I8794=\citep{i8794}, C805=\citep{c805} 
}
\end{deluxetable}

\clearpage

\begin{deluxetable}{lccc}
\tablecaption{P60 Photometry}
\tablewidth{0pt}
\tablehead{
\colhead{SN}  & 
\colhead{Filter} & 
\colhead{JD} &
\colhead{magnitude} 
}
\startdata
SN\,2004dk & V       &       2453219.8  & 16.99$\pm$ 0.03 \\
\nodata    & \nodata &       2453220.7  & 16.69$\pm$ 0.03 \\
\nodata    & \nodata &       2453223.8  & 15.88$\pm$ 0.02 \\
\nodata    & \nodata &       2453233.7  & 14.74$\pm$ 0.02 \\
\nodata    & \nodata &       2453234.7  & 14.69$\pm$ 0.02 \\
\nodata    & \nodata &       2453237.8  & 14.55$\pm$ 0.02 \\
\nodata    & \nodata &       2453242.7  & 14.50$\pm$ 0.02 \\
\nodata    & \nodata &       2453252.6  & 14.72$\pm$ 0.02 \\
\nodata    & \nodata &       2453258.6  & 14.91$\pm$ 0.03 \\
\nodata    & \nodata &       2453261.6  & 15.01$\pm$ 0.02 \\
\nodata    & \nodata &       2453266.6  & 15.17$\pm$ 0.02 \\
\enddata
\label{tab:allphot}
\tablenotetext{---}{\bf Note: \rm Only a portion of this table is shown here to demonstrate its form and content.  A machine-readable version of the full table is available online.}
\end{deluxetable}

\clearpage

\begin{deluxetable}{llccccr}
\normalsize
\tablecaption{Literature SNe Ibc Sample}
\tablehead{
\colhead{SN}  & \colhead{Host} & \colhead{Distance} & \colhead{Inclination} &  \colhead{$E(B-V)$} & \colhead{Spectral} & \colhead{Ref.\tablenotemark{\dagger}} \\
\colhead{} &    \colhead{Galaxy}     & \colhead{(Mpc)}  &  \colhead{(degrees)}  & \colhead{Galactic (mag)\tablenotemark{1}} & \colhead{Type} & \colhead{} 
}
\startdata
1994I    &  NGC 5194      & $7.1\pm 1.2$\tablenotemark{a} &  $53^{+1}_{-1}$ & 0.035  &  Ic & 1 \\ 
1999ex   &  IC 5179       & $46\pm 4$\tablenotemark{\dagger}	 &  $64^{+2}_{-3}$ & 0.020  &  Ib &   2 \\ 
2002ap   &  NGC 628       & $8.2\pm 1.2$\tablenotemark{\dagger} &  $25^{+3}_{-3}$ & 0.071   &  Ic-BL &	 3 \\ 
2003jd   &  MCG -01-59-21 & $77\pm 5$			 &  $42^{+8}_{-14}$ & 0.044  & Ic-BL &  4 \\ 
2004aw   &  NGC 3997      & $73\pm 5$			 &  $71^{+2}_{-2}$ & 0.021   & Ic &    5 \\
2007Y    &  NGC 1187      & $18\pm 2$\tablenotemark{\dagger}  &  $43^{+3}_{-3}$ & 0.022  & Ib  &   6 \\
2007gr   &  NGC 1058      & $9.3\pm 0.7$\tablenotemark{b} &  $22^{+6}_{-10}$ & 0.062 &  Ic &    7 \\
2007ru   &  UGC 12381     & $64\pm 5$                   &  $31^{+10}_{-23}$ & 0.259  &   Ic-BL &  8 \\
2008D    &  NGC 2770      & $30\pm 3$\tablenotemark{\dagger} &  $77^{+1}_{-1}$ & 0.023  &  Ib &   9 \\
2008ax   &  NGC 4490      & $8.7\pm 1.2$\tablenotemark{\dagger} &  $63^{+1}_{-1}$ & 0.022  &  IIb &   10 \\
\enddata

\tablenotetext{\dagger}{References for photometry: 1=\citet{rvh+96}; 2=\citet{shs+02}; 3=\citet{fps+03}; 4=\citet{vbc+08}; 5=\citet{tpm+06}; 6=\citet{smp+09}; 7=\citet{hvk+09}; 8=\citet{sta+09}; 9=\citet{sbp+08}; 10=\citet{pkc+08}}
\tablenotetext{1}{Milky Way extinction estimates adopted from \citet{sfd+98}.}
\tablenotetext{\dagger}{Redshift independent distance.}
\tablenotetext{a}{Distance is from \citet{tk06}}
\tablenotetext{b}{Distance is from \citet{shm+96}}
\label{tab:templates}
\end{deluxetable}

\clearpage

\begin{deluxetable}{lclllccc}
\normalsize
\tablecaption{Light curve Properties of the Literature SNe Ibc}
\tablehead{
\colhead{SN\tablenotemark{\dagger}} &  \colhead{Band} & \colhead{Peak Time} & \colhead{Peak Apparent} & \colhead{$\Delta m_{15}$} & \colhead{$E(B-V)$} & \colhead{$(V-R)_{\rm 10~days}$} & \colhead{Peak Absolute} \\
\colhead{} & \colhead{} & \colhead{(JD)} & \colhead{Magnitude} & \colhead{(mag)} & \colhead{Host (mag)\tablenotemark{1}} & \colhead{(mag)\tablenotemark{2}} & \colhead{Magnitude}
}
\startdata
1994I    &  $V$ & 2449451.44$\pm$0.03 &  12.91$\pm$0.02 & 1.67$\pm$0.05 & 0.42 & 0.19   & $-17.94\pm 0.39$ \\ 
  & $R$ & 2449451.92$\pm$0.04 &  12.65$\pm$0.02 & 1.40$\pm$0.06 &  \nodata & 0.20 & $-17.65\pm 0.39$ \\
1999ex   &  $V$ & 2451501.2$\pm$0.5   &  16.63$\pm$0.04 & 0.96$\pm$0.10 & 0.26 & 0.33   & $-17.53\pm 0.23$\\
  & $R$ & 2451502.7$\pm$0.5   &  16.26$\pm$0.02 & 0.81$\pm$0.05 & \nodata & 0.37 & $-17.68\pm 0.21$  \\
2002ap   & $V$ & 2452313.42$\pm$0.16 &  12.360$\pm$0.006 & 0.90$\pm$0.04 & 0.0 & 0.31 & $-17.43\pm 0.32$ \\
  & $R$ & 2452315.74$\pm$0.18 &  12.267$\pm$0.003 & 0.69$\pm$0.06 & \nodata & 0.35 & $-17.50\pm 0.32$ \\
2003jd   &  $V$ & 2452944.1$\pm$1.0   &  15.94$\pm$0.18 & 0.97$\pm$0.33 & 0.10 & 0.24  & $-18.93\pm 0.33$ \\
  & $R$ & 2452944.7$\pm$1.0   &  15.82$\pm$0.15 & 0.74$\pm$0.25 & \nodata & 0.26 & $-18.94\pm 0.30$  \\
2004aw    & $V$ & 2453091.1$\pm$0.6   &  17.30$\pm$0.03 & 0.59$\pm$0.14 & 0.35 & 0.30  & $-18.16\pm 0.18$ \\
  & $R$ & 2453095.0$\pm$0.6   &  16.90$\pm$0.03 & 0.39$\pm$0.11 & \nodata & 0.44 & $-18.27\pm 0.18$  \\
2007Y    &   $V$ & 2454165.6$\pm$0.5   &  15.336$\pm$0.011 & 1.07$\pm$0.05 & 0.09 & 0.24  & $-16.29\pm 0.20$ \\
  & $R^*$ & 2454166.5$\pm$0.5   &  15.322$\pm$0.015 & 0.72$\pm$0.05 & \nodata & 0.25 & $-16.22\pm 0.21$ \\
2007gr   &  $V$ & 2454338.5$\pm$1.1   &  12.91$\pm$0.01 & 0.95$\pm$0.11 & 0.03 & 0.34  & $-17.22\pm 0.17$ \\
  & $R$ & 2454341.0$\pm$1.7   &  12.76$\pm$0.02 & 0.78$\pm$0.13 & \nodata & 0.31 & $-17.30\pm 0.18$  \\
2007ru   &   $V$ & 2454439.2$\pm$0.5   &  15.95$\pm$0.015 &  0.75$\pm$0.07 & 0.01 & 0.22 & $-18.92\pm 0.17$\\
  & $R$ & 2454439.2$\pm$0.5   &  15.68$\pm$0.019 &  0.51$\pm$0.07 & \nodata & 0.22 & $-18.99\pm 0.17$ \\
2008D    &   $V$ & 2454491.7$\pm$0.5   &  17.40$\pm$0.02 & 0.55$\pm$0.08 & 0.59 & 0.22 & $-16.91\pm 0.26$  \\
  & $R^*$ & 2454491.7$\pm$0.5   &  16.84$\pm$0.04 & 0.55$\pm$0.13 & \nodata & 0.30 & $-17.00\pm 0.28$ \\
2008ax   &   $V$ & 2454549.5$\pm$0.5   &  13.51$\pm$0.02 & 0.94$\pm$0.05 & 0.28 & 0.22   & $-17.11\pm 0.33$ \\
  & $R^*$ & 2454341.0$\pm$1.7   &  12.76$\pm$0.02 & 0.78$\pm$ 0.13 & \nodata & 0.23 & $-17.05\pm 0.35$  \\
\enddata
\tablenotetext{\dagger}{The total extinction estimates, peak times and peak apparent magnitudes in the $V-$ and $R-$bands were compiled from the following references: 1994I=\citet{rvh+96}; 1999ex=\citet{shs+02}; 2002ap=\citet{fps+03}; 2003jd=\citet{vbc+08}; 2004aw=\citet{tpm+06}; 2007Y=\citet{smp+09}; 2007gr=\citet{hvk+09}; 2007ru=\citet{sta+09}; 2008D=\citet{sbp+08}; 2008ax=\citet{pkc+08}.}
\tablenotetext{1}{In all cases, the host galaxy extinction was inferred using
a Milky Way extinction law.}
\tablenotetext{2}{The $(V-R)_{10}$ values are estimated from the observed light curves after correcting for the total line-of-sight extinction.}
\tablenotetext{*}{$R-$band magnitudes estimated from transformation of $r-$band data using the transformations of \citet{stk+02}.}

\label{tab:templates_lc}
\end{deluxetable}

\clearpage

\begin{deluxetable}{lcllllcc}
\normalsize
\tablecaption{Light curve Properties of the P60 Sample}
\tablehead{
\colhead{SN} & \colhead{Band} & \colhead{Peak Time} & \colhead{Peak Apparent} & \colhead{$\Delta m_{15}$} & \colhead{$E(B-V)$} & \colhead{Peak Absolute} & \colhead{Group}\\
\colhead{} & \colhead{} & \colhead{(JD)} & \colhead{Magnitude\tablenotemark{\dagger}} & \colhead{(mag)} &  \colhead{Host (mag)} & \colhead{Magnitude} & \colhead{}
}
\startdata
2004dk   & $V$    & 2453238.0$\pm$0.5    & 14.76$\pm$0.1 &  0.56$\pm$0.12 & 0.18$\pm$0.11 & $-18.10\pm 0.59$ & G \\
         & $R$    & 2453242.5$\pm$0.5    & 14.42$\pm$0.1 &  0.45$\pm$0.13 &\nodata  & $-18.18\pm 0.51$ & \nodata \\
2004dn   & $V$    & 2453230.5$\pm$0.5    & 17.35$\pm$0.1 &  0.76$\pm$0.15 & 0.52$\pm$0.13  & $-17.95\pm 0.66$ & G\\
  & $R$    & 2453232.0$\pm$0.5    & 16.92$\pm$0.1 &  0.61$\pm$0.14 & \nodata & $-17.94\pm 0.56$ & \nodata \\
2004fe   & $V$    & 2453318.0$\pm$0.5    & 16.91$\pm$0.1 &  0.87$\pm$0.16 & 0.29$\pm$0.11 & $-18.35\pm 0.59$ & G\\
  & $R$    & 2453320.5$\pm$0.5    & 16.66$\pm$0.1 &  0.80$\pm$0.15 & \nodata & $-18.36\pm 0.51$ & \nodata \\
2004ff   & $V$    & \nodata              & $<$17.57      &  \nodata       & 0.27$\pm$0.15  & $<$-18.02 & S \\
&   $R$    & 2453316.0$\pm$1.0    & 17.18$\pm$0.2 &  0.85$\pm$0.25 &\nodata  & $-18.35\pm 0.60$ & \nodata \\ 
2004ge   & $V$    & \nodata              & $<$17.69      &  \nodata       & 0.88$\pm$0.15 & $<$-19.21 & S \\
&   $R$    & 2453341.5$\pm$1.0    & 17.09$\pm$0.2 &  0.55$\pm$0.27 & \nodata & $-19.31\pm 0.60$ & \nodata \\
2004gk   & $V$    & \nodata         & $<$13.98       &  \nodata       & $>$0.030 & $<$-17.20 & B \\
 & $R$    & \nodata         & $<$13.62       &  \nodata       & \nodata & $<$-17.86 & \nodata \\
2004gq   & $V$    & 2453361.0$\pm$0.5    & 15.32$\pm$0.1 &  0.78$\pm$0.22 & 0.18$\pm$0.11 & $-17.55\pm 0.90$ & G \\
  & $R$    & 2453361.5$\pm$0.5    & 14.99$\pm$0.1 &  0.60$\pm$0.22 & \nodata & $-17.69\pm 0.82$ & \nodata\\
2004gt   & $V$    & \nodata	    & $<$15.77       &  \nodata       & $>$0.046 & $<$-16.26 & B \\
  & $R$    & \nodata	    & $<$15.34        &  \nodata       & \nodata& $<$-17.34 & \nodata \\
2004gv   & $V$    & \nodata	    & $<$18.33	  &  \nodata       & $>$0.033 & $<$-16.27 & B\\
  & $R$    & \nodata   	    & $<$17.86        &  \nodata       &\nodata  & $<$-17.17 & \nodata \\
2005az    & $V$    & \nodata              & $<$16.11      &  \nodata       & 0.43$\pm$0.11 & $<$-18.23 & S \\
& $R$    & 2453476.0$\pm$1.0    & 15.55$\pm$0.2 &  0.42$\pm$0.22 & \nodata & $-18.45\pm 0.56$ & \nodata \\
2005eo   & $V$    & \nodata              & $<$17.71      &  \nodata       & 0.57$\pm$0.20 & $<$-18.90 & S \\  
& $R$    & 2453648.0$\pm$1.0    & 17.25$\pm$0.2 &  0.71$\pm$0.25 & \nodata & $-18.87\pm 0.75$ & \nodata \\
2005hg   & $V$    & 2453684.0$\pm$0.5    & 17.44$\pm$0.1 &  0.79$\pm$0.15 & 0.58$\pm$0.13 & $-19.36\pm 0.65$ & G\\
  & $R$    & 2453686.0$\pm$0.5    & 16.80$\pm$0.1 &  0.62$\pm$0.15 &\nodata  & $-19.47\pm 0.55$ & \nodata\\
2005kz   & $V$    & \nodata              & $<$18.15      &  \nodata       & 0.46$\pm$0.13  & $<$-18.75 & S \\
& $R$    & 2453715.0$\pm$1.0    & 17.61$\pm$0.2 &  0.41$\pm$0.29 & \nodata & $-18.89\pm 0.56$ & \nodata \\
2005la   & $V$    & \nodata         & $<$18.00        & \nodata        & $>$0.011 & $<$-16.44 & B\\
  & $R$    & \nodata         & $<$17.64        & \nodata       &\nodata  & $<$-17.10 & \nodata \\
2005mf   & $V$    & 2453736.0$\pm$0.5    & 18.00$\pm$0.1 &  0.78$\pm$0.19 & 0.38$\pm$0.12 & $-18.51\pm 0.62$ & G \\
  & $R$    & 2453737.5$\pm$0.5    & 17.65$\pm$0.1 &  0.66$\pm$0.19 & \nodata & $-18.54\pm 0.53$ & \nodata \\
2005nb   & $V$    & \nodata         & $<$17.60       & \nodata &  $>$0.037 & $<$-17.63 & B\\
  & $R$    & \nodata         & $<$17.05       & \nodata &\nodata  & $<$-18.15 & \nodata \\
2006F    & $V$    & \nodata              & $<$17.70      &  \nodata       &0.54$\pm$0.17  & $<$-18.27 & S \\
& $R$    & 2453754.0$\pm$1.0    & 17.00$\pm$0.2 &  0.74$\pm$0.24 &\nodata  & $-18.41\pm 0.65$ & \nodata \\
2006ab   & $V$    & \nodata              & $<$18.55      &  \nodata       &0.27$\pm$0.18  & $<$-17.93 & S \\
& $R$    & 2453784.5$\pm$2.0    & 17.58$\pm$0.2 &  0.56$\pm$0.21 & \nodata  & $-18.32\pm 0.67$ & \nodata \\
2006ck   & $V$    & \nodata         & $<$18.29       & \nodata & $>$0.028 & $<$-17.04 & B \\
 & $R$    & \nodata         & $<$17.68       & \nodata & \nodata & $<$-17.91 & \nodata \\
2006dn   & $V$    & \nodata              & $<$17.56      &  \nodata       &0.43$\pm$0.29  & $<$-18.36 & S \\
& $R$    & 2453932.0$\pm$1.0    & 17.07$\pm$0.2 &  0.44$\pm$0.26 &\nodata  & $-18.43\pm 0.99$ & \nodata \\
2006el   & $V$    & 2453982.5$\pm$1.0    & 17.61$\pm$0.2 &  0.73$\pm$0.25 & 0.19$\pm$0.12 & $-17.56\pm 0.79$ & G\\
  & $R$    & 2453984.5$\pm$1.0    & 17.26$\pm$0.2 &  0.57$\pm$0.24 & \nodata & $-17.67\pm 0.70$ & \nodata \\
2006fo   & $V$    & \nodata	    & $<$17.07 	  & \nodata & $>$0.029 & $<$-17.70 & B\\
  & $R$    & \nodata 	    & $<$16.65	  & \nodata & \nodata & $<$-18.25 & \nodata \\
2006jc   & $V$    & \nodata 	    & $<$14.74       & \nodata & $>$0.020 & $<$-18.00 & B\\
  & $R$    & \nodata         & $<$14.70       & \nodata & & $<$-18.47 & \nodata \\
2007C    & $V$    & 2454116.5$\pm$0.5    & 16.07$\pm$0.1 &  0.94$\pm$0.14 & 0.64$\pm$0.12 & $-18.04\pm 0.70$ & G \\
  & $R$    & 2454118.0$\pm$0.5    & 15.49$\pm$0.1 &  0.72$\pm$0.14 & \nodata & $-18.10\pm 0.61$ & \nodata \\
2007D    & $V$    & \nodata              & $<$18.15      &  \nodata       & 0.91$\pm$0.13 & $<$-20.54 & S \\
& $R$    & 2454120.0$\pm$1.0    & 17.09$\pm$0.2 &  0.67$\pm$0.29 &\nodata  & $-20.65\pm 0.55$ & \nodata \\
  
\enddata
\label{tab:p60_lc}
\tablenotetext{\dagger}{Peak apparent magnitude errors are dominated by the uncertainty associated with template fitting for the epoch and magnitude at maximum.}
\end{deluxetable}

\clearpage

\begin{deluxetable}{lccc}
\tablecaption{Physical Parameters of the SN Ejecta}
\tablewidth{0pt}
\tablehead{
\colhead{SN}  & 
\colhead{$\tau_{c}$} & 
\colhead{$M_{\rm ej,\odot}E_{K,51}^{-1/4}$} &
\colhead{$M_{\rm Ni}$} \\
\colhead{} &
\colhead{(days)} &
\colhead{$(M_{\odot})^{3/4}(10^{51}~\rm erg)^{-1/4}$} &
\colhead{($M_{\odot}$)}
}
\startdata
{\rm 1994I}  & $4.8^{+0.3}_{-0.3}$ & $0.60^{+0.04}_{-0.04}$ & $0.06^{+0.03}_{-0.03}$ \\
{\rm 1999ex} & $10^{+1}_{-1}$ & $1.3^{+0.1}_{-0.1}$ & $0.1^{+0.02}_{-0.02}$ \\
{\rm 2002ap} & $12^{+1}_{-1}$ & $1.5^{+0.1}_{-0.1}$ & $0.09^{+0.03}_{-0.03}$ \\
{\rm 2003jd} & $11^{+5}_{-3}$ & $1.4^{+0.6}_{-0.4}$ & $0.34^{+0.11}_{-0.08}$ \\
{\rm 2004aw} & $19^{+4}_{-3}$ & $2.3^{+0.5}_{-0.4}$ & $0.27^{+0.05}_{-0.05}$ \\
{\rm 2004dk} & $17^{+4}_{-3}$ & $2.1^{0.5}_{-0.4}$ & $0.23^{+0.14}_{-0.09}$ \\
{\rm 2004dn} & $13^{+3}_{-2}$ & $1.6^{+0.4}_{-0.3}$ & $0.16^{+0.11}_{-0.06}$ \\
{\rm 2004fe} & $10^{+2}_{-2}$ & $1.3^{+0.3}_{-0.2}$ & $0.19^{+0.11}_{-0.07}$ \\
{\rm 2004ff} & $10^{+4}_{-3}$ & $1.2^{+0.5}_{-0.3}$ & $0.18^{+0.13}_{-0.08}$ \\
{\rm 2004ge} & $14^{+9}_{-4}$ & $1.8^{+1.1}_{-0.5}$ & $0.59^{+0.44}_{-0.25}$ \\
{\rm 2004gq} & $13^{+6}_{-3}$ & $1.7^{+0.7}_{-0.4}$ & $0.13^{+0.13}_{-0.07}$ \\
{\rm 2005az} & $18^{+9}_{-5}$ & $2.2^{+1.1}_{-0.6}$ & $0.31^{+0.21}_{-0.13}$ \\
{\rm 2005eo} & $11^{+5}_{-3}$ & $1.4^{+0.6}_{-0.4}$ & $0.33^{+0.33}_{-0.17}$ \\
{\rm 2005hg} & $13^{+3}_{-2}$ & $1.6^{+0.4}_{-0.3}$ & $0.64^{+0.41}_{-0.26}$ \\
{\rm 2005kz} & $18^{+12}_{-6}$ & $2.2^{+1.5}_{-0.8}$ & $0.47^{+0.31}_{-0.19}$ \\
{\rm 2006F} & $11^{+4}_{-3}$ & $1.4^{+0.6}_{-0.4}$ & $0.21^{+0.17}_{-0.10}$ \\
{\rm 2006ab} & $14^{+6}_{-3}$ & $1.8^{+0.8}_{-0.4}$ & $0.23^{+0.20}_{-0.11}$ \\
{\rm 2006dn} & $17^{+10}_{-5}$ & $2.1^{+1.3}_{-0.7}$ & $0.30^{+0.44}_{-0.17}$ \\
{\rm 2006el} & $14^{+7}_{-4}$ & $1.7^{+0.9}_{-0.5}$ & $0.13^{+0.11}_{-0.06}$ \\
{\rm 2007C} & $11^{+2}_{-2}$ & $1.4^{+0.3}_{-0.2}$ & $0.16^{+0.12}_{-0.07}$ \\
{\rm 2007D} & $12^{+7}_{-4}$ & $1.5^{+0.8}_{-0.5}$ & $1.5^{+0.5}_{-0.5}$  \\
{\rm 2007Y\tablenotemark{\dagger}} &  $11^{+1}_{-1}$ & $1.4^{+0.1}_{-0.1}$ & $\lsim 0.05$ \\
{\rm 2007gr} & $10^{+2}_{-2}$ & $1.2^{+0.2}_{-0.2}$ & $0.07^{+0.01}_{-0.01}$ \\
{\rm 2007ru} & $15^{+2}_{-1}$ & $1.9^{+0.2}_{-0.2}$ & $0.45^{+0.08}_{-0.07}$ \\
{\rm 2008D}  & $14^{+0.3}_{-0.2}$ & $1.8^{+0.4}_{-0.3}$ & $0.07^{+0.02}_{-0.02}$ \\
{\rm 2008ax} & $10^{+1}_{-1}$ & $1.3^{+0.1}_{-0.1}$ & $0.05^{0.02}_{-0.03}$
\label{tab:params}
\enddata
\tablenotetext{\dagger}{The observed properties for SN\,2007Y reside at the edge of our model parameter grid so our derived physical parameters are uncertain; we quote an upper limit for $M_{\rm Ni}$.}
\end{deluxetable}

\clearpage

\begin{deluxetable}{llccccccc}
\tablecaption{Sample Averages}
\tablewidth{0pt}
\tablehead{
\colhead{SN Type}  & 
\colhead{$M_{V_{\rm peak}}$} & 
\colhead{$M_{R_{\rm peak}}$} &
%\colhead{$E(B-V)$} &
\colhead{$M_{\rm Ni}$} &
\colhead{$\tau_c$} & 
\colhead{$M_{\rm ej}^{3/4}E_K^{-1/4}$} &
\colhead{$M_{\rm ej}$\tablenotemark{\dagger}} &
\colhead{$E_K$\tablenotemark{\dagger}} \\
\colhead{} &
\colhead{(mag)} &
\colhead{(mag)} &
%\colhead{(mag)} &
\colhead{($M_{\odot}$)} &
\colhead{(days)} &
\colhead{(($10^{51}~{\rm erg})^{-1/4}(M_{\odot})^{3/4}$)} &
\colhead{($M_{\odot}$)} &
\colhead{($10^{51}~\rm erg$)} \\
}
\startdata
{\rm SNe Ib} & $-17.6\pm 0.9$ & $-17.9\pm 0.9$ & $0.20\pm 0.16$ & $13\pm 3$ & $1.7\pm 0.3$ & $2.0^{+1.1}_{-0.8}$ & $1.2^{+0.7}_{-0.5}$\\
{\rm SNe Ic} & $-18.0\pm 0.5$ & $-18.3\pm 0.6$ & $0.24\pm 0.15$ & $12\pm 4$ & $1.5\pm 0.4$ & $1.7^{+1.4}_{-0.9}$ & $1.0^{+0.9}_{-0.5}$\\
{\rm SNe Ic-BL} & $-18.3\pm 0.8$ & $-19.0\pm 1.1$ & $0.58\pm 0.55$ & $14\pm 3$ & $1.7\pm 0.4$ & $4.7^{+2.3}_{-1.8}$ & $11^{+6}_{-4}$ \\
{\rm Engine-driven SNe} & $-18.9\pm 0.3$ & $-18.9\pm 0.4$ & $0.40\pm 0.18$ & $12\pm 3$ & $1.5\pm 0.3$ & $3.6^{+2.0}_{-1.6}$ & $9.0^{+5.0}_{-4.0}$ 
%{\rm SNe Ibc} & $-17.9\pm 0.8$ & $-18.2\pm 0.9$ & \nodata & \nodata & $13\pm 3$ & $1.6\pm 0.4$ & \nodata & \nodata
\label{tab:peak}
\enddata
\tablenotetext{\dagger}{Typical photospheric velocities of $v_{\rm ph}=10,000~\rm km~s^{-1}$ are assumed for SNe Ib and Ic and $v_{\rm ph}=20,000~\rm km~s^{-1}$ for SNe Ic-BL and engine-driven SNe.}
\end{deluxetable}

\end{document}